\documentstyle[emulateapj]{article}
\def\kms{km s$^{-1}$}
\def\zabs{$z_{\rm abs}$}
\def\lya{Ly$\alpha$\ }

\begin{document}

\submitted{Accepted for publication in {\it The Astrophysical Journal}}.

\title{The Relationship Between Galaxies and Low Redshift Weak Lyman 
$\alpha$ Absorbers in the Directions of H 1821+643 and PG 
1116+215\altaffilmark{1,}\altaffilmark{2}}

\altaffiltext{1}{Based on observations obtained with the WIYN 
Observatory, which is a joint facility of the University of Wisconsin, Indiana 
University, Yale University, and the National Optical Astronomy 
Observatories.}

\altaffiltext{2}{Based on observations with the NASA/ESA 
{\it Hubble Space Telescope}, obtained at the Space Telescope Science 
Institute, which is operated by the Association of Universities for Research in 
Astronomy, Inc., under NASA contract NAS 5-2555.}

\author{Todd M. Tripp\altaffilmark{3,}\altaffilmark{4}, Limin 
Lu\altaffilmark{5,}\altaffilmark{6}, and Blair D. 
Savage\altaffilmark{3}.}

\altaffiltext{3}{Department of Astronomy, University of Wisconsin 
- Madison, 475 N. Charter St., Madison, WI 53706 - 1582, 
Electronic mail: tripp@astro.wisc.edu, 
savage@madraf.astro.wisc.edu}

\altaffiltext{4}{Current address: Princeton University Observatory, 
Peyton Hall, Princeton, NJ 08544}

\altaffiltext{5}{Astronomy Department, 105-24, California Institute 
of Technology, Pasadena, CA 91125, Electronic mail: 
ll@astro.caltech.edu}

\altaffiltext{6}{Hubble Fellow}

\begin{abstract}
To study the nature of low $z$ \lya absorbers in the spectra of QSOs, we 
have obtained high signal-to-noise UV spectra of H 1821+643 ($z_{\rm 
em}$ = 0.297) and PG 1116+215 ($z_{\rm em}$ = 0.177) with the 
Goddard High Resolution Spectrograph on the {\it Hubble Space 
Telescope}. The spectra have minimum S/N $\sim$ 70-100 and 3$\sigma$ 
limiting equivalent widths of 50-75 m\AA\ at a resolution of $\sim$150 \kms 
. Excluding lines within 3000 \kms\ of $z_{\rm em}$, we detect 26 \lya lines 
with $W_{\rm r} >$ 50 m\AA\ toward H 1821+643 and 13 toward PG 
1116+215 (comparable to the 13 \lya lines observed toward 3C 273 by 
Morris et al.\markcite{mor93}), which implies a density of 102$\pm$16 
lines per unit redshift for $W_{\rm r} >$ 50 m\AA\ and \zabs\ $<$ 0.28. The 
two-point velocity correlation function shows marginal evidence of clustering 
of \lya lines on $\sim$500 \kms\ scales, but only if the weakest lines are 
excluded.

We have also used the WIYN Observatory to measure galaxy redshifts in the 
$\sim \ 1^{\circ}$ fields centered on each QSO in order to study the 
relationship between the \lya absorbers and galaxies. We find 17 
galaxy-absorber pairs within projected distances of 1 $h_{75}^{-1}$ Mpc 
with velocity separations of 350 \kms\ or less. Monte Carlo simulations show 
that if the \lya lines are randomly distributed, the probability of observing this 
many close pairs on the two sight lines is 3.6$\times 10^{-5}$. We find that 
{\it all} galaxies with projected distances $\rho \leq$ 600 $h_{75}^{-1}$ 
kpc from the QSO sight lines have associated \lya absorbers within 1000 
\kms , and the majority of these galaxies have absorbers within 350 \kms . 
We also find that the \lya equivalent width is anticorrelated with the 
projected distance of the nearest galaxy out to at least $\rho \approx$ 600 
$h_{75}^{-1}$ kpc. For $\rho >$ 600$h_{75}^{-1}$ kpc, we find galaxies 
which do not have associated \lya lines, but nevertheless the anticorrelation 
persists if we select galaxies with $\rho \lesssim$ 2 $h_{75}^{-1}$ Mpc 
which are within 500 or 1000 \kms\ of a \lya absorber. This anticorrelation 
has a high significance but should be interpreted cautiously because there are 
potential selection biases which could lead to an artificial correlation. 
Statistical tests also show that the \lya absorbers are not randomly distributed 
with respect to the galaxies. Splitting the sample into roughly equal sets with 
$W_{\rm r} >$ 100 m\AA\ and $W_{\rm r} <$ 100 m\AA\ shows that {\it 
the weakest absorbers are not randomly distributed} either. Comparison of 
the nearest neighbor distances of the weaker and stronger absorbers suggests 
that the weakest absorbers are less closely associated with galaxies, but the 
difference is not yet statistically significant. We find several galaxy groups 
which do not have clearly associated \lya absorbers. However, given the 
projected distance of the nearest galaxy, we do not necessarily expect to find 
detectable \lya lines in these groups based on the equivalent width-projected 
distance anticorrelation. Furthermore, we find several counterexamples of 
comparable galaxy groups which {\it do} have associated \lya lines. As in 
previous studies, we find some \lya absorbers in regions apparently devoid of 
galaxies, although this may be due to the limited spatial extent and/or limited 
depth of the redshift survey. The equivalent width distributions of the 
absorbers apparently in voids and non-void absorbers are statistically 
indistinguishable, but the sample is small. We discuss the nature of the \lya 
absorbers in light of the new data. The observations are consistent with the 
hypothesis that many of the low redshift \lya absorption lines with rest 
equivalent widths in the range from 50 to $\sim$500 m\AA\ trace the overall 
gas distributions in the large scale structures of galaxies rather than the 
gaseous halos of individual galaxies. Other phenomena may also cause \lya 
absorption lines.
\end{abstract}

\section{Introduction}

The ubiquitous \ion{H}{1} Lyman $\alpha$ lines detected in the spectra of 
QSOs trace an important gaseous component of the universe. Typically {\it 
hundreds} of \lya forest lines are detected between the \lya and Ly$\beta$ 
QSO emission lines in a high resolution spectrum of a {\it single} high $z$ 
QSO (e.g., Hu et al.\markcite{hu95} 1995; Lu et al.\markcite{lu96} 1996; 
Kirkman \& Tytler\markcite{kt97} 1997), and recent computations suggest 
that these \ion{H}{1} absorbers contain a substantial fraction of the baryons 
in the universe at high and low redshifts (e.g., Shull et al.\markcite{ssp96} 
1996; Miralda-Escud\'{e} et al.\markcite{mira96} 1996; Rauch et 
al.\markcite{rau97} 1997; Weinberg et al.\markcite{wein97} 1997; Bi \& 
Davidsen\markcite{bidad97} 1997). Cosmological simulations of structure 
formation suggest that at high redshift, the \lya absorption lines trace the gas 
in large scale structures in between collapsed objects (Cen et 
al.\markcite{cen94} 1994; Petitjean, M\"{u}cket, \& 
Kates\markcite{peti95} 1995; Miralda-Escud\'{e} et al.\markcite{mira96} 
1996; Hernquist et al.\markcite{hern96} 1996; Zhang et al. 
1997\markcite{zh97},1998\markcite{zh98}). The \lya forest lines therefore 
provide a critical constraint on cosmological models as well as important 
clues about the formation and evolution of galaxies and large scale structures 
(i.e., voids, galaxy clusters, and galaxy superclusters). However, to make 
progress on this important topic and take advantage of the \lya lines as a 
sensitive probe of cosmic phenomena from $z$ = 0 to $z$ = 5, a better 
understanding of the {\it nature} of the absorbers is needed.

The advent of the {\it Hubble Space Telescope (HST)} has provided 
important new insights about the \lya absorbers and an opportunity to 
directly study the relationship between the \lya absorption lines and galaxies. 
Spectroscopy of 3C 273 with {\it HST} in the first cycle of observations 
revealed that there are considerably more Ly$\alpha$ lines at low redshift 
than expected based on a simple extrapolation of the observed evolution of 
the number of lines per unit redshift ($dN/dz$) at high $z$ (Morris et 
al.\markcite{mor91} 1991; Bahcall et al.\markcite{bah91} 1991). 
Subsequently, Morris et al.\markcite{mor93} (1993) carried out a redshift 
survey of galaxies brighter than $B \ \sim$ 19 within $\sim$1$^{\circ}$ of 
3C 273 to study the relationship of the absorbers with galaxies, and they 
found no galaxies within a projected distance of 230 $h_{80}^{-1}$ kpc of 
any of the thirteen lines toward 3C 273 with rest frame equivalent width 
$W_{\rm r}$ exceeding 50 m\AA . In fact, one \lya line in the Morris et 
al.\markcite{mor93} (1993) sample turned out to be 4.8 $h_{80}^{-1}$ 
Mpc (projected) from the nearest bright galaxy. However, based on various 
statistical tests, Morris et al.\markcite{mor93} (1993) conclude that the 
absorbers are {\it not} randomly distributed with respect to the galaxies 
though the absorber-galaxy correlation is not as strong as the galaxy-galaxy 
correlation. In a similar program, Stocke et al.\markcite{stoc95} (1995) and 
Shull, Stocke, \& Penton\markcite{ssp96} (1996) used the {\it HST} to 
search for \lya lines in the CfA and Arecibo redshift survey regions, and they 
find no galaxies within 450 $h_{75}^{-1}$ kpc of the 10-11 \lya absorbers 
they discovered, and 3 or 4 of their \lya lines are found in galaxy voids (the 
fourth is a tentative detection). Overall, however, the majority of the Stocke 
et al.\markcite{stoc95} and Shull et al.\markcite{ssp96} \lya absorbers are 
located within large scale galaxy structures supporting the statistical 
conclusions of Morris et al.\markcite{mor93} (1993).

In striking contrast to the results of Morris et al.\markcite{mor93} and 
Stocke\markcite{stoc95} et al., Lanzetta et al.\markcite{lanz95} (1995) 
have measured the redshifts of galaxies within 1$\farcm$3 of six QSOs 
observed for the {\it HST} QSO Absorption Line Key Project (Bahcall et 
al.\markcite{bah93} 1993), and they find that between 32($\pm$10)\% and 
60($\pm$19)\% of the \lya lines in their sample are associated with luminous 
galaxies within $\sim$160 $h_{100}^{-1}$ kpc of the QSO sight lines, 
which suggests that a significant fraction of the clouds arise in large gaseous 
halos of intervening galaxies. Furthermore, Lanzetta et al.\markcite{lanz95} 
report that the \lya line equivalent widths are anticorrelated with the 
intervening galaxy's projected distance from the QSO sight line. These 
conclusions appear to be in conflict with the findings of Morris et 
al.\markcite{mor93} (1993) and Stocke et al.\markcite{stoc95} (1995). 
However, it may be possible to reconcile these discordant results. The 
Lanzetta et al.\markcite{lanz95} sample is sensitive only to {\it strong} \lya 
lines with $W_{\rm r} \ >$ 300 m\AA\ while the samples of Morris et 
al.\markcite{mor93} (1993) and Stocke et al.\markcite{stoc95} (1995) are 
dominated by much weaker lines with $W_{\rm r} \ <$ 200 m\AA . Based 
on this line strength difference it has been suggested that there are two 
populations of \lya absorbers at low redshift: (1) strong \lya absorbers with 
$N$(\ion{H}{1}) $\gtrsim \ 10^{14}$ cm$^{-2}$ (corresponding to 
$W_{\rm r} >$ 300 m\AA\ for Doppler parameter $b \approx$ 35 \kms ) 
that mostly occur in large gaseous halos of luminous galaxies, and (2) 
weaker \lya absorbers with $N$(\ion{H}{1}) $\lesssim \ 5 \times 10^{13}$ 
cm$^{-2}$ which are less closely tied to galaxies and may in some cases be 
truly intergalactic (primordial?) gas clouds. Currently this suggestion cannot 
be rigorously tested, however, because the sample of weaker \lya lines is 
rather small; there are no more than 28 weak \lya lines in the combined 
samples of Morris et al.\markcite{mor93} (1993), Stocke et 
al.\markcite{stoc95} (1995), and Shull et al.\markcite{ssp96} (1996). 
Recently, Grogin \& Geller\markcite{gg98} (1998) have reanalyzed the 
Morris et al.\markcite{mor93} and Stocke/Shull et al. data with additional 
redshifts for 3C 273, and they conclude that the \lya absorbers on these sight 
lines are randomly distributed with respect to galaxies. Evidently the nature 
of the low z \lya absorbers is still an open question.

To elucidate the nature of low redshift {\it weak} Ly$\alpha$ absorbers, we 
initiated a program in 1995 to study the relationship between low $z$ 
Ly$\alpha$ absorption line systems and galaxies in the directions of three 
QSOs using the {\it HST} Goddard High Resolution Spectrograph (GHRS), 
the Space Telescope Imaging Spectrograph (STIS), and the WIYN 
multiobject fiber-fed spectrograph (HYDRA). To significantly improve the 
size of the sample of weak \lya lines, we designed the observations to obtain 
spectra adequate for detection of Ly$\alpha$ lines with $W_{\lambda}$ = 
50 m\AA\ at the 3$\sigma$ level. The WIYN HYDRA is used to measure 
the redshifts of bright galaxies in $\sim$1$^{\circ}$ fields centered on the 
QSOs. In this paper we present a study of the relationship between \lya 
clouds and galaxies in the directions of H 1821+643 ($z_{\rm em}$ = 
0.297) and PG 1116+215 ($z_{\rm em}$ = 0.177). These QSOs have been 
observed previously with the {\it HST} Faint Object Spectrograph (FOS, 
Bahcall et al.\markcite{bah92} 1992,1993\markcite{bah93}; Jannuzi et 
al.\markcite{jan98} 1998), but the FOS spectra are not adequate for 
detecting {\it weak} \lya lines; their 4.5$\sigma$ limiting equivalent widths 
are generally greater than 250 m\AA\ throughout most of the \lya cloud 
region. The third sight line for our program will be observed with STIS in 
Cycle 7.

The paper is organized as follows. In \S 2 we review the GHRS observations 
and data reduction and present the UV spectra. We also summarize the 
WIYN observations and data reduction in this section. We discuss the 
absorption line selection and identification in \S 3. The properties of the \lya 
clouds are examined in \S 4, and the relationship between the \lya absorbers 
and galaxies is analyzed in \S5. A discussion of the results and the nature of 
the \lya absorbers is given in \S 6. An overall summary of our investigation is 
provided in \S 7. Further details on this project can be found in 
Tripp\markcite{tri97} (1997). Throughout the rest of this paper we assume 
$H_{0}$ = 75 km s$^{-1}$ Mpc$^{-1}$ and $q_{0}$ = 0, and we rescale 
measurements from the literature to these values for the cosmological 
parameters.

\section{Observations and Data Reduction}

\subsection{GHRS Spectroscopy - H 1821+643}

H 1821+643 was observed for 5.1 hours with the GHRS on March 29, 
1996. Standard target acquisition procedures were used to center the target in 
the large science aperture (1.74''$\times$1.74''), and the QSO was observed 
with the G140L grating. These observations were obtained after the 
installation of COSTAR,\footnote{Corrective Optics Space Telescope Axial 
Replacement} and the spectra have resolution $R = \lambda /\Delta \lambda 
\approx$ 2000 with a spectral spread function approximately described by a 
narrow Gaussian profile containing $~\sim$70\% of the line profile area and 
a broader profile containing $\sim$30\% of the area (see Figure 4 in 
Robinson et al.\markcite{robin98} 1998). In velocity units the resolution 
(FWHM) ranges from 160 \kms\ at 1250 \AA\ to 130 \kms\ at 1540 \AA . 
Half-diode substepping was employed, which samples the GHRS spectral 
line spread function with $\sim$1.2 diodes/FWHM. With this substepping 
pattern (step pattern 4), $\sim$11\% of the observing time is used for 
measuring the background with the science diodes. For accurate wavelength 
calibration, a Pt-Ne lamp was observed briefly at 1.7 hour intervals. The 
grating was positioned to provide complete wavelength coverage from 
1252--1528 \AA . Combined with the earlier intermediate resolution (15-20 
\kms ) GHRS observations obtained by Savage et al.\markcite{ssl95} 
(1995), which included the 1231.7--1268.9 and 1521.2--1557.5 \AA\ 
wavelength ranges, this allowed us to search for all \lya clouds between 
\zabs\ = 0.013 and \zabs\ = 0.281. Since the QSO systemic redshift is 
$z_{\rm em}$ = 0.297, these spectra cover all but a small fraction of the 
\lya\ forest on this sight line. The FOS data obtained by Bahcall et 
al.\markcite{bah92} (1992) can be used to fill in the missing redshift range, 
but their spectrum is not adequate for detection of the \lya lines with 
$W_{\lambda} \ll$ 250 m\AA . Bahcall et al.\markcite{bah92} (1992) do 
not report detection of any \lya lines at \zabs\ $<$ 0.013, but they do detect a 
strong \lya line at \zabs\ = 0.297, and a strong Ly$\beta$ line is present at 
this redshift in our GHRS spectrum.

The signal-to-noise ratio of high quality GHRS observations is ultimately 
limited by fixed pattern noise (FPN) due to scratches and manufacturing 
marks on the detector window and response irregularities of the 
photocathode. The detector diodes also introduce fixed pattern noise due to 
diode-to-diode sensitivity differences, but this noise is effectively reduced by 
comb addition and is generally weaker than the FPN due to the photocathode 
and faceplate. To reduce the fixed pattern noise, comb addition and the ``FP-
SPLIT'' procedure were employed for the H 1821+643 observations. The FP-
SPLIT procedure breaks an observation into subexposures and tilts the 
grating slightly between subexposures so that the position of the spectrum on 
the photocathode is shifted. When the subexposures are coadded {\it in 
wavelength space}, the FPN is smeared out and hence reduced. Typically 
subexposures at four different grating tilts are obtained when the FP-SPLIT 
procedure is used, and S/N $\approx$ 150 can be achieved when the 
subexposures are coadded (Cardelli \& Ebbets\markcite{ce94} 1994). To 
obtain higher S/N, the fixed pattern noise spectrum (i.e., the flat field) must 
be derived from the FP-SPLIT subexposures and divided into the data. To 
prevent loss or degradation of data in case of {\it HST} glitches or orbital 
interrupts (c.f. Cardelli \& Ebbets\markcite{ce94}), sets of four FP-SPLIT 
subexposures were read out roughly every four minutes, and a large number 
of subexposures at each of the four FP-SPLIT positions were obtained over 
the course of five hours.

The preliminary data reduction was carried out at the GHRS 
computing facility at the Goddard Space Flight Center with the 
standard CALHRS software. This included spectrum extraction and 
conversion to count rates, corrections for paired pulse events, 
background subtraction, and wavelength calibration. Software 
developed at the University of Wisconsin was then used to merge 
the individual subexposures and explore the usefulness of the fixed 
pattern noise flatfielding. Due to non-repeatability of the grating 
carousel rotation, slight positioning errors occur every time the 
grating is rotated, and degradation of resolution occurs if it is 
assumed that the grating was rotated by exactly the instructed amount 
for merging of subexposures. To achieve the highest possible 
resolution, the centroids of well-detected and narrow absorption lines in the 
individual FP-SPLIT subexposures are calculated and used to 
determine the shifts for merging. Finally, the zero 
point in the wavelength scale was determined by comparing the 
wavelengths of Galactic interstellar absorption lines in the G140L 
spectrum to their known wavelengths derived from the high 
resolution interstellar absorption line study of Savage et al.\markcite{ssl95} 
(1995)\footnote{In some cases the Milky Way interstellar lines are 
significantly blended in our low resolution spectrum but are resolved 
in the higher resolution spectrum. These cases are not suitable for 
determining the $\lambda$ scale zero point and were not used for 
this purpose.}. Savage et al.\markcite{ssl95} also measured accurate 
wavelengths of the \ion{H}{1} Ly$\beta$ line at \zabs\ = 0.22489 and the 
\ion{O}{6} 1031.93 \AA\ line at \zabs\ = 0.22503 (see also Savage, Tripp, 
\& Lu\markcite{stl98} 1998), so these lines were also used to determine the 
mean wavelength zero point. We estimate that the H 1821+643 G140L 
observations have an absolute velocity uncertainty of $\sim$30 \kms , which 
corresponds to a redshift uncertainty of $\sim$0.0001.

The H 1821+643 G140L observation is close to the S/N regime where 
application of a FPN correction may be beneficial (see \S 3). After merging 
the data into four ``master'' FP-SPLIT subexposures (i.e., all FP-SPLIT 
subexposures at the same position on the photocathode combined) we used 
the method of Cardelli \& Ebbets\markcite{ce94} (1994) to derive and apply 
the FPN correction. The final FPN corrected spectrum of H 1821+643 is 
shown in Figure ~\ref{h1821grand} where the observed count rate is plotted 
against heliocentric wavelength (in \AA ). Note that at $\lambda \ <$ 1251.6 
\AA\ and at $\lambda \ >$ 1537.9 \AA , the spectrum falls off the end of the 
diode array in some of the FP-SPLIT subexposures and consequently the S/N 
is lower in these regions. Since these spectral regions have already been 
observed by Savage et al.\markcite{ssl95} (1995) at higher resolution, their 
only purpose in the G140L spectrum is to provide better constraints on the 
continuum placement. This is important at the short $\lambda$ end of the 
G140L spectrum where the continuum rises (see Figure ~\ref{h1821grand}); 
without the spectrum extention a significantly different continuum fit could 
be selected. The final S/N ranges from 100 to 160 per resolution element. 
The combined S/N and resolution are adequate for detection of 50 m\AA\ 
lines at the 3$\sigma$ level or better throughout the G140L spectrum.

\subsection{GHRS Spectroscopy - PG 1116+215}

PG 1116+215 was observed for 3.5 hours with the GHRS on 4 February 
1997. The observational setup and data reduction procedure were nearly 
identical to the H 1821+643 observation and reductions, but 
quarter-diode substepping (step pattern 5) was used to provide more 
on-target exposure time and less (but still ample) time spent measuring the 
background. After the PG 1116+215 observations were fully reduced 
(including the FPN correction), the data were binned to half-diode samples to 
improve the S/N and match the H 1821+643 data. The final FPN corrected 
and half-diode binned spectrum of PG 1116+215 is plotted in Figure 
~\ref{p1116grand}. While H 1821+643 was actually brighter than expected 
when we observed it with the GHRS G140L grating, PG 1116+215 was 
apparently fainter than expected based on previous {\it IUE} and {\it HST} 
measurements. Consequently, the count rate was lower than expected. The 
S/N ($\sim$ 75-160 per resolution element) of the PG 1116+215 GHRS 
G140L spectrum is adequate for detection of 75 m\AA\ lines at the 
3$\sigma$ level or better throughout the entire spectrum and at the 50 m\AA\ 
level for $\lambda >$ 1400 \AA . A special zero point velocity correction 
was not applied to the PG 1116+215 observations. The weak interstellar 
lines of \ion{S}{2} and \ion{Ni}{2} (see \S 3.1.2) have average heliocentric 
velocities of $\sim$17 \kms\ while \ion{H}{1} 21 cm emission in the 
direction of PG 1116+215 has an average heliocentric velocity of -20 \kms\ 
(Lockman \& Savage\markcite{loc95} 1995). Some of this difference could 
be due to the pronounced positive velocity wing of \ion{H}{1} emission seen 
in the 21 cm profile. We estimate that the PG 1116+215 G140L spectrum 
has an absolute velocity uncertainty of $\sim$50 \kms , which corresponds to 
a redshift uncertainty of $\sim$0.0002.

\subsection{WIYN Galaxy Redshift Measurements}

To study the relationship between the \lya absorbers and galaxies in the 
directions of H 1821+643 and PG 1116+215, we have conducted a survey 
of galaxy redshifts in the $\sim 1^{\circ}$ fields centered on the QSOs using 
the fiber-fed multiobject spectrograph (HYDRA) on the WIYN telescope. 
The galaxy targets were selected primarily from  (1) eleven 6\farcm 7 
$\times$ 6\farcm 7 images in Harris R clustered around H 1821+643 
acquired with the WIYN imager in July 1995, and (2) 1$^{\circ} \times 
1^{\circ}$ images obtained with the KPNO Schmidt telescope by B. Jannuzi 
\& R. Green (1995, private communication). For H 1821+643, a few 
additional targets from the POSS II database were observed to extend the 
angular coverage at the lowest redshifts. Objects in the WIYN and Schmidt 
images were selected and classified as stars or galaxies by B. Jannuzi and A. 
Tanner using FOCAS (Tyson \& Jarvis\markcite{tys79} 1979; 
Valdes\markcite{val82} 1982). Our original objective was to measure the 
redshifts of all galaxies brighter than $B \approx$ 19 in the 1$^{\circ}$ 
fields centered on the QSOs (estimates of completeness of the final sample 
are provided below). However, a galaxy redshift sample complete to $B 
\approx$ 19 will not provide many redshifts with $z \ \gtrsim$ 0.2. Since the 
H 1821+643 sight line probes \lya clouds out to \zabs\ = 0.297, we obtained 
the WIYN images for a deeper redshift survey near the center of the QSO 
field.

A full description of the WIYN observations and data reduction will be
provided in a separate paper (Tripp et al.\markcite{tri98} 1998); here we 
provide a summary. The observations were carried out on several observing 
runs between August 1995 and April 1997. For all runs we used the 
Simmons camera with the 400 line mm$^{-1}$ ``blue'' grating and a 
Tektronics 2048$\times$2048 CCD (T2KC). For most of the runs we used 
the red fibers which subtend 2'' on the sky. However, on one of the early runs 
we used the blue fibers which subtend 3''. This setup provided a resolution of 
4.6 \AA\ ($\sim$300 km s$^{-1}$) with the red fibers or 7.1 \AA\ ($\sim$ 
450 km s$^{-1}$) with the blue fibers. To include the \ion{Ca}{2} H \& K 
lines at $z$ = 0, the grating was positioned to provide spectra extending from 
3900 to 7100 \AA . We typically observed 60-75 candidate galaxies per 
setup with 10-30 fibers positioned on the sky. Three 50 minute exposures 
were obtained for each fiber configuration. Targets were given priority based 
on brightness and proximity to the QSO. Due to fiber positioning constraints, 
in some setups not all of the fibers could be placed on new targets, so 
previously observed galaxies were reobserved to improve the S/N. Spectra of 
objects observed in multiple setups were coadded to obtain the final redshift, 
but the individual spectra were also measured to test the reliability and 
uncertainties of the redshift measurements (see Tripp et 
al.\markcite{tri98}1998).

The data were reduced with IRAF\footnote{IRAF is distributed by the 
National Optical Astronomy Observatories, which are operated by AURA, 
Inc., under contract to the NSF.} using the DOHYDRA package. We then 
used the standard method of Tonry \& Davis\markcite{td79} (1979) to 
determine the galaxy redshifts by cross-correlating the target spectra with 
spectra of ``templates'', i.e., high S/N galaxy spectra with known redshifts, 
after masking out any emission lines or night sky residuals and passing both 
the target and the templates through a Fourier bandpass filter. We used six 
templates for the cross-correlation measurements. The first template is a 
spectrum of M32 observed with HYDRA with one of the central fibers in the 
slit (which have the highest throughput) and reduced in an identical fashion 
to the target data reductions. The heliocentric velocity of M32 is 
--209 \kms\ (de Vaucouleurs et al.\markcite{dv91} 1991). The other 
templates are actual targets in the H 1821+643 and PG 1116+215 fields for 
which we obtained high S/N spectra, and the heliocentric velocities of these 
templates were established by cross-correlation with the M32 spectrum. The 
templates were intentionally selected to have a range of redshifts to check the 
reliability of the cross-correlation redshifts in marginal cases. For emission 
line galaxies, the redshifts were measured using templates with absorption 
lines and night sky residuals masked out.

The galaxy redshifts measured in the fields of H 1821+643 and PG 
1116+215 are summarized in Tables ~\ref{h1821reds} and 
~\ref{p1116reds}, respectively. We also list in these tables the POSS II $J$ 
magnitudes (denoted $B_{J}$ since $J$ is roughly equivalent to $B$) and 
impact parameters (projected distances from the sight line) for each galaxy. 
The POSS II magnitudes have rms errors of $\sim$0.21 magnitudes (Weir, 
Djorgovski, \& Fayyad\markcite{weir95} 1995). Since CCD calibration of 
the POSS II magnitudes was not available for these fields, we determined the 
magnitude zero-point by comparison to the measurements of Kirhakos et 
al.\markcite{kir94} (1994). Absolute magnitudes calculated using the 
interstellar extinction corrections based on $E(B-V)$ from Lockman \& 
Savage\markcite{loc95} (1995) and the K-correction from 
Peebles\markcite{peeb93} (1993) are also provided in Tables 
~\ref{h1821reds} and ~\ref{p1116reds}. We have measured 98 redshifts in 
the field of H 1821+643 and 118 redshifts in the field of PG 1116+215. For 
H 1821+643, we have obtained 56 additional redshifts from Schneider et 
al.\markcite{sch92} (1992), Le Brun, Bergeron, \& 
Boiss\'{e}\markcite{leb96} (1996), and Bowen, Pettini, \& 
Boyle\markcite{bpb98} (1998), bringing the grand total to 154. These 
additional redshifts and the other quantities when known are listed at the end 
of Table ~\ref{h1821reds}. The listed redshifts are on a heliocentric basis. In 
the direction of H 1821+643, $z$(Galactocentric) - $z$(heliocentric) = 
0.0009, while toward PG 1116+215, $z$(Galactocentric) - $z$(heliocentric) 
= -0.0003.

Table ~\ref{complete1116} provides estimates of the completeness of the PG 
1116+215 galaxy redshift survey for various limiting magnitudes out to 
various radii based on the number of galaxy candidates in the Schmidt 
catalog and the number of redshifts actually obtained. It will be important to 
consider the impact of the angular coverage of the galaxy redshift survey (at 
low redshifts) and its limited depth (at high redshifts), so we also list in Table 
~\ref{complete1116} the redshifts at which $B_{\rm J}$ = 17, 18, 19, and 
20 correspond to a 1$L^{*}$ galaxy [using $M_{B}^{*}$ = --19.5 from 
Loveday et al.\markcite{love92} (1992)] and the physical scale 
corresponding to a 10', 20', and 30' radius at each of these redshifts. For 
example, based on the Schmidt catalog of targets within 20' of the QSO, we 
estimate that the PG 1116+215 survey is 86.9\% complete for $B_{\rm J} 
<$ 19.0. $B_{\rm J}$ = 19 corresponds to a 1$L^{*}$ galaxy, and 20' 
corresponds to 2.4 Mpc, at $z \ \approx$ 0.121. The PG 1116+215 target list 
contained very little stellar contamination; only six observed targets turned 
out to be stars. Unfortunately, the H 1821+643 catalog suffered much more 
severe stellar contamination; 64 observed candidates turned out to be stellar 
(see Tripp et al.\markcite{tri98} 1998). This stellar contamination was 
partly due to the low latitude and direction of the QSO ($l = 94\fdg 0, \ b = 
+27\fdg$4) which causes the sight line to pass over several spiral arms and 
through the warped part of the outer Milky Way (see Savage et 
al.\markcite{ssl95} 1995), but it was also partly due to the fact that we 
observed the H 1821+643 field first --- experience gained from the H 
1821+643 field was used to improve the star/galaxy separation procedure for 
PG 1116+215. Due to the large number of stars in the target list, estimates of 
the H 1821+643 survey completeness are more uncertain. If we remove 
objects which are observed to be stars and objects for which we did not get 
good spectra but are probably stars (based on magnitude and POSS II 
classification), then we estimate that within a 20' radius from the QSO, the H 
1821+643 redshift survey is roughly 72.4\% complete for $B_{\rm J} <$ 
18.0 and 50.8\% complete for $B_{\rm J} <$ 20.0. Our 
survey is substantially more complete than the previous surveys noted above 
because any redshifts from those surveys which we missed were added to our 
sample.

Table ~\ref{h1821reds} shows that redshifts have been obtained for a 
number of galaxies as faint as $B_{\rm J} \ \approx$ 21.0 in the H 
1821+643 field. Taking $M_{B}^{*}$ = --19.5 and the interstellar 
extinction and K-corrections described above, we find that $B_{\rm J} \ 
\approx$ 21.0 corresponds to a roughly 0.5$L^{*}$ galaxy at $z$ = 0.20. 
Therefore we have not measured redshifts for galaxies fainter than 
0.5$L^{*}$ beyond $z \sim$ 0.2. Furthermore, most of the galaxies at $z >$ 
0.2 for which we have obtained redshifts are in the rich galaxy cluster which 
hosts the QSO, so our redshift survey is severely incomplete for intervening 
galaxies at $z >$ 0.2. Consequently, in our analysis of the relationship 
between \lya absorbers and galaxies in \S 5, we shall consider a sample which 
only includes galaxies and absorbers with $z <$ 0.2.

\section{Absorption Line Selection and Identification}

Absorption lines in the H 1821+643 G140L spectrum were selected for 
further analysis based on the statistical significance of the observed 
equivalent width $W_{\lambda}$. The automated objective procedure 
described in Tripp et al.\markcite{tri96} (1996) was used to measure the 
equivalent widths. The errors in the equivalent widths account for continuum 
placement uncertainty as well as statistical noise. For the lines detected at the 
$\sim 3\sigma$ level\footnote{The significance of the $\sim 3\sigma$ lines 
typically increases to 4-5$\sigma$ if we only consider statistical noise.}, the 
reality of the feature was checked by visually inspecting the four FP-SPLIT 
subexposures; if the line is apparent in at least three of the subexposures, then 
the detection is probably reliable and the line was included in the line list. We 
have measured all of the equivalent widths with and without the fixed pattern 
noise correction, and for the most part, the FPN correction did not have a 
significant impact on the equivalent width measurements. However, for a few 
of the weakest lines the correction was important. Figure ~\ref{fpndemo} 
shows an example of a 3$\sigma$ absorption line which is completely 
removed by the FPN correction. Consequently, we elected to apply the 
correction to all lines in the final sample. Table ~\ref{h1821list} lists the 
observed equivalent widths of all reliably detected lines in the H 1821+643 
G140L spectrum. Table ~\ref{h1821list} also lists the heliocentric vacuum 
wavelength of the centroid of each line, the line identification, the redshift, 
and, for ISM lines, the heliocentric velocity of the line. For line identification 
and throughout this paper, we use wavelengths from 
Morton\markcite{mort91} (1991). Profile fitting was employed to extract 
the redshifts of the individual features using VPFIT (developed by J. Webb 
and R. Carswell, see Carswell et al.\markcite{cars91} 1991) and to infer the 
presence of additional components in some complex absorption blends (see 
below). Table ~\ref{p1116list} provides the same information for lines 
detected in the spectrum of PG 1116+215. The redshift uncertainties from 
VPFIT are typically $\sim$0.0001. Including the uncertainties in the velocity 
scale (see \S \S 2.1,2.2), we estimate that the total redshift uncertainties are 
$\sim$0.0002.


As noted above, the previous {\it HST} observations of H 1821+643 
obtained by Bahcall et al. (1992\markcite{bah92},1993\markcite{bah93}) 
and Savage et al.\markcite{ssl95} (1995) can be used to expand the redshift 
range probed in this study. Table ~\ref{g160mlist} summarizes the 
equivalent widths, identifications, and redshifts of extragalactic absorption 
lines relevant to this study detected in these previous observations. Note that 
while the Savage et al.\markcite{ssl95} observations have a limiting 
equivalent width similar to that of our G140L spectrum, $\sim$50 m\AA\ at 
the 3$\sigma$ level, their resolution is considerably better so we prefer the 
Savage et al. data in cases where the lines are also present in the G140L 
spectrum. 

\subsection{Galactic Interstellar Absorption Lines}

An important issue in the study of weak \lya lines is possible contamination 
of the extragalactic line sample by absorption lines due to the ISM of the 
Milky Way. While the strong ISM lines are easily identified (c.f. Table 4 in 
Morton, York, \& Jenkins\markcite{mort88} 1988), in spectra sensitive to 
50 m\AA\ equivalent widths, a significant number of weaker ISM resonance 
lines which are less familiar may be detected as well, and these can easily be 
misidentified as \lya clouds in low resolution spectra. Galactic interstellar 
lines identified in the GHRS spectra of H 1821+643 and PG 1116+215 are 
listed in Tables ~\ref{h1821list} and ~\ref{p1116list}, and absorption 
profiles of ISM lines of interest are plotted in Figures ~\ref{h1821ism} and 
~\ref{pg1116ism}. In the following two sections we justify the Galactic line 
identifications.

\subsubsection{H 1821+643}

In the wavelength range of the G140L spectrum shown in Figure 
~\ref{h1821grand}, weak interstellar lines of \ion{C}{1} and \ion{Ni}{2} 
are of particular concern. 

Neutral carbon is not the dominant ionization state of C in warm \ion{H}{1} 
clouds (the ionization potential of \ion{C}{1} is 11.3 eV), but nevertheless it 
has been detected in absorption on interstellar sight lines with low ion 
column densities comparable to the those observed toward H 1821+643. For 
example, Lu, Savage, \& Sembach\markcite{lu94} (1994) measure 
$W_{\lambda}$ = 104$\pm$17 m\AA\ for the Galactic \ion{C}{1} 
1560.31 \AA\ line at low velocity\footnote{  Lu et al.\markcite{lu94} 
(1994) also report detections of S II and weak C I absorption 
in a Galactic high velocity cloud in the direction of this active galaxy. The 
high velocity absorption is not included in the equivalent width quoted here.} 
in the spectrum of the Seyfert galaxy NGC 3783. The \ion{S}{2} equivalent 
widths in the low velocity interstellar gas toward NGC 3783 are comparable 
to the corresponding \ion{S}{2} equivalent widths observed toward H 
1821+643 (Savage et al.\markcite{ssl95} 1995), so comparable \ion{C}{1} 
lines may be expected toward H 1821+643. Our G140L H 1821+643 
spectrum does not cover the \ion{C}{1} 1560.31 \AA\ multiplet, but it does 
cover the \ion{C}{1} 1277.25, 1280.13, and 1328.83 \AA\ resonance lines. 
Since the value of $f\lambda $, the product of the oscillator strength and 
wavelength, of the 1277.25 \AA\ line is very similar to $f\lambda $ of the 
1560.31 \AA\ line (Morton\markcite{mort91} 1991), we might expect to see 
a roughly 100 m\AA\ \ion{C}{1} line due to the ISM in the H 1821+643 
spectrum, and indeed a 150 m\AA\ line is observed at 1277.21 \AA\ (see 
Figure ~\ref{h1821grand} and Table ~\ref{h1821list}). Figure 
~\ref{h1821ism} shows the normalized absorption profiles of the Galactic 
\ion{C}{1} 1277.25, 1280.13, and 1328.83 \AA\ lines (the 1280.13 \AA\ 
profile is in the same panel as the 1277.25 \AA\ line). For purposes of 
comparison, we also plot the Milky Way \ion{S}{2} 1253.81, \ion{Si}{2} 
1526.71, and \ion{C}{2} 1334.53 \AA\ lines from the G140L spectrum in 
Figure ~\ref{h1821ism}. The $f\lambda $-values of the 1280.13 and 
1328.83 \AA\ transitions are lower than the 1277.25 $f\lambda $-value by 
factors of 4.0 and 1.6 respectively. The 1280.13 \AA\ line is not detected, but 
this is not surprising since it has a lower oscillator strength. An absorption 
feature of moderate strength is present at the expected wavelength of 
\ion{C}{1} 1328.83 \AA , but it is blended with the strong \ion{H}{1} 
Ly$\beta$ line at \zabs\ = 0.297 (see Figure ~\ref{h1821ism}). It is possible 
that we have detected the Galactic \ion{C}{1} 1277.25 and 1328.83 \AA\ 
lines. However, the \ion{C}{1} 1277.25 \AA\ line is rather broad compared 
to the strong Galactic ISM lines (Figure ~\ref{h1821ism}). There are 
\ion{C}{1}* and \ion{C}{1}** transitions near the 1277.25 \AA\ line, but 
the strongest of these are only separated from \ion{C}{1} 1277.2 by 70 
\kms\ (see Morton\markcite{mort91} 1991), so this seems an unlikely 
explanation of the breadth of the 1277.21 \AA\ absorption line. This 
\ion{C}{1} line may be blended with a \lya cloud at \zabs\ $\approx$ 0.05. 
This is important because a cluster of galaxies is detected near this redshift in 
the WIYN survey (see \S 5.2). Higher resolution spectroscopy with STIS 
will be required to determine if \lya absorption actually contributes to the 
feature at 1277.21 \AA .

Singly ionized nickel also has several weak resonance lines which are likely 
to be detected in high S/N ISM spectra. In the wavelength range of the 
G140L H 1821+643 spectrum, the strongest \ion{Ni}{2} lines occur at 
1317.22, 1370.13, and 1454.84 \AA\ with f-values of 0.1458, 0.1309, and 
0.05954 (Morton\markcite{mort91} 1991). Figure ~\ref{h1821ism} shows 
the normalized absorption profiles of these \ion{Ni}{2} lines; all three of the 
lines are well-detected in the G140L spectrum of H 1821+643. From Figure 
~\ref{h1821ism} we see that the profiles of the \ion{Ni}{2} 1317.22 and 
1370.13 \AA\ lines appear to have complex component structure with a main 
component at the velocity of the other ISM lines and weaker absorption 
extending from roughly --400 \kms\ to +400 \kms . Since the weak 
component structure appears in both profiles, one is tempted to attribute this 
high velocity absorption to interstellar \ion{Ni}{2}. However, this weak high 
velocity absorption cannot be entirely due to \ion{Ni}{2} because it is not 
detected in the absorption profiles of strong ISM lines. For example, Savage 
et al.\markcite{ssl95} (1995) observed the interstellar \ion{Mg}{2} 
2796.35, 2803.53 \AA\ doublet at 15 \kms\ resolution, and the absorption 
profiles of these lines, which are among the strongest of the ISM lines, show 
no evidence of absorption with 
$v \ <$ --160 \kms\ or $v \ >$ +60 \kms\ (see Figure 2 in Savage et 
al.\markcite{ssl95}). Similarly, the strong \ion{C}{2} 1334.53 and 
\ion{Si}{2} 1526.71 \AA\ lines do not show high velocity interstellar gas of 
comparable strength (see Figure ~\ref{h1821ism}). What then is the source 
of the high velocity absorption near the \ion{Ni}{2} profiles? For 
\ion{Ni}{2} 1317.22 \AA , continuum placement may play a role. With the 
uncertainty of the continuum placement, the extended absorption near 
\ion{Ni}{2} 1317.33 \AA\ is marginally detected. The extended absorption 
near \ion{Ni}{2} 1370.13 \AA , on the other hand, is real. The \ion{Cu}{2} 
1367.95 \AA\ transition is close to the component at -400 \kms\ in the 
\ion{Ni}{2} 1370.13 \AA\ profile, but this identification cannot be correct 
because \ion{Cu}{2} has a much stronger line at 1358.77 \AA\ which is not 
apparent and because the wavelength doesn't match the Cu line well. Given 
the close proximity of this \ion{Ni}{2} line to the strong multicomponent 
\lya line at 1363.15 \AA , it is likely that the extra absorption near 
\ion{Ni}{2} 1370.13 \AA\ is extragalactic \ion{H}{1} \lya absorption. 
There are multiple absorption components near the \ion{Ni}{2} 1454.84 
\AA\ line as well, but these occur at high positive velocities with different 
relative strengths than the components near the other \ion{Ni}{2} lines (see 
Figure ~\ref{h1821ism}) and thus are not due to \ion{Ni}{2}.

The absorption line at 1414.41 \AA\ may have a contribution from 
Galactic \ion{Ga}{2} 1414.40 \AA . However, toward the well 
studied star $\zeta$ Oph with $N$(H) = 1.38$\times 10^{21}$ 
atoms cm$^{-2}$, this Ga line has $W_{\lambda}$ = 4 m\AA\ 
(Savage, Cardelli, \& Sofia\markcite{scs92} 1992). Since the line at 
1414.41 \AA\ in the H 1821+643 spectrum has $W_{\lambda}$ = 
63$\pm$10 m\AA, it is unlikely that this line is entirely due to \ion{Ga}{2}, 
but Ga is significantly depleted from the gas phase by dust in the $\zeta$ Oph 
interstellar cloud. If we assume that Ga is not depleted at all in the interstellar 
gas toward H 1821+643 and scale the $\zeta$ Oph 
equivalent width by the H 1821+643/$\zeta$ Oph hydrogen column 
densities, then we predict $W_{\lambda} \ \approx$ 12 m\AA\ for 
the \ion{Ga}{2} line in the H 1821+643 spectrum. This is a 
conservative upper limit since Ga probably is at least somewhat dust 
depleted on the H 1821+643 sight line. Therefore we conclude that 
the line at 1414.41 \AA\ is primarily due to extragalactic \ion{H}{1} 
with possibly a small contribution from \ion{Ga}{2}.

\subsubsection{PG 1116+215}

The interstellar lines in the GHRS G140L spectrum of PG 1116+215 are 
listed in Table ~\ref{p1116list} and many of them are displayed on a 
velocity basis in Figure ~\ref{pg1116ism}. Close inspection of these 
interstellar profiles reveals evidence of a Galactic high velocity cloud (HVC) 
at $\sim$200 \kms\ (see, e.g., the \ion{Si}{4} and \ion{Si}{2} profiles in 
Figure ~\ref{pg1116ism}). We have taken this HVC into account in our 
identification of the lines. 

In contrast to H 1821+643, the \ion{C}{1} 1277.25 \AA\ multiplet is not 
detected toward PG 1116+215. The difference may be due to the fact that the 
Galactic sight line  to PG 1116+215 contains a considerably lower column 
density of \ion{H}{1} than that toward H 1821+643: $N$(\ion{H}{1}) = 
1.40$\times 10^{20}$  versus 3.84$\times 10^{20}$ cm$^{-2}$ (Lockman 
\& Savage\markcite{loc95} 1995). We attribute the line found at 1328.46 
\AA\ to \lya rather than \ion{C}{1} 1328.83 because of the absence of the 
\ion{C}{1} 1277.25 multiplet which should be $\sim$1.6 times stronger.

The line at 1239.40 \AA\ with $W_{\lambda}$ = 170$\pm$24  lies close to 
the wavelengths of Galactic \ion{N}{5} 1238.80 and \ion{Mg}{2} 1239.93 
and 1240.39. However, we identify this feature as dominated by \lya 
absorption at \zabs\ = 0.1950 because there is no evidence for the weaker 
member of the \ion{N}{5} doublet at 1242.80 \AA\ (see Fig. 
~\ref{pg1116ism}) and the velocity centroid implied for the \ion{N}{5} 
absorption ($\sim$145 km s$^{-1}$) is larger than we would expect to 
observe given that the ISM \ion{Si}{4} 1393.76 and 1402.77 \AA\ lines are 
at +43 km s$^{-1}$. Also, the \ion{Mg}{2} 1239.93 and 1240.39 doublet 
lines have small f-values  and these lines typically have equivalent widths of 
~10 to  25 m\AA\ in the spectra of distant stars in the Milky Way halo 
(Savage \& Sembach\markcite{savsem96} 1996). Again, it would be useful 
to obtain a high resolution STIS spectrum to confirm the identification of this 
line and to properly remove any absorption due to interstellar \ion{N}{5} 
and \ion{Mg}{2}.

We identify the line at 1317.42 \AA\ as \ion{Ni}{2} 1317.21. With this 
identification we would expect to see a \ion{Ni}{2} line of comparable 
strength at 1370.13 \AA . While there is no statistically significant line 
detected at that  wavelength, a feature of low significance is evident in Figure 
~\ref{pg1116ism}. We believe the \ion{Ni}{2} 1317.21 identification is the 
correct one and that noise and/or continuum placement uncertainties have 
probably combined to cause the detection of one \ion{Ni}{2} line and the 
apparent weakness of the second.

The interstellar \ion{S}{2} triplet lines at 1259.52, 1253.81,  and 1250.58 
\AA\ (with f-values of 0.0162, 0.0109, and 0.00545, respectively) shown in 
Figure ~\ref{pg1116ism} are blended with various other absorption features. 
 The high S/N of the observations permits a clear separation of the 
overlapping absorption features. \ion{S}{2} 1259.52 blends with the very 
strong interstellar \ion{Si}{2} 1260.42 line and separate equivalent widths 
are given in Table ~\ref{p1116list}. \ion{S}{2} 1253.81 blends with a  
weaker line at 1255.00 \AA\ that we identify as a \lya feature at \zabs\ = 
0.03223.  \ion{S}{2} 1250.58 blends with a broader and stronger feature we 
identify as a \lya line at \zabs\ = 0.02845. This is evident from the strength of 
the 1250.30 \AA\ line; if this line is entirely due to Galactic \ion{S}{2} 
1250.58, then it should be a factor of $\sim$2.0 weaker than the \ion{S}{2} 
1253.81 \AA\ line. Instead we see from Figures ~\ref{p1116grand} and 
~\ref{pg1116ism} that the 1250.30 \AA\ line is clearly stronger than the the 
line at 1253.86 \AA . Also, the centroid of the 1250.30 \AA\ line is 
blueshifted with respect to the other ISM lines (see Figure 
~\ref{pg1116ism}). With these identifications the relative strengths and 
velocities of the three \ion{S}{2} lines from profile fitting are consistent with 
their origin in the Galactic ISM. The two \lya lines that blend with the 
\ion{S}{2} lines differ in strength by about a factor of two. We considered 
the possibility that these lines might be a metal line doublet but could not 
find any plausible identifications involving the stronger \lya systems we see 
in the spectrum of PG 1116+215. While the line at 1255.0 \AA\ may contain 
a contribution from \ion{S}{2} in the Galactic HVC, this feature cannot be 
entirely due to high velocity interstellar \ion{S}{2}. This is clear from 
comparison of the \ion{S}{2} 1250.58 and 1253.81 \AA\ lines; if the line at 
1255.0 \AA\ were mostly due to high velocity \ion{S}{2}, then it would be 
much stronger than observed in the \ion{S}{2} 1250.58 profile.

The other ISM lines found in the spectrum of PG 1116+215 are definite 
detections of lines known to be strong in the Galactic ISM. There are no 
known Galactic ISM lines close to the observed line at 1448.43 \AA . There 
are also no reasonable identifications of this feature as a metal line associated 
with one of the stronger \lya lines seen in the spectrum of PG 1116+215. 
Therefore this line is left as unidentified in Table ~\ref{p1116list}.

\subsection{Extragalactic Absorption Lines}

Heavy elements are detected in two extragalactic absorption line systems in 
the G140L spectrum of H 1821+643. We detect the \ion{O}{6} 1031.93, 
1037.62 \AA\ doublet and \ion{C}{3} 977.02 \AA\ in the associated 
absorber at \zabs\ = 0.2967, and the \ion{O}{6} doublet is also detected in 
the intervening absorption system at \zabs\ = 0.2250. These metal absorbers 
have been analyzed in a separate paper (Savage, Tripp, \& 
Lu\markcite{stl98} 1998). Note that the line at 1478.27 \AA\ could be 
identified as \ion{Si}{3} 1206.50 \AA\ associated with the \ion{O}{6} 
absorber at \zabs\ = 0.2250. However, this identification seems unlikely 
because neither \ion{Si}{2} nor \ion{Si}{4} lines are detected at this 
redshift with sensitive upper limits (see Savage et al.\markcite{stl98} 1998), 
and it is difficult to simultaneously satisfy the lower limits on the 
\ion{Si}{3}/\ion{Si}{2} and \ion{Si}{3}/\ion{Si}{4} ratios in photoionized 
or collisionally ionized gas. Therefore we attribute the line at 1478.27 \AA\ 
at least partially to extragalactic \lya\ absorption. Corresponding Ly$\beta$ 
absorption is detected in the G160M spectrum (see Table ~\ref{g160mlist}), 
but the redshifts derived from the Ly$\beta$ and \lya lines differ by 0.0008, 
so this \lya line is probably blended, perhaps with \ion{Si}{3} associated 
with the \ion{O}{6} absorber. Once again, higher resolution observations 
are needed to resolve this line identification ambiguity.

In the spectrum of PG 1116+215, a statistically significant line is detected at 
the expected wavelength of \ion{Si}{3} 1206.50 at \zabs\ = 0.13852, within 
$\sim$30 km s$^{-1}$ of the strong \lya line at \zabs\ = 0.13861. 
\ion{Si}{2} 1260.42 \AA\ is also marginally detected at this redshift.

The rest of the lines not identified as Galactic ISM lines or extragalactic 
metals (or \ion{H}{1} Ly$\beta$) are identified as \lya lines provided \zabs\ 
$\leq z_{\rm em}$. In a few cases, the $\chi ^{2}$ statistic from profile 
fitting with VPFIT is poor unless additional components are added to the fit. 
On this basis we infer the presence of two components in the line at 1384.16 
\AA\ in the spectrum of PG 1116+215. There are some indications that there 
is an additional \lya line (not listed in Table ~\ref{h1821list}) blended with 
interstellar  \ion{Si}{4} 1393.76 \AA\ toward H 1821+643 (see Tripp 
1997), but higher resolution observations are required to reliably identify this 
line.

\section{Lyman $\alpha$ Absorbers}

\subsection{Lyman $\alpha$ Line Density}

The high detection sensitivity of our GHRS spectra provides a rare 
opportunity to assess the line density, $dN/dz$ (the number of lines per unit 
redshift), of weak Ly$\alpha$ lines and its evolution at low redshift. In the 
following analysis we exclude from consideration the region within 3000 km 
s$^{-1}$ of the emission redshift of each quasar for the reasons given in \S 
5.2.

Combining the G140L spectrum obtained here with the G160M spectra of 
Savage et al.\markcite{ssl95} (1995), we detect a total of 26 Ly$\alpha$ 
lines with $W_{\rm r} \ > \ 50$ m\AA\ (19 with $W_{\rm r} \ > \ 75$ 
m\AA) in the spectrum of H 1821+643 over the wavelength region 
1233-1556 \AA, corresponding to a redshift path $\Delta z$ = 0.247 after 
applying a correction of $\Delta z\sim 0.019$ for the redshift path blocked 
out by strong ISM lines. This implies
\begin{displaymath}
\frac{dN}{dz} = 105\pm21
\end{displaymath}
for Ly$\alpha$ lines with $W_{\rm r} \ > \ 50$ m\AA, where the error is 
determined from $\sqrt{N}$ statistics.

A similar analysis for the 3C 273 sightline (Morris et al.\markcite{mor93} 
1993) with 13 Ly$\alpha$ lines with $W_{\rm r} \ > \ 50$ m\AA\ (8 with 
$W_{\rm r} \ > \ 75$ m\AA) over the wavelength region 1218-1382 \AA\ 
with a redshift path of 0.135 (including the two lines associated with the 
Virgo cluster and the line blended with Galactic \ion{Si}{4} 
absorption\footnote{Brandt et al (1997) have verified the reality of this line 
by detecting the H I Ly$\beta$ absorption at the same redshift.}) 
yields
\begin{displaymath}
\frac{dN}{dz} = 97\pm27
\end{displaymath}
for Ly$\alpha$ lines with $W_{\rm r} \ > \ 50$ m\AA.

For PG 1116+215, 13 Ly$\alpha$ lines with $W_{\rm r} \ > \ 50$ m\AA\ 
(10 with $W_{\rm r} \ > \ 75$ m\AA) are found in the G140L spectrum 
between 1226 and 1417 \AA, corresponding to $\Delta z=0.144$ after 
applying a correction of 0.013 for the redshift path blocked out by the ISM 
lines. The resulting line density is then
\begin{displaymath}
\frac{dN}{dz} > 90\pm 25
\end{displaymath}
for Ly$\alpha$ lines with $W_{\rm r} \ > \ 50$ m\AA, where the inequality 
results from the fact that not all spectral regions of the PG 1116+215 G140L 
spectrum have adequate S/N to detect Ly$\alpha$ lines with $W_{\rm r} \ = 
\ 50$ m\AA.

These estimates are in essential agreement with each other. Combining the H 
1821+643 and the 3C 273 sightlines, we find
\begin{displaymath}
\frac{dN}{dz} = 102\pm16
\end{displaymath}
for Ly$\alpha$ lines with $W_{\rm r} \ > \ 50$ m\AA\ at $0<z<0.28$. 
Similarly we find 
\begin{displaymath}
\frac{dN}{dz} = 71\pm12
\end{displaymath}
for Ly$\alpha$ lines with $W_{\rm r} \ > \ 75$ m\AA\ combining the data 
for all three sightlines.

The line density at $0<z<0.28$, $dN/dz=102\pm16$ for Ly$\alpha$ lines 
with $W_{\rm r} \ > \ 50$ m\AA , is consistent with the number of 
Ly$\alpha$ absorbers with $N$(\ion{H}{1})$>10^{13}$ cm$^{-2}$ at 
these redshifts predicted by Riediger, Petitjean, \& 
M\"ucket\markcite{rpm98} (1998) based on their study of Ly$\alpha$ cloud 
evolution in the context of a CDM model. Note that an absorber with 
$N$(\ion{H}{1})$=10^{13}$ cm$^{-2}$ would produce a Ly$\alpha$ 
absorption line with $W_{\rm r}$ = 50 m\AA\ if $b=35$ km s$^{-1}$. 
Recently, Shull\markcite{shull97} (1997) has reported preliminary results 
from analysis of 7 sight lines observed in {\it HST} Cycle 6. Based on these 
new spectra, which are reported to be adequate for detection of 20 m\AA\ 
lines, Shull\markcite{shull97} (1997) derives $dN/dz$ = 250$\pm$40. 
Evidently the number of lines detected continues to rise as the sensitivity 
improves. This is also predicted by Riediger et al. and may have important 
implications for the nature of the 
absorbers and the baryon content of the low $z$ \lya forest (see 
Shull\markcite{shull97} 1997 and discussion below).

\subsection{Redshift Evolution of Lyman $\alpha$ Line Density}

Previous studies of the redshift evolution of Ly$\alpha$ clouds at low 
redshift have been limited to relatively strong lines ($W_{\rm r} \ > \ 300$ 
m\AA, e.g., Weymann et al.\markcite{wey98} 1998). The redshift evolution 
of weaker Ly$\alpha$ lines is of considerable interest since, according to 
cosmological simulations of structure formation and evolution, they may 
arise from the tenuous gas in void regions and therefore may evolve 
differently from the stronger lines which mostly trace the gas distribution 
near galaxy structures (compare the $P_u$ and $P_s$ populations in Figure 
6 of Riediger et al.\markcite{rpm98} 1998). The distribution of Ly$\alpha$ 
lines in redshift is generally described by the expression
 \begin{equation}
\frac{dN}{dz} = \left( \frac{dN}{dz} \right) _{0} (1 + z)^{\gamma},
\end{equation}
where $(dN/dz)_0$ is the line density at $z=0$. Dividing the data for the 
above three sightlines into two redshift bins at $z<0.15$ and $z>0.15$ and 
considering only Ly$\alpha$ lines with $W_{\rm r}$ between 75 and 300 
m\AA, we find $dN/dz(z\simeq 0.075)=46\pm11$ and 
$dN/dz(z\simeq0.22)=62\pm21$. The data seem to indicate some evolution 
in the number density of the weak lines but the relatively small number of 
lines and the small redshift span of the current sample ($z=0-0.28$) make 
any determination of the evolution index $\gamma$ very uncertain. A formal 
fit to the equation above using the technique of Lu, Wolfe, \& 
Turnshek\markcite{luwolfe91} (1991) yields $\gamma=1.6\pm3.0$ for 
Ly$\alpha$ lines with $75<W_{\rm r}<300$ m\AA\ at $z=0-0.28$. A 
similar value is obtained if lines as weak as 50 m\AA\ are included.

\subsection{Clustering of Weak Lyman $\alpha$ Lines}

A striking aspect of Figure ~\ref{h1821grand} is the complex component 
structure of the strong Ly$\alpha$ lines. To allow the reader to inspect this 
more closely, the normalized absorption profiles of the four strongest 
Ly$\alpha$ lines in the spectrum of H 1821+643 are shown on expanded 
scales in Figure ~\ref{lyapros}, and the rest equivalent widths of the 
Ly$\alpha$ lines are plotted versus redshift in Figure ~\ref{lyastick}. 
Though there is some confusion from adjacent unrelated lines, we see from 
Figures ~\ref{lyapros} and ~\ref{lyastick} that the absorption profiles of 
three out of the four strongest \ion{H}{1} Ly$\alpha$ lines contain complex 
component structure with a main strong component and several partially 
resolved weaker outlying components spanning 1000-1500 km s$^{-1}$. 
Even the Ly$\alpha$ line at $z_{\rm abs}=0.16990$ which does not show 
resolved weak components has an asymmetric profile which is evidence for 
unresolved components. Similarly, two out of the three strongest \ion{H}{1} 
Ly$\alpha$ lines seen toward PG 1116+215 show evidence of multiple 
components (see Table ~\ref{p1116list}). This seems to suggest that there is 
some clustering of weak Ly$\alpha$ lines at low redshift. Clustering of 
strong Ly$\alpha$ lines ($W_{\rm r} \gtrsim 300$ m\AA ) at low $z$ has 
been reported by Lanzetta et al.\markcite{lanz96} (1996) and 
Ulmer\markcite{ulm96} (1996), but clustering of weak Ly$\alpha$ lines has 
not been detected previously.

To quantitatively measure the degree to which the Ly$\alpha$ lines are 
clustered, we use the standard technique involving the two-point velocity 
correlation function, 
\begin{equation}
\xi (\Delta v) = \frac{N_{\rm obs}(\Delta v)}{N_{\rm ran}(\Delta v)} - 1,
\end{equation}
where $N_{\rm obs}(\Delta v)$ is the number of observed line pairs with 
velocity separation $\Delta v$ and $N_{\rm ran}(\Delta v)$ is the number of 
pairs expected in a random distribution. If the clouds are not clustered on 
velocity scale $\Delta v$, then $\xi(\Delta v)=0$. To expand our Ly$\alpha$ 
line sample, we have combined our data with those of Morris et al (1993) for 
3C 273, Stocke et al (1995) and Shull et al (1996) for I Zw 1, Mrk 335, 
Mrk 421, and Mrk 501, and Bruhweiler et al.\markcite{bru93} (1993) and 
Savage et al.\markcite{ssl97} (1997) for PKS 2155$-$304. All these spectra 
were obtained with the GHRS and have enough sensitivity to detect 
Ly$\alpha$ lines as weak as 75 m\AA\ throughout the spectra, except the 
G140L spectrum of PKS 2155$-$304 obtained by Bruhweiler et al (1993) 
which has somewhat lower sensitivity and is adequate for detection of lines 
with $W_{\rm r} > 100$ m\AA.  Figure ~\ref{twop}a shows the two-point 
correlation function for lines with $W_{\rm r} > 75$ m\AA. No evidence for 
clustering on any velocity scale is evident. Figure ~\ref{twop}b shows the 
two-point correlation function for lines with $W_{\rm r} > 100$ m\AA, 
which shows a marginal signal ($\sim 3\sigma$) at $\Delta v\sim 500$ km 
s$^{-1}$, and the signal may extend up to $\sim 1000$ 
km s$^{-1}$. The number of lines in the sample becomes too small for this 
analysis if a higher equivalent width limit is imposed. Clearly a much larger 
sample of lines is required to further explore the clustering properties of weak 
Ly$\alpha$ lines.

\section{The Relationship Between \lya Absorbers and Galaxies}

In this section we examine the relationship between \lya absorbers and 
galaxies. We begin with the sample definitions and nearest neighbor 
distances (\S 5.1). We then present some individual cases of interest and 
examine the correlations between the \lya line equivalent widths and the 
nearest galaxy impact parameters and distances in \S 5.2. Finally, in \S 5.3 
we carry out a statistical analysis of the nature of the absorber-galaxy 
relationship.

In Figures ~\ref{pie1821} and ~\ref{pie1116} we overplot the \lya line 
redshifts (indicated with vertical lines) on RA and Dec slices showing the 
galaxy redshifts (circles); the length of the line indicates the absorption line 
equivalent width as shown in the figure legend. Figure ~\ref{histreds} shows 
the number of galaxies in the survey as a function of redshift in 1000 \kms\ 
bins with the absorber redshifts plotted at the top of each panel.

\subsection{\lya Absorber Samples}

For the \lya absorption lines we consider the following samples: (1) a ``total'' 
sample which includes all \lya lines listed in Tables ~\ref{h1821list} - 
~\ref{g160mlist}, (2) a ``complete'' sample which only includes \lya lines 
with \zabs\ $\leq$ 0.20 but is otherwise identical to the total sample, and (3) 
a ``1 Mpc complete'' sample which only includes lines with 0.048 $\leq$ 
\zabs\ $\leq$ 0.20. The second sample is defined because the H 1821+643 
galaxy redshift survey is severely incomplete at $z \ \gtrsim$ 0.20 (see \S 
2.3). The ``1 Mpc complete'' sample is defined to ensure good spatial 
coverage with a high degree of galaxy redshift survey completeness out to at 
least 1 Mpc from the QSO (c.f., Table ~\ref{complete1116}). At \zabs\ 
$\lesssim$ 0.048, a 20' radius corresponds to less than 1 Mpc, so nearest 
galaxies could be missed because they are outside of the field covered by the 
redshift survey. We will compare results derived from the ``complete'' and ``1 
Mpc complete'' samples to evaluate the impact of this bias. For reasons given 
in \S 5.2, in all samples we exclude lines within 3000 \kms\ of the QSO 
redshift.

Following Morris et al.\markcite{mor93} (1993) and Stocke et 
al.\markcite{stoc95} (1995), we further delineate the samples based on the 
assumption used to calculate the radial distances between absorbers and 
galaxies. We either (a) assume the radial distance between a given absorber 
and galaxy is given by the Hubble law (referred to as the ``pure Hubble flow'' 
sample), or (b) assume the absorber and galaxy are at the same radial 
distance if $\mid \Delta v \mid \ \leq$ 300 \kms\ and calculate the radial 
distance between the absorber and galaxy from ($\mid \Delta v \mid - 
300)/H_{\rm 0}$ for $\mid \Delta v \mid \ >$ 300 \kms\ (the ``perturbed 
Hubble flow'' sample). The perturbed sample is intended to account for 
motions from, e.g., galaxy rotation, which cause departures from the Hubble 
flow. The projected (i.e., transverse) distance between clouds and galaxies, 
also referred to as the impact parameter, is calculated assuming $H_{\rm 
0}$ = 75 \kms\ Mpc$^{-1}$ and $q_{\rm 0}$ = 0. Tables ~\ref{h1821near} 
and ~\ref{p1116near} list the projected and three-dimensional distances from 
the \lya absorbers to their nearest galaxies, for the pure Hubble flow sample, 
along with the absorber-galaxy velocity separation.

\subsection{Relationship with Individual Galaxies and Specific Galaxy 
Structures}

We now turn to the relationship between galaxies and \lya absorbers implied 
by our study, and we begin with inferences drawn from direct inspection of 
the data.

{\it Individual Galaxies.} One of the major conclusions of the \lya cloud 
study of Lanzetta et al.\markcite{lanz95} (1995) is that a large fraction 
($\sim$32-60\% ) of the {\it strong} lines originate in individual luminous 
galaxies within projected distances of $\sim$200 kpc from the line of sight. 
More recently, Lanzetta et al.\markcite{lanz97} (1997) have suggested that 
the fraction is even larger and that possibly all of the absorbers arise in such 
galaxies. Furthermore, Lanzetta et al.\markcite{lanz95} claim that the \lya 
equivalent width is anticorrelated with impact parameter, which indicates 
that there is a physical association between the galaxies and the absorbers 
and that the density of the gaseous halo decreases with increasing distance 
from the galaxy. This anticorrelation has been contested by Le Brun et 
al.\markcite{leb96} (1996) and Bowen, Blades, \& Pettini\markcite{bbp96} 
(1996), but Chen et al.\markcite{chen98} (1998) argue that Le Brun et 
al.\markcite{leb96} and Bowen et al.\markcite{bbp96} have effectively 
diluted the anticorrelation by including galaxy-absorber pairs at large impact 
parameters (up to $\sim$ 5 Mpc) which are unlikely to be physically 
associated.

We can test the conclusions of Lanzetta et al.\markcite{lanz95} because our 
galaxy samples include 42 galaxies with impact parameters $\rho \ \leq$ 600 
kpc (see Tables ~\ref{h1821reds} and ~\ref{p1116reds}). However, we 
should exclude any galaxies and absorption lines which have \zabs\ $\sim 
z_{\rm em}$ because (1) the absorbers with \zabs\ $\sim \ z_{\rm em}$ are 
now recognized to be a special class of absorption systems; in some cases 
there is clear evidence that these absorbers are close to the QSO nucleus 
(e.g., Hamann et al.\markcite{ham97a} 1997a,b\markcite{ham97b}; 
Barlow \& Sargent\markcite{bs97} 1997), (2) these absorbers can be 
significantly photoionized by the QSO itself (the ``proximity effect''), and (3) 
in the case of H 1821+643, the QSO is located in a rich galaxy cluster (see 
Figure ~\ref{pie1821} and Hall, Ellingson, \& Green\markcite{hall97} 
1997) which is a special environment. For consistency with Lanzetta et 
al.\markcite{lanz95} and Chen et al.\markcite{chen98}, we exclude objects 
within 3000 \kms\ of the QSO redshift. This leaves us with 20 galaxies with 
$\rho \ \leq$ 600 kpc in the fields of H 1821+643 and PG 1116+215.

First consider the closer galaxies: if we find a galaxy with an impact 
parameter of 200 kpc or less, do we find a strong \lya line at that redshift? 
Our sample is small -- we only have three galaxies with impact parameters of 
200 kpc or less -- but in all three cases we find a strong \lya line with 
$W_{\rm r} \ >$ 300 m\AA\ within 150 \kms\ of the galaxy redshift. 
Furthermore, as we consider larger impact parameters, we continue to find 
nearby \lya absorbers. In fact, all galaxies within projected distances of 600 
kpc from the sight lines have associated \lya absorbers with $\mid \Delta 
v\mid \ = \ \mid c(z_{\rm abs} \ - \ z_{\rm gal})/(1 + z_{\rm mean})\mid \ <$ 
1000 \kms\ ($z_{\rm mean}$ is the mean of \zabs\ and $z_{\rm gal}$). In 
some cases there are several galaxies near the redshift of the \lya cloud; 
following Chen et al.\markcite{chen98} (1998) and Lanzetta et 
al.\markcite{lanz95} (1995), we first assume that the galaxy with the 
smallest impact parameter is the most likely to be associated with the 
absorber (we consider an alternate selection criterion below). This reduces 
our list of 20 close galaxies to 12 close galaxies. We also treat \lya lines 
within 350 \kms\ of each other as a single absorber. Given these assumptions, 
we find 12 galaxy-absorber pairs with $\rho \leq$ 600 kpc and $\mid \Delta 
v \mid \leq$ 1000 \kms\ (7 pairs toward H 1821+643 and 5 pairs toward PG 
1116+215). We have used $\mid \Delta v \mid \leq$ 1000 \kms\ because this 
is the selection criterion originally employed by Lanzetta et 
al.\markcite{lanz95} (1995). While there are situations which could lead to 
such high velocity gas flows in galaxy clusters (e.g., Roettiger, Stone, \& 
Mushotzky\markcite{roet98} 1998; Burns\markcite{burns98} 1998), in 
general $\mid \Delta v \mid \leq$ 1000 \kms\ is a rather large velocity 
interval which corresponds to many Mpc in an unperturbed Hubble flow. 
Therefore one might reasonably wonder if these 12 galaxy-absorber pairs are 
coincidental alignments of randomly distributed \lya lines which occasionally 
happen to be close to a galaxy. To address this question, we have used Monte 
Carlo simulations (see \S 5.3) to randomly distribute \lya lines along the H 
1821+643 and PG 1116+215 sightlines. Based on 10000 Monte Carlo 
trials, we find that the probability of 7 chance galaxy-absorber pairs toward 
H 1821+643 and 5 chance pairs toward PG 1116+215 is 4.7$\times 10^{-
3}$. Furthermore, nine of these 12 observed galaxy-absorber pairs have 
$\rho \leq$ 600 kpc and $\mid \Delta v \mid \leq$ 340 \kms , and the 
probability of this occuring if the \lya clouds are randomly distributed is 
1.3$\times 10^{-3}$, again based on 10000 Monte Carlo trials. Taking into 
account the uncertainties in the galaxy and cloud redshifts, it seems 
reasonable to attribute galaxy-absorber velocity differences of $\sim$350 
\kms\ to normal galaxy motions such as rotation.

A similar corroborating result is obtained from the 3C 273 data obtained by 
Morris et al.\markcite{mor93} (1993). There are five 
galaxies\footnote{Actually there are eight galaxies with $\rho \ <$ 600 kpc 
in the 3C 273 sample, but four of these are in Virgo at roughly the same 
redshift with $<z>$ = 0.0065, and again we use the galaxy with the smallest 
impact parameter.} in the sample of Morris et al.\markcite{mor93} with 
impact parameters of 600 kpc or less, and in all five cases there are \lya 
clouds detected within 500 \kms . In four out of the five 3C 273 close 
galaxy-absorber pairs, the \lya absorber is within 250 \kms\ of the galaxy 
redshift. In all, there are 17 galaxies with $\rho \leq$ 600 kpc along the three 
sight lines (3C 273, H 1821+643, and PG 1116+215); every one of them 
has a corresponding \lya absorption line with $\Delta v <$ 1000 \kms , and 
14 out of the 17 galaxies are within 500 \kms\ of a \lya line. This is a strong 
indication that \lya absorbers are somehow associated with luminous galaxies 
but does not necessarily indicate that the absorption lines originate in the 
large gaseous halos of the identified individual luminous galaxies. The \lya 
lines could be due to undetected faint dwarf galaxies which are clustered with 
the observed luminous galaxies, or they could be due to broadly distributed 
gas within the large scale structures where luminous galaxies are found, for 
example.

Since in some cases there are several galaxies close to a line of sight at a 
given \lya absorber redshift (see Figures ~\ref{pie1821} and 
~\ref{pie1116}), it isn't clear that the galaxy with the smallest impact 
parameter is the most likely one (or the only one) to be associated with the 
absorber (as assumed above). In some cases the galaxy with the smallest 
impact parameter has a larger velocity separation from the absorber than 
another galaxy at a somewhat larger impact parameter. An alternative 
criterion for identifying galaxy-absorber pairs is to match absorbers with the 
galaxy with the smallest {\it three-dimensional} distance. Assuming the 
perturbed Hubble flow described above, this selection method leads to the 
identification of 17 galaxy-absorber pairs within projected distances of 1 
Mpc with $\mid \Delta v \mid <$ 350 \kms\ in the H 1821+643 and PG 
1116+215 sight lines. These galaxy-absorber pairs are summarized in Table 
~\ref{galx600}. Monte Carlo simulations indicate that the probability of 
drawing this many pairs from a randomly distributed \lya population is 
3.6$\times 10^{-5}$.

We now turn to the equivalent width - impact parameter anticorrelation 
which Lanzetta et al.\markcite{lanz95} use to argue that the strong \lya lines 
are truly due to luminous {\it individual} galaxies. Does this anticorrelation 
extend to larger impact parameters? Lanzetta et al.\markcite{lanz95} (1995) 
and Chen et al.\markcite{chen98} (1998) could not detect such an extension 
because their FOS \lya line measurements generally do not have sufficient 
sensitivity to detect the weak absorption lines, but our data are adequate for 
this purpose. To explore this question, we have combined the H 1821+643 
and PG 1116+215 data with the 3C 273 data from Morris et 
al.\markcite{mor93} (1993) and, for the stronger lines at smaller impact 
parameters, the data from Chen et al.\markcite{chen98} (1998). From these 
papers we have selected all galaxy-absorber pairs with impact parameters of 
600 kpc or less and $\Delta v \leq$ 1000 \kms . The results are plotted in 
Figure ~\ref{rhocor}. This figure shows that the Lanzetta et al. 
\markcite{lanz95} (1995) anticorrelation does appear to extend to larger 
impact parameters. The Spearman rank-order correlation coefficient for the 
data in Figure ~\ref{rhocor} is $r_{s}$ = -0.762, and with 44 data points, 
this indicates that the data are anticorrelated at the $7.6\sigma$ significance 
level.\footnote{According to Equation 14.6.2 in Press et 
al.\markcite{prss92} (1992). The sum squared difference of ranks statistic 
indicates that the significance of the anticorrelation is 5.0$\sigma$.} Fitting a 
power-law of the form
\begin{equation}
{\rm log} \ W_{\rm r} = \alpha \ {\rm log} \ \rho + C,
\end{equation}
where $\rho$ is the galaxy impact parameter and $C$ is a constant, to the 
data in Figure ~\ref{rhocor} yields $\alpha$ = -0.80$\pm$0.10 and $C$ = 
4.23$\pm$0.20 where $W_{\rm r}$ is in m\AA\ and $\rho$ is in kpc (fit 
shown as a solid line in Figure ~\ref{rhocor}). Based on the strong line data 
alone, Chen et al.\markcite{chen98} derive $\alpha$ = -0.93$\pm$0.13, in 
agreement within the 1$\sigma$ uncertainties. 

In Figure ~\ref{rhocor}, we have only included galaxies with impact 
parameters of 600 kpc or less. If we increase the impact parameter cutoff, we 
continue to find galaxies apparently associated with \lya lines, be we also 
find some galaxies which do not have associated \lya absorption lines, within 
$\Delta v <$ 1000 \kms , of the strength expected from equation (3). For 
example, a galaxy at $z_{\rm gal}$ = 0.18967 in the H 1821+643 field is at 
a projected distance of 635 kpc, and therefore from eqn. (3) we expect to find 
a line with $W_{\rm r} \ \approx$ 95 m\AA\ near this redshift, which should 
be easily detected. From Figure ~\ref{h1821grand}, it is clear that no \lya 
line of this strength is present at this redshift (a weak feature is apparent at 
the expected wavelength, but it is lower than 3$\sigma$ significance and 
much weaker than $W_{\rm r} \ \approx$ 95 m\AA ). From the H 
1821+643, PG 1116+215, and 3C 273 sight lines, we find 15 
cases\footnote{Again, for groups of galaxies we only consider the galaxy 
with the smallest impact parameter.} of galaxies with 600 $< \rho <$ 2200 
kpc that do not have unambiguously associated \lya lines within 1000 \kms . 
However, five of these cases are at redshifts where the associated \lya line 
could be hidden within strong ISM lines or extragalactic lines. The remaining 
10 cases which do not have associated \lya lines are plotted as upper limits in 
Figure ~\ref{bigrhocor}, which also shows all galaxies with $\rho <$ 2200 
kpc that do have \lya lines within 1000 \kms . Remarkably, using the sample 
shown in Figure ~\ref{bigrhocor}, we find that the anticorrelation persists at 
high significance levels out to $\rho \approx$ 2200 kpc. There is a lot of 
scatter in the data shown in Figure ~\ref{bigrhocor}, so the cases where we 
do not find associated \lya lines may simply be cases where the line is weak 
enough to fall below our detection threshold; the upper limits for these cases 
are consistent with the anticorrelation. If we select only galaxy-absorber pairs 
with $\rho \ \leq$ 2200 kpc and $\Delta v \ <$ 500 \kms , the anticorrelation 
is still strong because there are only five galaxy-absorber pairs in Figure 
~\ref{bigrhocor} with 500 $< \Delta v <$ 1000 \kms\ (the triangles encased 
in squares in Figure ~\ref{bigrhocor}).

This strong anticorrelation is another indication that the absorbers and 
galaxies shown in Figure ~\ref{bigrhocor} are in some way physically 
related and may have important implications regarding the nature of the \lya 
absorbers. We discuss possible interpretations of the anticorrelation in \S 6. 
However, it is important to emphasize that there are some potential selection 
biases which could artificially tighten the anticorrelations shown in Figures 
~\ref{rhocor} and ~\ref{bigrhocor}. For example, the current galaxy redshift 
surveys only probe relatively bright galaxies (see Table 
~\ref{complete1116}), so there could be fainter galaxies at smaller impact 
parameters than those plotted in Figure ~\ref{bigrhocor} that might be 
revealed by a deeper redshift survey. Discovery of such galaxies would 
modify the appearance of Figure ~\ref{rhocor} and weaken the 
anticorrelation by moving points from the lower right corner of Figure 
~\ref{bigrhocor} to the lower left corner of the plot. Going deeper might also 
reveal close galaxies which do not have corresponding \lya absorption. Along 
these lines, Linder\markcite{lind98} (1998) has used simulations to argue 
that most \lya absorbers could be due to faint galaxies rather than the 
observable luminous galaxies at the same redshift. Some deep searches near 
the lowest redshift \lya lines have not found such faint galaxies (e.g., Morris 
et al.\markcite{mor93} 1993; van Gorkom et al.\markcite{vang93} 1993; 
Rauch et al.\markcite{rwm96} 1996), but van Gorkom et 
al.\markcite{vang96} (1996) and Hoffman et al.\markcite{hoff98} (1998) 
have located faint dwarf galaxies at the redshifts of a few low $z$ \lya lines. 
It will be important to do more deep searches for faint galaxies to establish 
whether or not the anticorrelation in Figure ~\ref{rhocor} is real.

Another potential bias in Figure ~\ref{bigrhocor} (and in the analyses of 
Lanzetta et al.\markcite{lanz95} and Chen et al.\markcite{chen98}) may be 
introduced by the selection of {\it only} \lya lines which are known to be 
within a certain projected distance and velocity of a galaxy. Due to these 
selection criteria, there are many \lya lines in our sight lines and in the Chen 
et al.\markcite{chen98} (1998) sight lines which are not plotted in Figure 
~\ref{bigrhocor}. In some cases, these unplotted \lya lines may not have 
known galaxies nearby simply because the appropriate galaxy redshift survey 
is incomplete. Alternatively, these lines may not have known nearby galaxies 
because the nearest galaxy is outside of the angular extent of the redshift 
survey. If this is the case, then these missing galaxy-absorber pairs would 
have relatively large impact parameters and would fill in the upper right 
corner of Figure ~\ref{bigrhocor}, and this would weaken the anticorrelation. 
To illustrate the possible impact of this selection effect, we plot on the far 
right side of Figure ~\ref{bigrhocor} some of the missing lines from the Chen 
et al.\markcite{chen98} (1998) sample (using equivalent widths reported by 
Jannuzi et al.\markcite{jan98} 1998 and Bahcall et al.\markcite{bah93} 
1993) and the missing lines from our sight lines. From this we see that the 
missing lines could indeed substantially dilute the anticorrelation. To check 
to see if this selection bias is occuring, more galaxy redshift measurements 
are needed.

{\it Specific Galaxy Structures.} A visual inspection of Figures 
~\ref{pie1821}, ~\ref{pie1116}, and ~\ref{histreds} and Tables 
~\ref{h1821near} and ~\ref{p1116near} yields the following information on 
the relationship between the \lya clouds and galaxy structures.

First, there are no unambiguously detected \lya lines in the vicinity of the 
prominent galaxy clusters at $z \ \approx$ 0.050 and 0.190 in the direction 
of H 1821+643. The galaxy survey of Morris et al.\markcite{mor93} (1993) 
revealed a similar cluster of galaxies at $z \approx$ 0.08 with no associated 
\lya absorbers in the direction of 3C 273. This may indicate that \lya 
absorbers are destroyed (or are not created in the first place) in this type of 
environment. However, do we expect to see absorbers in these clusters based 
on the impact parameters of the nearest galaxies? As already noted above, in 
the cluster at $z \ \approx$ 0.190, there is a galaxy at a projected distance of 
635 kpc and we should see an absorber there based on equation (3). In the 
cluster at $z \ \approx$ 0.050, the nearest galaxy has an impact parameter of 
590 kpc, so we might expect a \lya line at that redshift with $W_{\rm r} \ 
\approx$ 100 m\AA . In this case a line with the right strength is 
well-detected at the expected wavelength, but its identification is ambiguous 
since this absorption may be due to \ion{C}{1} in the Milky Way ISM (see 
\S 3.1.1). Morris \& van den Bergh\markcite{mvdb} (1994) have shown that 
a substantial fraction of \lya lines at low $z$ could be due to material 
stripped out of galaxies in tidal interactions, and they point out that in this 
model \lya absorbers would be less likely to be found in rich clusters because 
(1) there are fewer gas-rich spirals, and (2) tidal debris should be less 
common due to the higher velocity dispersion of rich clusters. Furthermore, 
rich galaxy clusters contain hot intracluster media, so \lya clouds may be 
fully ionized in this environment. Therefore it is not necessarily surprising to 
find that \lya absorbers may avoid rich clusters. However, more sight lines 
crossing rich clusters must be studied to give this suggestive result statistical 
significance. We note that there is a strong \lya cloud in the rich cluster 
which hosts H 1821+643, but since this is a \zabs\ $\approx \ z_{\rm em}$ 
absorber, this is a special case. This cloud may be quite close to the QSO 
nucleus or gas ejected by the QSO, for example (see Savage et 
al.\markcite{stl98} 1998). Given the very large number of galaxies in this 
rich cluster (see Figure ~\ref{pie1821}), it is noteworthy that the \lya 
absorption only shows one strong component, albeit at the FOS G130H 
resolution (FWHM $\approx$ 250 \kms ).

Examination of Figures ~\ref{pie1821} - ~\ref{histreds} reveals other 
smaller galaxy groups which do not have associated \lya lines within $\Delta 
v$ = 1000 \kms\ such as the group at $z$ = 0.072 toward H 1821+643. 
However, such inspection also shows several galaxy groups which {\it do} 
have associated \lya absorbers, e.g., the group at $z$ = 0.121 toward H 
1821+643 or the groups at $z$ = 0.060, 0.138, and 0.166 toward PG 
1116+215. Evidently the \lya absorbers do not always avoid the regions of 
higher galaxy concentration. This was known previously from the results of 
Stocke et al.\markcite{stoc95} (1995) and Shull et al.\markcite{ssp96} 
(1996) who showed that the majority of their \lya lines are found within large 
scale structures even though their sight lines probe substantial paths through 
galaxy void regions, but it is still possible that the \lya lines avoid the regions 
of {\it highest} galaxy density, i.e. rich clusters.

Second, as found for other sight lines by Morris et al.\markcite{mor93} 
(1993), Stocke et al.\markcite{stoc95} (1995), and Shull et 
al.\markcite{ssp96} (1996), our study reveals some \lya absorbers 
apparently in galaxy voids at very large distances from the nearest galaxy in 
the survey (see Tables ~\ref{h1821near} and ~\ref{p1116near} and Figures 
~\ref{pie1821} - ~\ref{pie1116}). At \zabs\ $\gtrsim$ 0.20, absorbers 
apparently in voids should not be taken too seriously because at these 
redshifts our galaxy survey is very incomplete. At lower redshifts, our survey 
is deep enough to reveal faint galaxies. Consider the examples of \lya lines at 
large distances from galaxies at \zabs\ = 0.04105 toward H 1821+643 and at 
\zabs\ = 0.01639 toward PG 1116+215. At $z$ = 0.04105, $B_{\rm J}$ = 
19.0 corresponds to 0.14$L^{*}$, while at $z$ = 0.01639, $B_{\rm J}$ = 
19.0 corresponds to 0.02$L^{*}$. Therefore based on the completeness 
estimates from \S 2.3, there is a good chance that we would have identified a 
galaxy as bright as the LMC near these void absorbers if such galaxies were 
present. However, at these redshifts the limited angular extent of the galaxy 
redshift survey (see Table ~\ref{complete1116}) is a source of concern; in 
these cases a galaxy at a projected distance of one to a few Mpc might be 
missed because it is outside of the region surveyed. Therefore more redshift 
measurements with broader angular coverage are needed to establish that 
these are truly void absorbers.

Figures 9-11 give a visual impression that the absorption lines in voids tend 
to be weaker. To explore this possibility, we have divided (somewhat 
arbitrarily) our complete perturbed Hubble flow sample into ``void'' and 
``non-void'' samples, the non-void absorbers defined by having 
three-dimensional nearest neighbor distances of 3 Mpc or less and the void 
absorbers defined by nearest neighbor distances greater than 3 Mpc (we have 
also included the Morris et al.\markcite{mor93} and Stocke/Shull et 
al.\markcite{ssp96} data in this exercise). We then compared the equivalent 
width distributions. This division at 3 Mpc is not entirely arbitrary though 
because it correctly separates the \lya lines which Stocke et 
al.\markcite{stoc95} (1995) and Shull et al.\markcite{ssp96} (1996) define 
as void and non-void absorbers on different grounds. Figure ~\ref{voidnot} 
compares the resulting equivalent width distributions of the void and 
non-void absorbers. While it does appear that the strongest lines tend to be 
non-void absorbers (which is not surprising given the discussion above), the 
two distributions are similar, and a KS test indicates that there is a 15\% 
probability that they were drawn from the same parent distribution. 
Therefore there is no statistically significant difference between the 
equivalent width distribution of the void and non-void absorbers in the 
present sample. Applying the same test to the ``1 Mpc complete'' sample, 
which is less prone to the problems discussed in the previous paragraph, 
yields a very similar result. In this case we find a 14\% probability that the 
void and non-void absorbers were drawn from the same equivalent width 
distribution.

\subsection{Statistical Relationships}

In \S 5.2 we addressed the question: if there is a galaxy within some impact 
parameter, do we find a \lya absorber near that redshift? The other obvious 
question to ask about our data is: given that we have detected a \lya absorber 
at redshift $z$, is there a galaxy or galaxy structure nearby? Toward PG 
1116+215, 69\% of the \lya absorbers in the total sample have nearest 
neighbor galaxies within three-dimensional distances of 2 Mpc, while toward 
H 1821+643, 47\% of the absorbers have galaxies within 2 Mpc, assuming 
the perturbed Hubble flow\footnote{In their full sample, Stocke et 
al.\markcite{stoc95} (1995) found that 70\% of the \lya lines are within 2 
Mpc of a galaxy.}. In the ``1 Mpc complete'' sample, 75\% of the PG 
1116+215 absorbers and 53 \% of the H 1821+643 absorbers are within 2 
Mpc of the nearest galaxy.

Specifically, we consider the following issues. (1) Previous studies have 
indicated that overall, the \lya absorbers are not randomly distributed with 
respect to galaxies. Is this confirmed by our data? Recently this conclusion 
has been challenged by Grogin \& Geller\markcite{gg98} (1998). (2) As 
noted in \S 1, there is considerable interest in comparing the relationship 
between the weak \lya absorbers and galaxies to that between the strong \lya 
absorbers and galaxies. Is there a statistically significant difference? Stocke 
et al.\markcite{stoc95} (1995) split their sample into weak and strong 
absorbers and found that the weak lines are statistically consistent with a 
random distribution. However, Stocke et al.\markcite{stoc95} emphasize 
that their sample of weak clouds is rather small. We can check this result 
with a larger sample.

Following the previous studies of Morris et al.\markcite{mor93} (1993) and 
Stocke et al.\markcite{stoc95} (1995), we use Monte Carlo nearest neighbor 
tests to examine these issues. To do this, we randomly distributed absorbers 
along the H 1821+643, PG 1116+215, and 3C 273 sight lines with a line 
density in accordance with Eqn. (1); we tried various values of $\gamma$ 
ranging from 0.0 to 0.5 and found no significant changes in the results. The 
number of absorbers randomly distributed on each sight line was set equal to 
the number in the observed \lya sample, and we ran 1000 simulations for 
each QSO. We then determined the three-dimensional nearest neighbor 
distance for each real absorber-galaxy, random absorber-galaxy, and galaxy-
galaxy pair as discussed above and plotted their cumulative nearest neighbor 
distributions. The results for the total and complete samples are shown in 
Figure ~\ref{cumdist} (see \S 5.1 for the sample definitions). The cumulative 
percentage on the y-axis in this figure indicates the fraction of objects 
(absorbers, galaxies, and random absorbers) with nearest neighbor distances 
equal to or less than the corresponding value on the x-axis. The results of 
Kolmogorov-Smirnov (KS) tests which compare the nearest neighbor 
distributions for all of the samples defined in \S 5.1 are summarized in Table 
~\ref{kstests}. Column 6 of Table ~\ref{kstests} shows that {\it the 
probability that the \lya clouds are drawn from a random distribution is 
extremely small for all samples,} in agreement with the previous conclusions 
of Morris et al.\markcite{mor93} (1993) and Stocke et al.\markcite{stoc95} 
(1995). The ``complete'' and ``1 Mpc complete'' samples yield similar results, 
which indicates that the limited spatial coverage of the redshift survey at low 
$z$ does not have a severe impact. Table ~\ref{kstests} also shows that for 
the complete sample, the galaxy-galaxy and galaxy-\lya distributions are 
statistically difficult to distinguish. However, this does not necessarily 
indicate that the \lya absorbers are clustered as strongly as galaxies in general 
because the galaxy clustering implied by the galaxy-galaxy nearest neighbor 
distribution in Figure ~\ref{cumdist} is underestimated due to 
incompleteness of the redshift surveys.

We have also divided our complete sample into ``weaker'' and ``stronger'' 
absorbers based on the \lya line equivalent widths. Figure ~\ref{cumweak} 
compares the resulting nearest neighbor distribution of the \lya lines with 
$W_{\rm r} >$ 100 m\AA\ to the nearest neighbor distribution of the lines 
with $W_{\rm r} <$ 100 m\AA . We selected 100 m\AA\ as the cutoff 
between the weak and strong absorbers because this divides the sample 
roughly in half (28 weak vs. 21 strong absorbers) and because this is the 
cutoff used by Stocke et al.\markcite{stoc95} (1995). Table ~\ref{ksweak} 
lists the KS probabilities that the weak and strong absorbers are drawn from 
a random distribution. Figure ~\ref{cumweak} and Table ~\ref{ksweak} 
clearly show that {\it neither the weak absorbers nor the strong absorbers are 
randomly distributed with respect to the galaxies in the survey.} Figure 
~\ref{cumweak} also shows that the weaker absorbers are difficult to 
distinguish from the stronger absorbers, at least based on the nearest neighbor 
distribution with 100 m\AA\ delineating weak lines from strong. However, 
previous studies indicate that the cutoff between the hypothesized two 
populations of \lya absorbers (see \S 1) may be closer to 300 m\AA\ than 100 
m\AA . Unfortunately, we have very few \lya lines in our sample with 
$W_{\rm r} \ >$ 300 m\AA , so we cannot apply the same test using 300 
m\AA\ as the cutoff. However, the sparse data that we do have suggest that 
there is a difference between absorbers with $W_{\rm r} \ >$ 300 m\AA\ and 
weaker absorbers, as expected from the anticorrelation discussed in \S 5.2. 
For example, four out of six \lya lines in the H 1821+643 and PG 1116+215 
spectra with $W_{\rm r} \ >$ 300 m\AA\ (not including lines within 3000 
\kms\ of the QSO redshift) have a nearby galaxy with $\rho \ <$ 200 kpc and 
$\Delta v <$ 300 \kms , and the fifth has a galaxy at $\rho$ = 373 kpc and  
$\mid \Delta v \mid$ = 246 \kms . The only \lya line with $W_{\rm r} \ >$ 
300 m\AA\ that does not have a nearby galaxy is at $z >$ 0.2 toward H 
1821+643 where the galaxy redshift survey is very incomplete. In contrast, 
out of the 41 remaining \lya lines with $W_{\rm r} \ <$ 300 m\AA , only 
two have $\rho \ <$ 200 kpc and $\Delta v <$ 300 \kms , and these are both 
within 350 \kms\ of a strong \lya line with $W_{\rm r} \ >$ 300 m\AA\ (see 
Tables ~\ref{h1821near} and ~\ref{p1116near}). If we divide the sample 
into lines with $W_{\rm r} \ >$ 200 m\AA\ and $W_{\rm r} \ <$ 200 m\AA 
, we obtain the nearest neighbor distributions shown in Figure 
~\ref{weak200} and the statistical results in the lower half of Table 
~\ref{ksweak}. While Figure ~\ref{weak200} does suggest that there is a 
difference between the weaker and stronger lines, the difference has a 
marginal statistical significance; a KS test indicates that there is a 6.5\% 
probability that the weaker and stronger absorbers in Figure ~\ref{weak200} 
are drawn from the same distribution. However, there are only 9 lines in the 
sample with with $W_{\rm r} \ >$ 200 m\AA .  Clearly more observations 
are needed.

\section{The Nature of the Low Redshift Ly$\alpha$ Absorbers}

In this section we discuss the implications of our results for the relationship 
between \lya absorbers and galaxies. To facilitate the discussion, we first 
summarize the main results of our study in the context of earlier studies:

(1) The distribution of \lya absorbers in space is not random but rather is 
correlated with the distribution of galaxies as found previously (Morris et 
al.\markcite{mor93} 1993; Lanzetta et al.\markcite{lanz95} 1995; Stocke 
et al.\markcite{stoc95} 1995; Lanzetta et al.\markcite{lanz97} 1997). A 
new result of this work is that, because of our larger sample of weak \lya 
absorption lines, we find that even the very weak absorbers ($W_{\rm r} <$ 
100 m\AA ) are correlated with the galaxy distribution. On the other hand, 
the correlation between galaxies and absorbers is not as strong as the 
correlation between galaxies themselves (Morris et al.\markcite{mor93} 
1993; Mo \& Morris\markcite{mom94} 1994; Stocke et 
al.\markcite{stoc95} 1995; Lanzetta et al 1997). There appears to be 
marginal evidence that the weak absorption lines are less correlated with the 
galaxy distribution than the strong absorption lines, but the difference is not 
yet significant.

(2) Earlier work by Lanzetta and co-workers (Lanzetta et 
al.\markcite{lanz95} 1995, 1997\markcite{lanz97}) showed that luminous 
galaxies at impact parameters $<$ 200 kpc frequently have corresponding 
\lya absorption lines. They also suggested that galaxies at impact parameters 
greater than 200 kpc are almost never associated with corresponding \lya 
absorption lines. However, we find that all galaxies within 600 kpc of the 
quasar sightlines do have corresponding \lya absorption with $\Delta v <$ 
1000 \kms , and a significant number of galaxies at 600 $< \ \rho \ <$ 2200 
kpc have corresponding \lya absorption lines with $\Delta v <$ 1000 \kms . 
Furthermore, most of these galaxy-absorber pairs are separated by less than 
$\sim$350 \kms . These different conclusions result mostly from the fact that 
the \lya absorption lines tend to be weak for galaxies with large impact 
parameters, and the FOS spectra used by Lanzetta et al.\markcite{lanz97} 
(1997) are not sufficiently sensitive to detect these weak lines.

(3) It was shown by Lanzetta and co-workers (Lanzetta et 
al.\markcite{lanz95} 1995,1997\markcite{lanz97}; Chen et 
al.\markcite{chen98} 1998) that the equivalent widths of the \lya absorption 
lines are anticorrelated with the impact parameters to galaxies for impact 
parameter up to 200 kpc and $\Delta v$ up to 500 \kms . We find that the 
same relation extends to impact parameters as large as $\sim$2 Mpc for 
$\Delta v <$ 1000 \kms . In addition, there is no noticeable increase in the 
scatter around the mean anticorrelation relation going from small $\rho$ to 
large $\rho$ (see Figure ~\ref{bigrhocor}), while increasingly large scatter 
with increasing $\rho$ might be expected if the absorbers and the galaxies 
are physically related at small impact parameters but are less physically 
related at larger projected distances. However, there are potential selection 
effects which can artificially tighten this anticorrelation, and 
follow-up studies are needed to determine whether or not it is real.

(4) Some prominent groups of galaxies do not have clearly associated \lya 
lines. However other galaxy groups do have associated \lya absorbers. 
Evidently \lya absorbers do not always avoid regions of higher galaxy 
density, but they may avoid the densest regions.

(5) Dividing the \lya absorbers into ``void'' and ``non-void'' absorbers based 
on whether or not there is a galaxy within 3 Mpc in 3-dimensional space, 
assuming a perturbed Hubble flow for the determination of radial distance, 
we find that the equivalent width distributions of void and non-void 
absorbers are statistically indistinguishable.

There are two main competing interpretations of the relationship between 
galaxies and \lya absorbers at low redshifts. Lanzetta and co-workers (e.g., 
Lanzetta et al.\markcite{lanz95} 1995,1997\markcite{lanz97}; Chen et 
al.\markcite{chen98} 1998) have suggested that low redshift \lya absorbers 
(mostly strong absorption lines with $W_{\rm r} >$ 300 m\AA ) are 
physically associated with individual galaxies which are surrounded by 
extended gaseous envelopes of $\sim$200 kpc radius of unknown geometry 
which give rise to the observed \lya absorption. Two main arguments are 
cited in support of the physical association interpretation: (1) the 
galaxy-absorber cross-correlation function appears to be strong for $\rho <$ 
200 kpc and $\Delta v <$ 500 \kms\ and is much weaker at larger impact 
parameters and velocity separations (Lanzetta et al.\markcite{lanz97} 
1997); (2) there is an anticorrelation between absorber equivalent width and 
impact parameter for galaxy-absorber pairs that are likely to be physically 
associated (Chen et al.\markcite{chen98} 1998), where ``physically 
associated'' pairs refer to those with galaxy-absorber cross-correlation 
significantly above zero (essentially galaxy-absorber pairs with $\rho <$ 200 
kpc and $\Delta v <$ 500 \kms ). The latter point suggests that the density of 
the gas distribution near galaxies drops off with increasing galactocentric 
radius. Some independent support for this picture is provided by Zaritsky et 
al.\markcite{zari97} (1997) who conclude, based on the velocity dispersion 
of satellites around isolated spiral galaxies, that the dark matter halos of 
spirals have radii $>$ 200 kpc.

However, there are difficulties with the gaseous halo interpretation (see, e.g., 
\S 5 in Stocke et al.\markcite{stoc95} 1995). It is difficult to understand how 
galaxies would be able to maintain such large gaseous halos --- such halos 
would probably be weakly bound and vulnerable to disruption by 
interactions in groups and clusters. Our finding that the equivalent width 
versus impact parameter anticorrelation extends to impact parameters as 
large as $\sim$600 kpc and may extend to several Mpc further aggrevates 
the problem and challenges the Lanzetta et al.\markcite{lanz95} 
interpretation. In addition, there are several instances in our sample and 
elsewhere (e.g., Rauch et al.\markcite{rwm96} 1996) in which a group of 
galaxies within a few 100 \kms\ of each other with separations comparable to 
the impact parameter to the nearest galaxy appear to be responsible for the 
same \lya absorption. As noted by Rauch et al.\markcite{rwm96} (1996), the 
``cloud size'' or ``coherence length'' of \lya absorbers inferred by Dinshaw et 
al.\markcite{dins95} (1995,1997\markcite{dins97}) from common \lya 
lines detected toward QSO pairs, $\sim$700 kpc, is much larger than the 
transverse separation between galaxies so it is doubtful that the large cloud 
size could refer to absorption produced by a single galactic halo or disk. 
Rather, the large transverse sizes and small velocity differences of common 
\lya absorbers measured by Dinshaw et al.\markcite{dins95} appear to favor 
spatially coherent and rather quiescent gaseous structures on scales 
significantly larger than the virial radii of individual galaxies.

An alternative to the Lanzetta et al.\markcite{lanz95} interpretation is that, 
rather than being associated with individual galaxies, the \lya absorbers trace 
the large scale structures in which both galaxies and gas reside. Recent 
cosmological simulations of structure formation involving gas 
hydrodynamics suggest a picture in which galaxies are surrounded and 
connected by filamentary or sheet-like gaseous structures. Such structures 
have properties which are consistent with the \lya absorption lines at $z$ = 2-
4, as has been shown by a number of recent studies (e.g., Cen et 
al.\markcite{cen94} 1994; Petitjean, M\"{u}cket, \& 
Kates\markcite{peti95} 1995; Miralda-Escud\'{e} et al.\markcite{mira96} 
1996; Hernquist et al.\markcite{hern96} 1996; Zhang et al. 
1997\markcite{zh97},1998\markcite{zh98}). In these simulations, the 
density of the gas distribution drops with increasing distance from galaxies, 
and often even void regions contain enough gas to produce \lya absorption 
lines, albeit with much lower \ion{H}{1} column densities. It is reasonable 
to expect that the overall structure of the gas distribution at $z$ = 0 will 
remain similar to that at $z \ >$ 2, although the amount of gas in void 
regions and in the filamentary structures should decrease as the gas ``drains'' 
through the filaments into galaxies.

The association of \lya absorbers with large scale structures appears to be 
consistent with all of the observational evidence obtained so far for low $z$ 
\lya absorbers. The distribution of absorbers is obviously correlated with that 
of galaxies because they trace the same large scale structures, but clearly the 
absorber-galaxy correlation function will be weaker than the 
galaxy-galaxy correlation function. Since most (if not all) galaxies are likely 
to be engulfed by diffuse gas in the filamentary structures, an absorption line 
is likely to be found when a galaxy is near the QSO sightline (result (2) 
above). The anticorrelation between absorber equivalent width and impact 
parameter and the persistence of such an anticorrelation to very large impact 
parameters can also be understood in analogy with the simulations at large 
redshift; the density of the gas distribution is expected to decrease on average 
as one moves away from galaxies. However, since the gas distribution is not 
regular and smooth, a large scatter in this anticorrelation relation is expected, 
as observed (see Figures ~\ref{rhocor} and ~\ref{bigrhocor}). Chen et 
al.\markcite{chen98} (1998) attemped to reduce the scatter in the equivalent 
width vs impact parameter relation by incorporating various galaxy 
properties as a secondary parameter, such as galaxy luminosity, mean surface 
brightness, effective radius, disk-to-bulge ratio, redshift, and shape of the gas 
distribution; only marginal improvement in the relation was found even in 
the best case when galaxy B-band luminosity was incorporated into the 
relation\footnote{The  significance of the anticorrelation improved from 3.4-
4.5$\sigma$ to 4.2-5.9$\sigma$, depending on the statistical test applied.}. 
The apparent insensitivity of this relation to galaxy properties is naturally 
explained if the absorption arises from the overall large scale structure rather 
than in individual galaxies. 

Some absorption lines should also occur when the quasar sightlines pass 
through regions in the large scale structure that are relatively far (many Mpc) 
away from galaxies, where there may still be gas either in filaments or in the 
voids. This is consistent with result (5), namely, the equivalent width 
distributions of void and non-void absorbers are statistically 
indistinguishable. However, the absorption lines arising from regions in 
space far away from galaxies should, to some degree, be weaker on average. 
A much larger \lya line sample is required to search for such a trend.

A third interpretation of the \lya absorbers is that the lines are associated with 
faint dwarf galaxies which may be able to generate large ($\sim$100 kpc) 
gaseous halos via supernova-driven winds (Wang\markcite{wang95} 1995; 
Nath \& Trentham\markcite{nath97} 1997). At low $z$, this interpretation 
has gained support from the 21 cm imaging study of van Gorkom et 
al.\markcite{vang96} (1996) which has located gas-rich dwarf galaxies at 
the redshift of some low $z$ \lya lines. There are some potential problems 
with this interpretation. First, the dwarf galaxies are usually more than 
$\sim$100 kpc from the sight lines. Several sensitive imaging searches for 
very faint galaxies in the vicinity of the two lowest-redshift \lya absorbers 
toward 3C 273 have failed to find faint galaxies within 100 kpc of the sight 
line (e.g., Morris et al.\markcite{mor93} 1993; van Gorkom et 
al.\markcite{vang93} 1993; Rauch et al.\markcite{rwm96} 1996). By 
searching a somewhat larger field, Salzer\markcite{salz92} (1992) and 
Hoffman et al.\markcite{hoff98} (1998) have found two dwarf galaxies at 
projected distances of $\sim$200 kpc near the lowest redshift 3C 273 
absorber. Hoffman et al.\markcite{hoff98} (1998) have measured the 
rotation curve of one of these dwarfs, and they show that extension of the 
observed rotation curve cannot explain the velocity of the associated \lya 
absorber (see their Figure 6), but this may not be surprising if the \lya line is 
due to a galactic wind. A second criticism of the dwarf galaxy model of \lya 
absorbers stems from the recent study of Martin\markcite{martin98} (1998). 
Martin\markcite{98} (1998) has explored the likelihood that dwarf galaxies 
will drive galactic winds by surveying the expansion rates and sizes of their 
H$\alpha$ shells. She concludes that ``many of the shells will breakthrough 
the surrounding \ion{H}{1} layer supersonically, but the projected 
expansion speeds are typically less than the lower limits on the escape 
velocity.'' Martin\markcite{martin98} notes that in a few objects in her 
sample (e.g., NGC 1569, see also Heckman et al.\markcite{heck95} 1995), 
an unbound galactic wind is likely, but since she has exclusively selected 
dwarfs with prominent shells and only $\sim$25\% of dwarf galaxies show 
prominent H$\alpha$ filaments (Hunter, Hawley, \& 
Gallagher\markcite{hunt93} 1993), her study seems to suggest that the 
majority of nearby dwarf galaxies will not drive winds. Since many recent 
studies have shown that galaxy clusters and large scale structures contain 
large numbers of dwarf galaxies (e.g., Phillipps et al.\markcite{phip98} 
1998, and references therein) and since Gallagher, Littleton, \& 
Matthews\markcite{gall95} (1995) and Phillipps et al.\markcite{phip98} 
have shown that in these structures the distribution of the dwarfs tends 
to be more spatially extended than that of the giant galaxies, it seems 
more likely that \lya lines with dwarfs 100-200 kpc away are due to gas 
in the large scale structures where the dwarfs are found rather than the 
dwarfs themselves.

The hypothesis that many low redshift \lya absorption lines with $W_{\rm 
r}$ in the range from 50 to $\sim$500 m\AA\ trace the overall gas 
distribution in the large scale structures rather than gaseous halos of 
individual galaxies is physically plausible and is consistent with existing 
observations. However, it is probably true that a variety of phenomena cause 
\lya absorption lines in QSO spectra. The gaseous disks and halos of 
individual galaxies will certainly produce \lya absorption if the QSO sight 
line passes close enough, and these will probably be stronger lines. 
Additional studies at low redshift should ultimately allow the determination 
of all sites of origin for this important tracer of cosmic phenomena.

\section{Summary}

We have obtained high signal-to-noise UV spectra of the QSOs H 1821+643 
($z_{em}$ = 0.297) and PG 1116+215 ($z_{em}$ = 0.177) with the 
GHRS in order to study the nature of low $z$ \ion{H}{1} \lya absorbers and 
their relationship with galaxies. The spectra have S/N ranging from 75 to 
160 per 150 \kms\ resolution element and permit the detection of weak \lya 
lines. Toward H 1821+643 and PG 1116+215 we find 26 and 13 \lya lines, 
respectively, with 3$\sigma$ rest equivalent widths $W_{\rm r} >$ 50 
m\AA \ (not including lines within 3000 \kms\ of $z_{\rm em}$). The 
previous best studied sight line for \lya lines as weak as 50 m\AA , 3C 273, 
showed 13 \lya lines with $W_{\rm r} >$ 50 m\AA\ at 3$\sigma$ 
significance or better (Morris et al.\markcite{mor93} 1993). We have also 
measured 98 and 118 galaxy redshifts in the $\sim \ 1^{\circ}$ fields 
centered on H 1821+643 and PG 1116+215 with the multiobject 
spectrograph on the WIYN Telescope. With 56 additional redshifts from the 
literature, we have a total of 154 redshifts in the H 1821+643 field. Within a 
20' radius from the QSO, we estimate that our H 1821+643 and PG 
1116+215 redshift surveys are respectively 72.4\% and 85.0\% complete for 
$B_{\rm J} <$ 18.0 and 50.8\% and 68.5\% complete for $B_{\rm J} <$ 
20.0.

From our new data combined with data from previous studies we obtain the 
following results.

(1) The number of \lya lines detected per unit redshift with $W_{\rm r} >$ 
50 m\AA\ is $dN/dz = 102\pm 16$ at $z <$ 0.28.

(2) Several of the stronger \lya profiles show complex multiple component 
structure, but there is no compelling evidence of clustering in the two-point 
velocity correlation function of the whole sample. However, if we exclude 
the weakest lines ($W_{\rm r} <$ 100 m\AA ), the two-point correlation 
function shows a marginal signal (3$\sigma$) on velocity scales of 
$\sim$500 \kms .

(3) All galaxies within projected distances of 600 $h_{75}^{-1}$ kpc have 
associated \lya absorbers with $\Delta v <$ 1000 \kms , and the \lya 
equivalent widths are anticorrelated at high significance with the projected 
distance of the nearest galaxy out to $\rho \approx$ 600 $h_{75}^{-1}$ kpc. 
Further studies are needed to establish whether this anticorrelation is real or 
due to selection effects. For $\rho > 600 h_{75}^{-1}$ kpc, we find galaxies 
which do not have associated \lya lines. However, the anticorrelation 
continues to $\rho \approx 2 h_{75}^{-1}$ Mpc for galaxies within 500 or 
1000 \kms\ of a \lya absorber. Most of these galaxy-absorber pairs have 
$\Delta v \lesssim$ 350 \kms .

(4) We find three prominent galaxy clusters which do not have associated 
strong \lya absorbers. However, we also find galaxy groups which {\it do} 
have associated \lya absorbers.

(5) As in previous studies, we find \lya lines in voids at large distances from 
the nearest galaxies, but this may be due to the limited spatial coverage (at 
low $z$) or limited depth (at high $z$) of the galaxy redshift survey. 
Statistically, the equivalent width distributions of the void and non-void 
absorbers are indistinguishable, but the sample is small.

(6) Statistical tests show that the \lya absorbers are not randomly distributed 
with respect to the galaxies. Splitting the sample into roughly equal groups of 
weak ($W_{\rm r} <$ 100 m\AA ) and strong absorbers shows that the weak 
absorbers are  also not randomly distributed. Comparison of the nearest 
neighbor distances of the weak and strong absorbers suggests that the weak 
absorbers may be less closely associated with galaxies than the strong 
absorbers, but with the small sample this difference is not yet statistically 
significant.

(7) The observations are consistent with the hypothesis that many of the low 
redshift \lya absorbers with rest equivalent widths in the range from 50 to 
$\sim$500 m\AA\ trace the overall gas distributions in the large scale 
structures in which galaxies reside rather than the gaseous halos of individual 
galaxies. However, it is likely that not all \lya lines are produced in this way. 
Some of the stronger \lya lines may arise in the halos of individual galaxies.

\acknowledgments
The galaxy redshift survey employed in this paper is a collaboration with 
Buell Jannuzi and coworkers, and we thank this group, especially Buell and 
Angelle Tanner, for allowing us to use these redshifts in advance of 
publication. We acknowledge Michael Rauch for helpful discussions, and we 
thank R. Carswell for sharing his profile fitting software VPFIT. We also 
thank Jay Gallagher, Ed Jenkins, and especially John Stocke for valuable 
comments on the manuscript. Our observational program has utilized data 
from the {\it Hubble Space Telescope} and the WIYN 3.5m optical 
observatory on Kitt Peak. We very much appreciate the work of the many 
people who helped create and operate these facilities. B.D.S. and T.M.T. 
appreciate support from NASA through Grant GO-06499.02-95A. L.L. 
appreciates support through NASA funded Hubble Fellowship HF 1062.01-
94A.

Jason Cardelli made many contributions to the success of the Goddard High 
Resolution Spectrograph on {\it HST}, and some of Jason's work played an 
important role in this paper. Sadly, Jason died suddenly on May 14, 1996 at 
the age of 40. We dedicate this paper to Jason.

\clearpage
\begin{deluxetable}{llccccc}
\tablewidth{0pc}
\tablecaption{Redshifts of Galaxies in the Field of H 
1821+643\label{h1821reds}}
\tablehead{\ \ \ \ \ RA &\ \ \ \ \ Dec & $z_{\rm gal}$\tablenotemark{a} & 
$\sigma _{z}$\tablenotemark{a} & POSS II $B_{\rm J}$\tablenotemark{b} 
& $M_{B}$\tablenotemark{c} & $\rho$\tablenotemark{d} \nl
\multicolumn{2}{c}{(J2000)} & \ & (km s$^{-1}$) & \ & \ & (Mpc)}
\startdata
\multicolumn{7}{c}{\underline{\ \ \ \ \ \ \ \ \ \ \ \ \ \ \ Redshifts from Tripp et 
al. (1998)\ \ \ \ \ \ \ \ \ \ \ \ \ \ \ }} \nl
18:17:12.35 & 64:33:56.0 & 0.02089 &  12 &\nodata&\nodata& 0.788\nl
18:21:41.17 & 63:51:38.0 & 0.02404 &  17 & 17.4 & -17.8 &  0.783\nl
18:21:27.36 & 64:12:11.3 & 0.02752 &  12 & 20.6 & -14.9 &  0.277\nl
18:23:25.35 & 64: 8:34.2 & 0.05003 &  48 & 15.6 & -21.2 &  0.832\nl
18:21:49.15 & 64:37:49.0 & 0.05024 &  36 & 15.8 & -21.0 &  0.936\nl
18:22:31.72 & 64:35:29.9 & 0.05051 &  49 & 16.7 & -20.2 &  0.838\nl
18:18:14.76 & 64:18:23.5 & 0.05059 &  25 & 18.2 & -18.6 &  1.322\nl
18:21:52.71 & 64:39: 0.1 & 0.05062 &  16 & 17.5 & -19.4 &  1.006\nl
18:16:24.99 & 64: 5:11.9 & 0.05062 &  18 & 18.9 & -18.0 &  2.147\nl
18:22:41.28 & 64:10:27.8 & 0.05067 &  12 & 18.0 & -18.8 &  0.613\nl
18:22:50.89 & 64: 9:20.2 & 0.05079 &  55 &\nodata&\nodata& 0.696\nl
18:24:21.41 & 64: 6:26.3 & 0.05090 &  48 & 15.5 & -21.3 &  1.161\nl
18:23:42.37 & 64:32:54.2 & 0.05231 &  35 & 17.9 & -19.0 &  0.943\nl
18:24:33.29 & 64:16:14.5 & 0.05245 &  31 & 17.9 & -19.0 &  0.987\nl
18:16:31.37 & 64:43:54.6 & 0.06553 &  36 & 16.9 & -20.5 &  2.916\nl
18:22:30.84 & 64: 4:13.7 & 0.07103 &  30 & 17.6 & -20.0 &  1.251\nl
18:24:33.51 & 64: 6:21.0 & 0.07149 &  34 & 18.9 & -18.7 &  1.663\nl
18:20:56.89 & 64:27:54.5 & 0.07170 &  27 & 18.3 & -19.3 &  0.736\nl
18:27:46.25 & 64:14:28.2 & 0.07177 &  73 & 17.0 & -20.6 &  2.886\nl
18:25:46.50 & 64: 7:56.7 & 0.07192 &  44 & 18.4 & -19.2 &  2.107\nl
18:25:43.04 & 64:53:37.3 & 0.07608 &  29 & 18.4 & -19.3 &  3.246\nl
18:18:17.51 & 64:34:37.3 & 0.07843 &  37 & 18.7 & -19.1 &  2.242\nl
18:26:16.03 & 63:54: 5.7 & 0.08343 &  28 & 18.2 & -19.7 &  3.334\nl
18:24:29.24 & 64: 1: 5.2 & 0.08432 &  37 & 18.8 & -19.1 &  2.224\nl
18:23:38.83 & 64:41:47.7 & 0.08796 &  41 &\nodata&\nodata& 2.151\nl
18:22:29.42 & 64:41: 6.7 & 0.08822 &  40 & 17.7 & -20.3 &  1.881\nl
18:23:44.86 & 64:42:47.0 & 0.08834 &  45 & 18.5 & -19.5 &  2.266\nl
18:19:44.77 & 64:23: 7.3 & 0.08846 & 100 & 17.7 & -20.3 &  1.319\nl
18:20:42.96 & 64:19:45.8 & 0.08930 &  67 & 19.4 & -18.7 &  0.739\nl
18:18:35.52 & 64:14:22.3 & 0.09532 &  57 & 18.1 & -20.1 &  2.201\nl
18:25:52.36 & 64:40: 7.5 & 0.09574 &  57 & 17.8 & -20.4 &  3.105\nl
18:24:28.05 & 64:16:42.7 & 0.09701 &  66 & 18.4 & -19.8 &  1.651\nl
18:23:29.55 & 64:40:22.6 & 0.10614 &  96 & 18.8 & -19.7 &  2.351\nl
18:24:40.32 & 64:34:54.8 & 0.10642 &  35 & 17.8 & -20.6 &  2.413\nl
18:21:36.59 & 64:45:23.6 & 0.10682 &  32 & 17.8 & -20.7 &  2.658\nl
18:20: 7.54 & 65: 4: 0.3 & 0.10689 &  38 & 17.4 & -21.1 &  4.804\nl
18:25:53.45 & 64:19:19.7 & 0.11156 &  79 & 18.8 & -19.8 &  2.839\nl
18:20: 8.65 & 64:21: 4.2 & 0.11208 & 102 & 20.2 & -18.3 &  1.308\nl
18:25: 4.45 & 64:25:22.9 & 0.11957 &  99 & 18.9 & -19.8 &  2.445\nl
18:21: 0.88 & 64: 3:55.7 & 0.12120 & 105 &\nodata&\nodata& 2.112\nl
18:22: 2.65 & 64:21:39.3 & 0.12154 &  70 & 18.3 & -20.5 &  0.144\nl
18:21:22.38 & 64:28:29.0 & 0.12224 &  34 & 18.5 & -20.2 &  1.046\nl
18:23:58.19 & 64:26:52.3 & 0.12258 &  76 & 17.3 & -21.5 &  1.740\nl
18:21:20.26 & 64: 3:44.8 & 0.13774 &  59 & 18.5 & -20.5 &  2.291\nl
18:21:57.42 & 63:52:44.3 & 0.14220 & 206 &\nodata&\nodata& 3.782\nl
18:28:15.51 & 64:33:39.5 & 0.14228 &  70 & 18.3 & -20.8 &  5.816\nl
18:22: 4.38 & 64: 8:38.5 & 0.15465 &  59 & 18.9 & -20.4 &  1.741\nl
18:22:49.76 & 64:19:28.6 & 0.16376 & 100 &\nodata&\nodata& 0.882\nl
18:16:24.55 & 64: 2: 6.2 & 0.16556 &  94 & 18.2 & -21.2 &  6.238\nl
18:16:43.64 & 64:34:56.2 & 0.16564 &  39 & 17.9 & -21.6 &  5.635\nl
18:21:36.61 & 64:21:25.0 & 0.17086 &  74 & 19.1 & -20.4 &  0.373\nl
18:19:24.36 & 64: 7:53.6 & 0.17919 & 100 & 18.4 & -21.2 &  3.414\nl
18:19:55.22 & 64:35: 8.4 & 0.18012 &  56 & 18.9 & -20.7 &  3.213\nl
18:21:14.82 & 64:12:16.4 & 0.18504 &  35 & 20.9 & -18.7 &  1.592\nl
18:20:24.54 & 63:56:31.0 & 0.18532 &  97 & 19.4 & -20.2 &  4.377\nl
18:20:21.11 & 63:59: 1.2 & 0.18547 &  39 & 19.3 & -20.4 &  4.022\nl
18:19:45.65 & 64:30:52.5 & 0.18596 &  95 & 19.1 & -20.5 &  2.945\nl
18:20:25.56 & 64:23:59.0 & 0.18599 &  52 & 21.0 & -18.7 &  1.760\nl
18:20:25.99 & 64:14:57.5 & 0.18742 &  63 &\nodata&\nodata& 1.925\nl
18:20:43.29 & 65: 0:15.1 & 0.18921 &  40 & 18.6 & -21.1 &  6.885\nl
18:22:25.14 & 64:22:46.2 & 0.18967 &  40 & 19.9 & -19.8 &  0.635\nl
18:22:53.12 & 64:55:24.0 & 0.19117 &  56 & 19.1 & -20.7 &  6.060\nl
18:21:18.06 & 63:53:47.8 & 0.19161 &  39 & 18.7 & -21.0 &  4.667\nl
18:22:29.69 & 64:23: 8.2 & 0.19162 &  53 & 19.1 & -20.6 &  0.745\nl
18:22:20.96 & 64:26:52.2 & 0.19212 &  35 & 18.1 & -21.6 &  1.167\nl
18:23:59.02 & 64:25:58.7 & 0.19213 &  59 & 18.8 & -21.0 &  2.450\nl
18:22:14.04 & 64: 3: 5.4 & 0.20057 &  54 & 18.9 & -21.0 &  3.134\nl
18:21:38.82 & 64:20:31.7 & 0.22650 &  32 & 19.8 & -20.3 &  0.388\nl
18:23:47.60 & 64: 3:16.8 & 0.24433 &  49 & 19.6 & -20.7 &  4.340\nl
18:21:56.28 & 64:22:50.4 & 0.24435 &  64 & 20.2 & -20.1 &  0.462\nl
18:22:19.11 & 64:18:43.5 & 0.24568 &  86 & 20.8 & -19.5 &  0.625\nl
18:20:31.50 & 64:20:24.0 & 0.25147 &  72 & 19.8 & -20.5 &  1.950\nl
18:22:10.26 & 64:17:15.6 & 0.26669 &  80 &\nodata&\nodata& 0.795\nl
18:24: 4.97 & 63:52:38.0 & 0.27548 & 116 & 19.1 & -21.5 &  7.001\nl
18:20:31.98 & 64:22:19.7 & 0.27889 & 114 & 20.1 & -20.4 &  2.119\nl
18:22:30.01 & 64:13:16.4 & 0.27987 &  55 & 20.6 & -19.9 &  1.845\nl
18:23:50.02 & 63:52: 3.2 & 0.28420 & 100 & 19.5 & -21.1 &  7.115\nl
18:22: 0.23 & 64:18:48.9 & 0.28787 &  62 & 21.1 & -19.5 &  0.419\nl
18:21:48.70 & 64:24:20.4 & 0.28858 &  78 & 20.9 & -19.7 &  0.891\nl
18:21:39.29 & 64:22: 4.8 & 0.28880 &  92 & 20.5 & -20.2 &  0.564\nl
18:21:34.40 & 64:20:31.0 & 0.29044 &  82 & 21.4 & -19.2 &  0.574\nl
18:21:53.32 & 64:20:23.7 & 0.29056 &  83 &\nodata&\nodata& 0.109\nl
18:21:58.42 & 64:18:24.1 & 0.29116 &  89 & 20.9 & -19.7 &  0.512\nl
18:21:19.97 & 64:22:57.1 & 0.29216 &  53 & 20.2 & -20.5 &  1.088\nl
18:22:19.84 & 64:23:36.1 & 0.29305 &  79 & 20.7 & -20.0 &  0.905\nl
18:21:44.81 & 64:24:11.2 & 0.29359 &  97 & 20.3 & -20.4 &  0.896\nl
18:22:33.10 & 64:18:50.8 & 0.29481 &  54 & 21.1 & -19.6 &  1.000\nl
18:21:58.69 & 64:26:45.7 & 0.29552 &  84 & 21.3 & -19.4 &  1.448\nl
18:21:37.00 & 64:23:28.6 & 0.29606 & 101 & 20.7 & -20.0 &  0.850\nl
18:22:27.25 & 64:18:20.3 & 0.29639 &  74 &\nodata&\nodata& 0.933\nl
18:22: 3.51 & 64:23: 0.9 & 0.29766 &  80 & 19.8 & -20.9 &  0.592\nl
18:22:12.61 & 64:26:32.9 & 0.29828 &  66 & 20.2 & -20.5 &  1.460\nl
18:21:29.76 & 64:27:42.3 & 0.30025 &  50 & 21.0 & -19.7 &  1.828\nl
18:21:55.72 & 64:20: 3.1 & 0.30294 &  55 &\nodata&\nodata& 0.137\nl
18:24:45.87 & 64:33: 8.2 & 0.30512 &  85 & 19.9 & -20.8 &  5.302\nl
18:21:39.49 & 64:15:20.8 & 0.30529 &  50 & 20.7 & -20.0 &  1.343\nl
18:22:11.36 & 64:28:50.5 & 0.33069 & 144 & 20.8 & -20.1 &  2.121\nl
18:20:33.20 & 64:22:13.1 & 0.33233 &  40 & 20.3 & -20.7 &  2.344\nl
\multicolumn{7}{c}{\underline{\ \ \ \ \ \ \ \ \ \ \ \ \ \ \ Redshifts from Schneider 
et al. (1992)\ \ \ \ \ \ \ \ \ \ \ \ \ \ \ }} \nl
18:21:54.40 & 64:20: 9.3 & 0.22560 & 150 & 19.5 & -20.6 &  0.105\nl
18:21:55.98 & 64:21: 0.8 & 0.29160 & 150 & 19.8 & -20.9 &  0.101\nl
18:21:59.36 & 64:19:49.2 & 0.29360 & 300 & 21.4 & -19.2 &  0.191\nl
18:21:54.23 & 64:20:13.4 & 0.29860 & 300 &\nodata&\nodata& 0.117\nl
18:21:52.80 & 64:20:43.9 & 0.30060 & 300 & 21.3 & -19.4 &  0.118\nl
18:21:54.97 & 64:21:17.1 & 0.30310 & 300 & 19.0 & -21.7 &  0.174\nl
\multicolumn{7}{c}{\underline{\ \ \ \ \ \ \ \ \ \ \ \ \ \ \ Redshifts from Le Brun 
et al. (1996)\ \ \ \ \ \ \ \ \ \ \ \ \ \ \ }} \nl
18:22:20.16 & 64:21:46.7 & 0.17850 &\nodata & 20.8 & -18.8 &  0.447\nl
18:22: 4.57 & 64:20:52.5 & 0.28440 &\nodata &\nodata&\nodata& 
0.193\nl
18:21:30.35 & 64:20:44.8 & 0.29090 &\nodata & 20.8 & -19.8 &  0.677\nl
18:22:10.61 & 64:20:29.0 & 0.29170 &\nodata &\nodata&\nodata& 
0.339\nl
18:21:47.72 & 64:20: 8.9 & 0.29200 &\nodata &\nodata&\nodata& 
0.262\nl
18:21:55.01 & 64:20: 4.7 & 0.29300 &\nodata &\nodata&\nodata& 
0.134\nl
18:21:51.21 & 64:20:52.1 & 0.29380 &\nodata &\nodata&\nodata& 
0.164\nl
18:22: 1.07 & 64:20:29.0 & 0.29500 &\nodata &\nodata&\nodata& 
0.102\nl
18:22:28.30 & 64:20:39.5 & 0.29530 &\nodata &\nodata&\nodata& 
0.791\nl
18:21:34.30 & 64:21: 0.9 & 0.29660 &\nodata &\nodata&\nodata& 
0.592\nl
18:22:13.61 & 64:20:14.5 & 0.29690 &\nodata &\nodata&\nodata& 
0.427\nl
18:21:44.59 & 64:21:18.5 & 0.29700 &\nodata & 19.9 & -20.8 &  0.363\nl
18:21:17.91 & 64:21:23.5 & 0.29770 &\nodata &\nodata&\nodata& 
1.022\nl
18:21:27.64 & 64:21:59.6 & 0.29980 &\nodata &\nodata&\nodata& 
0.828\nl
18:22:20.16 & 64:20:39.8 & 0.30000 &\nodata &\nodata&\nodata& 
0.590\nl
18:22:31.94 & 64:19:34.7 & 0.30100 &\nodata &\nodata&\nodata& 
0.927\nl
18:22:34.80 & 64:20:51.6 & 0.30100 &\nodata &\nodata&\nodata& 
0.970\nl
18:21:57.96 & 64:20:44.1 & 0.30130 &\nodata & 20.9 & -19.8 &  0.037\nl
18:22: 6.59 & 64:19:50.2 & 0.30180 &\nodata &\nodata&\nodata& 
0.303\nl
\multicolumn{7}{c}{\underline{\ \ \ \ \ \ \ \ \ \ \ \ \ \ \ Redshifts from Bowen et 
al. (1998)\ \ \ \ \ \ \ \ \ \ \ \ \ \ \ }} \nl
18:23:16.01 & 64: 4:52.9 & 0.02754 & \nodata & 16.6 & -18.9 &  0.550\nl
18:20: 2.47 & 64:18:53.0 & 0.02788 & \nodata & 18.5 & -17.0 &  0.390\nl
18:22:36.11 & 64:45:44.9 & 0.04970 & \nodata &\nodata&\nodata& 
1.370\nl
18:22: 7.61 & 64:39:32.4 & 0.05039 & \nodata & 18.1 & -18.7 &  1.033\nl
18:24:34.11 & 64:19: 4.1 & 0.05050 & \nodata & 17.2 & -19.7 &  0.931\nl
18:23:34.34 & 64:18:35.1 & 0.05099 & \nodata & 17.8 & -19.0 &  
0.590\nl
18:21:21.38 & 64:49:40.4 & 0.05683 & \nodata & 18.5 & -18.5 &  
1.784\nl
18:22:47.01 & 63:59:28.6 & 0.07201 & \nodata & 17.3 & -20.3 &  
1.646\nl
18:23:41.16 & 64: 1: 8.3 & 0.07249 & \nodata & 17.8 & -19.8 &  1.710\nl
18:26: 1.05 & 64:23:16.4 & 0.07623 & \nodata & 17.7 & -20.0 &  2.106\nl
18:17:26.18 & 64:19:43.7 & 0.08210 & \nodata & 18.6 & -19.3 &  
2.492\nl
18:22: 9.95 & 64: 9:36.9 & 0.08410 & \nodata & 18.3 & -19.6 &  0.960\nl
18:19:10.44 & 64: 7:27.4 & 0.09458 & \nodata & 18.6 & -19.5 &  2.152\nl
18:25:26.87 & 64:36:39.1 & 0.09517 & \nodata & 18.4 & -19.8 &  
2.678\nl
18:25: 4.56 & 64:40:35.8 & 0.09583 & \nodata & 18.4 & -19.8 &  2.761\nl
18:24:33.84 & 64:38:51.2 & 0.09605 & \nodata & 17.4 & -20.8 &  
2.421\nl
18:24:40.55 & 64: 2:45.1 & 0.09681 & \nodata & 17.5 & -20.8 &  2.471\nl
18:25:20.57 & 64:16:35.7 & 0.09697 & \nodata & 18.8 & -19.5 &  
2.201\nl
18:20:53.45 & 64:19:37.1 & 0.11155 & \nodata & 18.6 & -20.0 &  
0.773\nl
18:22:40.55 & 64: 9:43.7 & 0.12050 & \nodata & 18.5 & -20.2 &  1.402\nl
18:18: 2.25 & 64:14: 4.2 & 0.12103 & \nodata & 17.9 & -20.8 &  3.124\nl
18:25: 8.16 & 64: 9:17.3 & 0.13853 & \nodata & 18.5 & -20.5 &  3.139\nl
18:21: 8.73 & 63:52: 9.0 & 0.15039 & \nodata & 18.4 & -20.8 &  4.111\nl
18:19:34.62 & 64:42:21.0 & 0.16410 & \nodata & 19.3 & -20.1 &  
4.054\nl
18:18:28.71 & 64:38: 3.6 & 0.16550 & \nodata & 18.9 & -20.5 &  4.362\nl
18:19:40.64 & 64:30:38.5 & 0.17955 & \nodata & 18.7 & -20.9 &  
2.916 \tablebreak 
18:18:17.71 & 64:16:15.9 & 0.18002 & \nodata & 18.2 & -21.4 &  
3.962\nl
18:21:35.27 & 64:25:24.0 & 0.18900 & \nodata &\nodata&\nodata& 
0.911\nl
18:23:15.55 & 64:33: 5.6 & 0.19188 & \nodata & 19.4 & -20.3 &  2.596\nl
18:21:25.61 & 64: 2:11.0 & 0.19406 & \nodata & 18.2 & -21.5 &  3.253\nl
18:21: 0.73 & 64:37:54.3 & 0.26650 & \nodata & 19.0 & -21.5 &  4.016\nl
\enddata
\tablenotetext{a}{Weighted mean and weighted uncertainty based on 
redshifts measured with 2-3 templates (see \S 2.3). For redshifts from the 
literature, uncertainties are listed when provided in the original papers. All 
redshifts are heliocentric.}
\tablenotetext{b}{For a few objects, no magnitude is available from the 
POSS II database. In most cases, this indicates that the object is too faint to 
be detected in the POSS II survey. However, in some cases this may be due 
to misclassification in the POSS II database or close proximity to a bright 
star.}
\tablenotetext{c}{Absolute magnitude calculated using interstellar extinction 
corrections based on $E(B-V)$ from Lockman \& Savage (1995) and the K-
correction from Peebles(1993), $K$ = 2.5 log $(1+z)$.}
\tablenotetext{d}{Impact parameter (i.e., projected distance to sight line). 
The QSO coordinates (J2000) are RA = 18:21:57.2, Dec = +64:20:36.}
\end{deluxetable}

\clearpage
\begin{deluxetable}{llccccc}
\tablewidth{0pc}
\tablecaption{Redshifts of Galaxies in the Field of PG 
1116+215\label{p1116reds}}
\tablehead{\ \ \ \ \ RA &\ \ \ \ \ Dec & $z_{\rm gal}$\tablenotemark{a} & 
$\sigma _{z}$\tablenotemark{a} & POSS II $B_{\rm J}$\tablenotemark{b} 
& $M_{B}$\tablenotemark{c} & $\rho$\tablenotemark{d} \nl
\multicolumn{2}{c}{(J2000)} & \ & (km s$^{-1}$) & \ & \ & (Mpc)}
\startdata
11:20:46.06 & 21:11:15.2 & 0.02053 &  42 & 17.6 & -17.1 &  0.557\nl
11:19:16.67 & 20:48:49.3 & 0.02077 &  34 & 13.8 & -20.9 &  0.715\nl
11:21: 0.27 & 21:20:13.8 & 0.02122 &  26 & 14.4 & -20.3 &  0.622\nl
11:18:44.44 & 21:33:51.6 & 0.02128 &  34 & 15.1 & -19.6 &  0.374\nl
11:18:45.81 & 21:28:27.9 & 0.02131 &  54 & 16.3 & -18.4 &  0.254\nl
11:19: 1.68 & 21:46: 6.3 & 0.02140 &  37 & 17.1 & -17.6 &  0.647\nl
11:18:22.07 & 21:30: 4.3 & 0.02153 &  48 & 18.2 & -16.6 &  0.371\nl
11:20:38.52 & 21:11:36.8 & 0.02156 &  45 & 18.0 & -16.8 &  0.541\nl
11:18:46.12 & 21: 1:55.3 & 0.02612 &  37 & 18.9 & -16.2 &  0.531\nl
11:18:43.36 & 21:27:23.6 & 0.03235 &  44 & 17.1 & -18.5 &  0.359\nl
11:20:38.81 & 21: 0:49.8 & 0.03744 &  35 & 17.1 & -18.8 &  1.153\nl
11:21:29.42 & 21:35:32.2 & 0.04107 &  51 & 16.6 & -19.5 &  1.643\nl
11:19: 9.67 & 21: 2:43.2 & 0.04108 &  47 & 17.9 & -18.2 &  0.746\nl
11:19:24.29 & 21:10:30.3 & 0.05916 &  40 & 16.6 & -20.4 &  0.601\nl
11:18:33.69 & 21:13: 0.8 & 0.05963 &  68 & 17.2 & -19.7 &  0.655\nl
11:19:59.04 & 21:31:53.8 & 0.05976 &  20 & 18.1 & -18.9 &  1.096\nl
11:21:25.04 & 20:58:37.3 & 0.05986 &  43 & 17.9 & -19.1 &  2.420\nl
11:16:55.43 & 21: 8:59.8 & 0.06003 &  51 & 15.6 & -21.4 &  2.094\nl
11:21:24.08 & 21:14: 2.0 & 0.06016 &  38 & 16.8 & -20.1 &  2.050\nl
11:19:43.67 & 21:26:52.0 & 0.06055 &  45 & 16.6 & -20.4 &  0.717\nl
11:21: 9.19 & 21:40:43.0 & 0.06056 &  38 & 18.6 & -18.4 &  2.275\nl
11:21:29.82 & 21:37: 7.4 & 0.06076 &  71 & 18.0 & -19.0 &  2.417\nl
11:19:42.06 & 21:26:10.7 & 0.06134 &  36 & 19.2 & -17.8 &  0.677\nl
11:21:32.33 & 20:42: 2.3 & 0.06750 &  87 & 17.2 & -20.0 &  3.568\nl
11:21:35.88 & 21:26:52.5 & 0.07017 &  56 & 17.9 & -19.4 &  2.587\nl
11:21:39.67 & 20:58:11.5 & 0.07048 &  75 & 16.9 & -20.5 &  3.039\nl
11:18: 5.82 & 20:59:42.8 & 0.08076 &  24 &\nodata&\nodata& 2.046\nl
11:19:52.05 & 21:30: 2.7 & 0.08247 &  53 & 17.4 & -20.3 &  1.256\nl
11:20:39.86 & 21: 0: 4.4 & 0.08353 &  42 & 18.0 & -19.7 &  2.470\nl
11:18:54.00 & 21: 5:38.1 & 0.08369 &  27 & 17.6 & -20.1 &  1.216\nl
11:18:23.11 & 21: 7:17.4 & 0.08373 &  50 & 16.8 & -20.9 &  1.385\nl
11:20:30.00 & 21:31:31.7 & 0.08442 &  49 & 16.7 & -21.0 &  1.960\nl
11:18:51.61 & 21: 4:42.6 & 0.09181 &  91 & 17.5 & -20.4 &  1.417\nl
11:19: 0.09 & 21: 6:39.8 & 0.09195 &  35 & 16.7 & -21.2 &  1.200\nl
11:19:10.35 & 21: 4:31.2 & 0.09282 &  33 & 17.2 & -20.7 &  1.398\nl
11:21:40.80 & 21:29: 0.7 & 0.10225 &  77 & 18.2 & -19.9 &  3.776\nl
11:21:53.61 & 20:50:37.8 & 0.10252 &  38 & 17.4 & -20.7 &  4.946\nl
11:17:12.60 & 21:35:39.9 & 0.11236 &  52 & 17.9 & -20.5 &  3.522\nl
11:19: 5.59 & 21:34: 9.2 & 0.11492 &  73 & 17.6 & -20.8 &  1.690\nl
11:19: 2.35 & 21:34:58.2 & 0.11542 &  48 & 17.4 & -20.9 &  1.796\nl
11:17:21.62 & 21:28:38.6 & 0.11767 &  44 & 17.6 & -20.9 &  3.087\nl
11:20: 3.85 & 21:43:48.5 & 0.11786 &  61 & 18.1 & -20.3 &  3.212\nl
11:17:31.41 & 20:38:53.6 & 0.13213 & 143 & 17.3 & -21.4 &  5.923\nl
11:20: 9.44 & 21: 8:28.8 & 0.13398 &  36 & 18.7 & -20.0 &  2.303\nl
11:19:13.61 & 21:34:44.7 & 0.13450 &  48 & 17.9 & -20.8 &  2.008\nl
11:19: 5.08 & 21:15: 3.0 & 0.13736 &  70 & 18.6 & -20.2 &  0.572\nl
11:19: 6.67 & 21:18:28.3 & 0.13814 &  50 & 18.4 & -20.4 &  0.127\nl
11:18:50.00 & 21:16:11.0 & 0.13845 &  28 & 19.8 & -19.0 &  0.711\nl
11:18:46.04 & 21: 6:11.2 & 0.13861 &  30 & 19.2 & -19.6 &  1.879\nl
11:19:47.84 & 21:15:44.7 & 0.13863 &  40 & 18.1 & -20.7 &  1.301\nl
11:20:58.83 & 20:39:38.2 & 0.14004 &  82 & 19.2 & -19.6 &  6.336\nl
11:21:26.73 & 20:41:58.9 & 0.14021 &  65 & 18.5 & -20.4 &  6.615\nl
11:19:44.75 & 21:32:11.3 & 0.14058 &  38 &\nodata&\nodata& 2.068\nl
11:21:10.72 & 20:39:17.9 & 0.14072 &  60 & 19.5 & -19.3 &  6.609\nl
11:18:57.41 & 21:44:42.3 & 0.14096 &  45 & 18.5 & -20.3 &  3.442\nl
11:17:35.31 & 21:26:34.1 & 0.14481 &  46 & 18.2 & -20.7 &  3.159\nl
11:18: 5.00 & 21: 9:47.8 & 0.14549 &  53 & 18.4 & -20.5 &  2.438\nl
11:21:16.99 & 20:51:34.1 & 0.14616 &  59 & 18.2 & -20.7 &  5.664\nl
11:20:20.93 & 20:48:28.0 & 0.14652 &  66 & 19.0 & -19.9 &  4.888\nl
11:20: 8.52 & 20:51:14.5 & 0.14656 &  63 & 18.2 & -20.7 &  4.360\nl
11:19:59.62 & 21:12: 9.8 & 0.14695 &  79 & 18.6 & -20.4 &  1.930\nl
11:18: 0.70 & 21:23:50.2 & 0.14828 &  40 & 19.4 & -19.5 &  2.313\nl
11:19:16.56 & 21: 1:50.6 & 0.15010 &  63 & 18.1 & -20.8 &  2.490\nl
11:18: 1.86 & 20:56:26.6 & 0.15263 &  41 & 18.0 & -21.0 &  3.978\nl
11:19:26.00 & 21:12:30.7 & 0.16362 &  35 & 18.0 & -21.2 &  1.200\nl
11:20:27.74 & 21:10:57.0 & 0.16431 &  34 & 19.7 & -19.5 &  3.084\nl
11:20:20.20 & 21: 9:53.2 & 0.16444 &  67 & 20.3 & -18.9 &  2.920\nl
11:20:21.95 & 21: 6:43.4 & 0.16478 &  54 & 18.6 & -20.6 &  3.241\nl
11:20: 9.85 & 21:21:30.6 & 0.16487 &  20 & 18.9 & -20.3 &  2.203\nl
11:20:13.05 & 21: 9:44.7 & 0.16496 &  64 & 18.6 & -20.6 &  2.720\nl
11:19: 3.12 & 21:11: 2.6 & 0.16535 &  69 & 17.6 & -21.6 &  1.281\nl
11:20:52.18 & 21:21:56.1 & 0.16540 & 110 & 19.3 & -19.9 &  3.716\nl
11:20:24.16 & 20:58:54.0 & 0.16552 &  60 & 17.8 & -21.4 &  4.132\nl
11:19:18.07 & 21:15: 3.5 & 0.16581 &  51 & 18.3 & -20.8 &  0.733\nl
11:18:34.78 & 21:25:37.0 & 0.16604 &  64 & 18.9 & -20.3 &  1.555\nl
11:19:19.33 & 21:12:42.4 & 0.16627 &  63 & 18.9 & -20.3 &  1.084\nl
11:20:14.66 & 20:44:30.1 & 0.16645 &  38 & 18.2 & -21.0 &  5.863\nl
11:18:43.84 & 20:44: 6.7 & 0.16660 &  72 & 18.7 & -20.5 &  5.499\nl
11:19:23.36 & 21: 2:27.7 & 0.16661 &  71 & 18.6 & -20.6 &  2.649\nl
11:18:46.74 & 20:43:36.2 & 0.16696 & 134 & 18.7 & -20.5 &  5.570\nl
11:18:59.62 & 20:39: 3.1 & 0.16712 &  36 & 18.7 & -20.5 &  6.230\nl
11:20: 1.20 & 21:35:54.4 & 0.16934 &  71 & 18.6 & -20.7 &  3.220\nl
11:18: 4.50 & 21:35:55.0 & 0.16995 &  47 & 18.4 & -20.8 &  3.500\nl
11:19:19.27 & 21: 5: 7.0 & 0.17305 &  30 & 19.3 & -20.0 &  2.287\nl
11:19:25.97 & 21: 6:47.9 & 0.17311 &  44 & 18.5 & -20.8 &  2.087\nl
11:19:53.62 & 21: 7:29.0 & 0.17321 &  89 & 18.7 & -20.6 &  2.510\nl
11:18:55.59 & 21: 3:56.6 & 0.17365 &  36 & 18.6 & -20.7 &  2.494\nl
11:18:53.56 & 21: 2:42.6 & 0.17503 &  53 & 18.4 & -20.9 &  2.719\nl
11:18:17.00 & 21:33:47.9 & 0.17522 &  52 & 18.4 & -20.9 &  3.023\nl
11:19:27.74 & 21: 6:36.9 & 0.17523 &  58 & 18.3 & -21.0 &  2.156\nl
11:21:23.89 & 21: 7:59.2 & 0.17532 &  41 & 19.0 & -20.3 &  5.373\nl
11:18:17.59 & 20:57:40.0 & 0.17532 &  57 & 18.9 & -20.4 &  3.964\nl
11:19:31.52 & 21:18: 6.5 & 0.17585 &  34 & 18.9 & -20.4 &  0.876\nl
11:18:29.05 & 21:39:46.0 & 0.17605 &  91 & 18.7 & -20.6 &  3.615\nl
11:19:23.63 & 21: 7: 3.6 & 0.17625 &  49 & 19.1 & -20.2 &  2.051\nl
11:18:51.14 & 21:35:16.3 & 0.17640 & 100 & 18.8 & -20.5 &  2.659\nl
11:18:41.96 & 21:28:28.4 & 0.17784 &  71 & 18.7 & -20.6 &  1.799\nl
11:18: 0.53 & 21:36:44.8 & 0.17895 &  53 & 18.7 & -20.6 &  3.846\nl
11:18:34.04 & 21:31:35.3 & 0.17915 &  71 & 18.2 & -21.2 &  2.400\nl
11:17:59.33 & 21:36:29.6 & 0.17928 &  99 & 18.6 & -20.7 &  3.852\nl
11:19:34.11 & 21:37:20.7 & 0.17997 &  74 & 18.7 & -20.7 &  3.111\nl
11:19:16.68 & 21:18:58.3 & 0.18018 &  36 & 19.3 & -20.1 &  0.309\nl
11:18:35.21 & 21:48:31.9 & 0.18083 &  85 & 18.1 & -21.3 &  4.974\nl
11:20:24.89 & 20:50:32.3 & 0.18821 &  57 &\nodata&\nodata& 5.734\nl
11:20:35.98 & 21: 5:38.0 & 0.18884 &  64 & 18.9 & -20.6 &  4.166\nl
11:20:54.43 & 20:42:16.8 & 0.20676 &  77 & 19.7 & -20.0 &  8.104\nl
11:20:12.33 & 21:50:47.0 & 0.21159 &  56 & 18.4 & -21.3 &  6.447\nl
11:18: 0.69 & 21: 7: 2.9 & 0.21189 &  61 & 19.8 & -20.0 &  3.717\nl
11:20:25.71 & 21: 2:36.0 & 0.21236 & 114 & 18.7 & -21.0 &  4.556\nl
11:19:47.13 & 21:28:56.1 & 0.21409 &  65 & 19.2 & -20.5 &  2.458\nl
11:21:44.40 & 20:56:12.0 & 0.21418 &  88 & 18.6 & -21.2 &  8.048\nl
11:20:42.03 & 20:39:46.6 & 0.21465 &  59 & 19.1 & -20.7 &  8.453\nl
11:19:45.21 & 21:31: 9.9 & 0.22931 &  63 & 18.9 & -21.0 &  2.870\nl
11:19:18.91 & 21:39:52.4 & 0.25950 &  68 & 18.8 & -21.3 &  4.450\nl
11:19:10.61 & 21:43: 2.3 & 0.25982 &  67 & 19.3 & -20.8 &  5.106\nl
11:21:23.82 & 21:33:49.3 & 0.27284 &  62 & 19.2 & -21.1 &  7.706\nl
11:20:32.81 & 21:25:42.8 & 0.27577 &  39 & 19.5 & -20.8 &  4.619\nl
11:20:20.17 & 21:36: 1.3 & 0.27807 &  37 & 18.9 & -21.4 &  5.317\nl
\enddata
\tablenotetext{a}{Weighted mean and weighted uncertainty based on 
redshifts measured with 2-3 templates (see \S 2.3). All redshifts are 
heliocentric.}
\tablenotetext{b}{For a few objects, no magnitude is available from the 
POSS II database. In most cases, this indicates that the object is too faint to 
be detected in the POSS II survey. However, in some cases this may be due 
to misclassification in the POSS II database or close proximity to a bright 
star.}
\tablenotetext{c}{Absolute magnitude calculated using interstellar extinction 
corrections based on $E(B-V)$ from Lockman \& Savage (1995) and the K-
correction from Peebles(1993), $K$ = 2.5 log $(1+z)$.}
\tablenotetext{d}{Impact parameter (i.e., projected distance to sight line). 
The QSO coordinates (J2000) are RA = 11:19:08.7, Dec = +21:19:18.}
\end{deluxetable}

\clearpage
\begin{deluxetable}{cccccccc}
\tablewidth{0pc}
\tablecaption{PG 1116+215 Galaxy Redshift Survey 
Completeness\label{complete1116}}
\tablehead{ \ & \ & 
\multicolumn{2}{c}{\underline{\ \ \ \ \ \ 10' Radius\ \ \ 
\ \ \ }} & 
\multicolumn{2}{c}{\underline{\ \ \ \ \ \ 20' Radius\ \ \ 
\ \ \ }} & 
\multicolumn{2}{c}{\underline{\ \ \ \ \ \ 30' Radius\ \ \ 
\ \ \ }} \nl
Magnitude & $z(L*)$\tablenotemark{a} & 
Completeness\tablenotemark{b} & Scale & 
Completeness\tablenotemark{b} & Scale & 
Completeness\tablenotemark{b} & Scale \nl
Limit & \ & \ & (Mpc) & \ & (Mpc) & \ & (Mpc)}
\startdata
$B_{\rm J} \leq$ 17.0 & 0.048 & 100\% & 0.5  & 100\% & 1.0 
& 83.3\% & 1.6 \nl
$B_{\rm J} \leq$ 18.0 & 0.076 & 85.7\% & 0.8 & 85.0\% & 
1.6 & 81.1\% & 2.4 \nl
$B_{\rm J} \leq$ 19.0 & 0.121 & 88.2\% & 1.2 & 86.9\% & 
2.4 & 78.1\% & 3.6 \nl
$B_{\rm J} \leq$ 20.0 & 0.191 & 70.4\% & 1.7 & 68.5\% & 
3.4 & 63.8\% & 5.1 \nl
\enddata
\tablenotetext{a}{Redshift at which the limiting magnitude 
corresponds to a 1 $L^{*}$ galaxy.}
\tablenotetext{b}{Percentage of targets for which 
redshifts were obtained.}
\end{deluxetable}

\clearpage
\begin{deluxetable}{cllllc}
\tablewidth{0pc}
\tablecaption{Absorption Lines Detected in the GHRS G140L Spectrum of 
H 1821+643\tablenotemark{a}\label{h1821list}}
\tablehead{
 Wavelength & $W_{\lambda}\pm \sigma _{W}$ & \ &
Identification & $z_{\rm abs}$ & $v_{\rm ISM}$ \nl
 (\AA ) & (m\AA ) & \ & \ & \ & (km s$^{-1}$)}
\startdata
1253.66 & 153$\pm$11 & \ & \ion{S}{2} \ 1253.81 & 0.0 & -51 \nl 
1256.43 & 477$\pm$12 & \ & \ion{H}{1} \ 1025.72 & 0.22484 & \ \nl 
1260.15 & 1379$\pm$18\tablenotemark{b} & (214) & \ion{S}{2} \ 
1259.52 & 0.0 & -51 \nl 
    \   &       \     & (1105)& \ion{Si}{2} \ 1260.42 & 0.0 & -68 \nl 
    \   &       \     & (243)& \ion{H}{1} \ 972.54 & 0.29678 & \ \nl 
1264.11 & 157$\pm$14 & \ & \ion{O}{6} \ 1031.93 & 0.22501 & \  \nl 
1266.92 & 246$\pm$12 & \ & \ion{C}{3} \ 977.02 & 0.29676 & \  \nl 
1271.00 & 107$\pm$10 & \ & \ion{O}{6} \ 1037.62 & 0.22501  & \ \nl 
1277.21 & 100$\pm$15 & \ & \ion{C}{1} \ 1277.46\tablenotemark{c} & 
0.0 & +21\nl 
1284.82 & 92$\pm$15 & \ & \ion{H}{1} \ 1215.67 & 0.05704 & \ \nl 
 1293.55 & 66$\pm$16  & \ & \ion{H}{1} \ 1215.67 & 0.06432 & \ \nl 
 1297.39 & 170$\pm$14 & \ & \ion{H}{1} \ 1215.67 & 0.06722 & \ \nl 
 1302.04\tablenotemark{d} & 921$\pm$16 & \ & \ion{O}{1} \ 1302.17 & 
0.0 & -29 \nl 
 1304.08\tablenotemark{d} & 542$\pm$11 & \ & \ion{Si}{2} \ 1304.37 & 
0.0 & -68 \nl 
 1317.20 & 161$\pm$19 & \ & \ion{Ni}{2} \ 1317.22 & 0.0 & -46 \nl 
 1323.93 & 51$\pm$12  & \ & \ion{H}{1} \ 1215.67 & 0.08910 & \ \nl 
 1328.96 & 72$\pm$10  & \ & \ion{C}{1} \ 1329.34 & 0.0 & +21 \nl 
 1330.09 & 422$\pm$13 & \ & \ion{H}{1} \ 1025.72 & 0.29677 & \ \nl 
 1334.25 & 873$\pm$10 & \ & \ion{C}{2} \ 1334.53 & 0.0 & -63 \nl 
 1335.43 & 174$\pm$9  & \ & \ion{C}{2}* \ 1335.71 & 0.0 & -31 \nl 
 1338.60 & 214$\pm$16 & \ & \ion{O}{6} \ 1031.93 & 0.29676 & \ \nl 
 1346.01 & 123$\pm$14 & \ & \ion{O}{6} \ 1037.62 & 0.29676 & \ \nl 
 1351.19 & 73$\pm$11  & \ & \ion{H}{1} \ 1215.67 & 0.11152 & \ \nl 
 1361.16 & 114$\pm$10 & \ & \ion{H}{1} \ 1215.67 & 0.11974 & \ \nl
 1363.15 & 719$\pm$19\tablenotemark{b} & (438)& \ion{H}{1} \ 
1215.67& 0.12123& \ \nl 
    \    &       \    & (396)& \ion{H}{1} \ 1215.67& 0.12157& \ \nl 
 1366.23 & 39$\pm$13  & \ & \ion{H}{1} \ 1215.67 & 0.12385 & \ \nl 
 1368.44 & 49$\pm$15  & \ & \ion{H}{1} \ 1215.67 & 0.12566 & \ \nl 
 1369.86 & 89$\pm$15  & \ & \ion{Ni}{2} \ 1370.13 & 0.0 & -46 \nl
 1370.91 & 46$\pm$15  & \ & \ion{H}{1} \ 1215.67 & 0.12781 & \ \nl 
 1393.40 & 452$\pm$20 & \ & \ion{Si}{4} \ 1393.76 & 0.0 & -75 \nl 
 1395.08 & 263$\pm$17 & \ & \ion{H}{1} \ 1215.67 & 0.14760 & \ \nl 
 1402.26 & 200$\pm$21 & \ & \ion{Si}{4} \ 1402.77 & 0.0 & -75 \nl 
 1407.17 & 79$\pm$17  & \ & \ion{H}{1} \ 1215.67 & 0.15727 & \ \nl 
 1414.41 & 63$\pm$10  & \ & \ion{H}{1} \ 1215.67 & 0.16350 & \ \nl 
 1422.21 & 612$\pm$20 & \ & \ion{H}{1} \ 1215.67 & 0.16990 & \ \nl 
 1433.46 & 84$\pm$13  & \ & \ion{H}{1} \ 1215.67 & 0.17915 & \ \nl 
 1435.09 & 89$\pm$15  & \ & \ion{H}{1} \ 1215.67 & 0.18049 & \ \nl 
 1454.67 & 86$\pm$20  & \ & \ion{Ni}{2} \ 1454.84 & 0.0 & -46 \tablebreak
 1456.33 & 69$\pm$18  & \ & \ion{H}{1} \ 1215.67 & 0.19794 & \ \nl 
 1457.67 & 35$\pm$14  & \ & \ion{H}{1} \ 1215.67: & 0.19905 & \ \nl 
 1470.44 & 39$\pm$15  & \ & \ion{H}{1} \ 1215.67: & 0.20961 & \ \nl 
 1473.03 & 126$\pm$17 & \ & \ion{H}{1} \ 1215.67 & 0.21176 & \ \nl 
 1474.87 & 586$\pm$22 & \ & \ion{H}{1} \ 1215.67 & 0.21325 & \ \nl 
 1478.27 & 177$\pm$18 & \ & \ion{H}{1} \ 1215.67\tablenotemark{e} & 
0.21577 & \ \nl 
1488.90 & 905$\pm$27 & \ & \ion{H}{1} \ 1215.67 & 0.22484 & \ \nl
 1490.56 & 343$\pm$25 & \ & \ion{H}{1} \ 1215.67 & 0.22621 & \ \nl 
 1492.61 & 87$\pm$14  & \ & \ion{H}{1} \ 1215.67 & 0.22782 & \ \nl 
 1506.05 & 57$\pm$19  & \ & \ion{H}{1} \ 1215.67 & 0.23864 & \ \nl 
 1509.05 & 98$\pm$20  & \ & \ion{H}{1} \ 1215.67 & 0.24132 & \ \nl 
 1513.28 & 98$\pm$20  & \ & \ion{H}{1} \ 1215.67 & 0.24514 & \ \nl 
 1526.39 & 902$\pm$32 & \ & \ion{Si}{2} \ 1526.71 & 0.0 & -51 \nl
 1529.75 & 144$\pm$42 & \ & \ion{H}{1} \ 1215.67 & 0.25822 & \ \nl
 1533.36 & 229$\pm$43 & \ & \ion{H}{1} \ 1215.67 & 0.26163 & \ \nl
\enddata
\tablenotetext{a}{Notes: Column 1 lists the vacuum heliocentric wavelength 
of 
the line centroid. The observed equivalent width measured as described 
in Tripp et al. (1996) is given in column 2 and the line identification in 
column 3. Column 4 provides the redshift measured with VPFIT from the 
G140L 
spectrum. For lines due to the Galactic ISM, the Heliocentric line velocity
is listed in column 5. A colon in column 3 indicates that the line is a 
probable detection but requires confirmation.}
\tablenotetext{b}{Strongly blended line. The individual equivalent widths 
derived from profile parameters measured with VPFIT are listed in 
parentheses 
following the total equivalent width of the blend.}
\tablenotetext{c}{This ISM line may be blended with an extragalactic 
\ion{H}{1} 
Ly$\alpha$ line (see text \S 3.1.1), which could cause this ISM line to have
a more positive velocity than the other ISM lines.}
\tablenotetext{d}{Contaminated by geocoronal \ion{O}{1} emission.}
\tablenotetext{e}{The identification of this line is ambiguous; this could be
\ion{Si}{3} 1206.5 associated with the \ion{O}{6} absorber at $z_{\rm 
abs}$ 
= 0.22503. However, for reasons noted in the text (\S 3.1.2, see also Savage 
et al. 1998), we believe the absorption is dominated by \ion{H}{1} 
Ly$\alpha$.}
\end{deluxetable}

\clearpage
\begin{deluxetable}{cllllc}
\tablewidth{0pc}
\tablecaption{Absorption Lines Detected in the GHRS G140L Spectrum of 
PG 1116+215\tablenotemark{a}\label{p1116list}}
\tablehead{
 Wavelength & $W_{\lambda}\pm \sigma _{W}$\tablenotemark{b} & \ &
Identification & $z_{\rm abs}$ & $v_{\rm ISM}$ \nl
 (\AA ) & (m\AA ) & \ & \ & \  & (km s$^{-1}$)}
\startdata
1235.41 & 112$\pm$29 & \     & \ion{H}{1} \ 1215.67  & 0.01639 & \nl
1239.40 & 170$\pm$24 & \     & \ion{H}{1} \ 1215.67\tablenotemark{b} 
& 0.01950& \nl
1250.30 & 322$\pm$28\tablenotemark{c} & (173) & \ion{H}{1} \ 
1215.67  & 0.02845& \nl
    \   &    \       & (72)  & \ion{S}{2} \ 1250.58  & 0.0 & +14\nl
1253.86 & 189$\pm$23 & \     & \ion{S}{2} \ 1253.81  & 0.0 & +14\nl
1255.00 & 95$\pm$17  & \     & \ion{H}{1} \ 1215.67  & 0.03223& \nl
1260.51 & 1077$\pm$28\tablenotemark{c,d}& (196) & \ion{S}{2} \ 
1259.52  & 0.0 & +14\nl
    \   &    \       & (851) & \ion{Si}{2} \ 1260.42 & 0.0 & +60\nl
1265.66 & 171$\pm$37 & \     & \ion{H}{1} \ 1215.67  & 0.04125& \nl
1287.46 & 166$\pm$24 & \     & \ion{H}{1} \ 1215.67  & 0.05905& \nl
1289.55 &  66$\pm$22 & \     & \ion{H}{1} \ 1215.67 & 0.06079& \nl
1302.30 & 467$\pm$33 & \     & \ion{O}{1} \ 1302.17  & 0.0 & +35\nl
1304.51 & 466$\pm$30 & \     & \ion{Si}{2} \ 1304.37 & 0.0 & +60\nl
1314.42 & 119$\pm$19 & \     & \ion{H}{1} \ 1215.67  & 0.08118& \nl
1317.42 & 147$\pm$28 & \     & \ion{Ni}{2} \ 1317.21 & 0.0 & +26 \nl
1328.46 & 70$\pm$23  & \     & \ion{H}{1} \ 1215.67 & 0.09279& \nl
1335.00 & 1048$\pm$35\tablenotemark{c}& (845) & \ion{C}{2} \ 
1334.53  & 0.0 & +61\nl
    \   &    \       & (176) & \ion{C}{2}* 1335.71   & 0.0 & +54\nl
1360.48 & 132$\pm$25 & \     & \ion{H}{1} 1215.67    & 0.11910& \nl
1373.69 & 91$\pm$16   & \     & \ion{Si}{3} 1206.50   & 0.13852& \nl
1375.70 & 61$\pm$15  & \     & \ion{H}{1} 1215.67    & 0.13159& \nl
1384.16 & 543$\pm$32\tablenotemark{c} & (509) & \ion{H}{1} 1215.67 
   & 0.13861& \nl
    \   &   \        & (96)  & \ion{H}{1} 1215.67    & 0.13862& \nl
1393.83 & 484$\pm$36\tablenotemark{d} & \     & \ion{Si}{4} 1393.76   
& 0.0 & +43\nl
1402.88 & 228$\pm$29\tablenotemark{d} & \     & \ion{Si}{4} 1402.77   
& 0.0 & +43\nl
1417.78 & 865$\pm$24\tablenotemark{c} & (177) & \ion{H}{1} 1215.67 
   & 0.16547 & \nl
    \   &   \        & (121) & \ion{H}{1} 1215.67    & 0.16616& \nl
    \   &   \        & (314) & \ion{H}{1} 1215.67   & 0.16697& \nl
1426.81 & 308$\pm$15 & \     & \ion{H}{1} 1215.67    & 0.17366& \nl
1435.10 & 46$\pm$15  & \     & \ion{Si}{2} 1260.42:  & 0.13858& \nl
1448.43 & 67$\pm$14  & \     & UID                  &  &  \nl
\enddata
\tablenotetext{a}{Notes: Column 1 lists the vacuum heliocentric wavelength 
of 
the line centroid. The observed equivalent width measured as described 
in Tripp et al. (1996) is given in column 2 and the line identification in 
column 3. Column 4 provides the redshift measured with VPFIT from the 
G140L 
spectrum. For lines due to the Galactic ISM, the Heliocentric line velocity
is listed in column 5. A colon in column 3 indicates that the line is a 
probable detection but requires confirmation.}
\tablenotetext{b}{As discussed in the text (\S 3.1.2), the ISM \ion{N}{5} 
1238.80, \ion{Mg}{2} 1239.93, and \ion{Mg}{2} 1240.39 \AA\ lines are 
all located near this feature.  However, as argued in \S 3.1.2, we believe this 
absorption is dominated by an extragalactic Ly$\alpha$ line at \zabs\ = 
0.01950.}
\tablenotetext{c}{Strongly blended line. The individual equivalent widths 
derived from profile parameters measured with VPFIT are listed in 
parentheses following the total equivalent width of the blend.}
\tablenotetext{d}{The Milky Way lines of \ion{Si}{2} 1260.42 \AA\ and 
\ion{Si}{4} 1393.76, 1402.77 \AA\ contain substructure implying the 
detection of a high velocity cloud at $v_{\rm ISM} \approx$ 200 \kms . This 
cloud was not detected in the wideband \ion{H}{1} 21 cm observations of 
Lockman \& Savage (1995).}
\end{deluxetable}

\clearpage
\begin{deluxetable}{ccllc}
\tablewidth{0pc}
\tablecaption{H 1821+643 Extragalactic Absorption Line Measurements 
from the Literature\tablenotemark{a}\label{g160mlist}}
\tablehead{
Wavelength & $W_{\lambda}\pm \sigma _{W}$ &
Identification & $z_{\rm abs}$ & Reference\tablenotemark{b} \nl
(\AA ) & (m\AA ) & \ & \  & \ }
\startdata
1244.466 & 183$\pm$23 & \ion{H}{1} 1025.72 & 0.21326$\pm$0.00003 
& 1 \nl
1245.500 & 297$\pm$23 & \ion{H}{1} 1215.67 & 0.02454$\pm$0.00001 
& 1 \nl
1247.858 & 052$\pm$16 & \ion{H}{1} 1025.72 & 0.21656$\pm$0.00004 
& 1 \nl
1256.400 & 459$\pm$27 & \ion{H}{1} 1025.72 & 0.22489$\pm$0.00001 
& 1 \nl
1261.124 & 227$\pm$16 & \ion{H}{1} 972.54  & 0.29674$\pm$0.00002 
& 1 \nl
1264.138 & 206$\pm$18 & \ion{O}{6} 1031.93 & 0.22503$\pm$0.00001 
& 1 \nl
1265.579 & 047$\pm$14 & \ion{H}{1} 1215.67 & 0.04105$\pm$0.00002 
& 1 \nl
1266.917 & 247$\pm$19 & \ion{C}{3} 977.02  & 0.29672$\pm$0.00001 
& 1 \nl
1529.486 & 168$\pm$21 & \ion{H}{1} 1215.67 & 0.25814$\pm$0.00003 
& 1 \nl
1533.635 & 206$\pm$17 & \ion{H}{1} 1215.67 & 0.26156$\pm$0.00002 
& 1 \nl
1539.773 & 206$\pm$22 & \ion{H}{1} 1215.67 & 0.26660$\pm$0.00002 
& 1 \nl
1576.93  & 680$\pm$10       & \ion{H}{1} 1215.67 & 0.2972   & 2 \nl
\enddata
\tablenotetext{a}{Notes -- Vacuum heliocentric wavelength is listed in 
column 1. The observed equivalent widths are listed in column 2, and 
columns 3 and 4 provide the line identifications and redshifts.}
\tablenotetext{b}{  References: (1) Savage, Sembach, \& 
Lu\markcite{ssl95} (1995). These measurements are from GHRS G160M 
observations with a resolution of $\sim$18 km s$^{-1}$ and a 3$\sigma$ 
detection limit of $\sim$50 m\AA ; (2) Bahcall et al.\markcite{bah93} 
(1993). These measurements are from FOS observations with a resolution of 
$\sim$300 km s$^{-1}$ with a 3$\sigma$ detection limit of $\sim$ 150 
m\AA\ for $\lambda >$ 1550 \AA .}
\end{deluxetable}

\clearpage
\begin{deluxetable}{cccccccc}
\tablewidth{0pc}
\tablecaption{Nearest Galaxies to H 1821+643 Lyman $\alpha$ 
Absorbers\label{h1821near}}
\tablehead{
$z_{\rm abs}$ & $W_{\rm r}$\tablenotemark{a} & 
Distance\tablenotemark{b} & Projected & $\Delta v$\tablenotemark{c} & 
RA(gal) & Dec(gal) & $z_{\rm gal}$ \nl
    \         & (m\AA )       & (Mpc)    & Distance\tablenotemark{b} \ (Mpc) & 
(km s$^{-1}$) & (2000) & (2000) & \nl
}
\startdata
0.02454 &  290 &   2.103 &   0.783 &   146 & 18:21:41.17 & 63:51:38.0 
& 0.02404\nl
0.04105 &   45 &  33.136 &   1.370 & -2481 & 18:22:36.11 & 64:45:44.9 
& 0.04970\nl
0.05704 &   87 &   1.954 &   1.784 &    60 & 18:21:21.38 & 64:49:40.4 & 
0.05683\nl
0.06432 &   62 &   5.405 &   2.916 &  -341 & 18:16:31.37 & 64:43:54.6 & 
0.06553\nl
0.06722 &  159 &   6.985 &   2.916 &   475 & 18:16:31.37 & 64:43:54.6 
& 0.06553\nl
0.08910 &   47 &   1.043 &   0.739 &   -55 & 18:20:42.96 & 64:19:45.8 & 
0.08930\nl
0.11152 &   66 &   0.780 &   0.773 &    -8 & 18:20:53.45 & 64:19:37.1 & 
0.11155\nl
0.11974 &  102 &   2.520 &   2.445 &    46 & 18:25: 4.45 & 64:25:22.9 & 
0.11957\nl
0.12123 &  391 &   1.122 &   0.144 &   -83 & 18:22: 2.65 & 64:21:39.3 & 
0.12154\nl
0.12157 &  353 &   0.180 &   0.144 &     8 & 18:22: 2.65 & 64:21:39.3 & 
0.12154\nl
0.12385 &   35 &   4.871 &   1.740 &   339 & 18:23:58.19 & 64:26:52.3 & 
0.12258\nl
0.12566 &   44 &  11.164 &   1.740 &   821 & 18:23:58.19 & 64:26:52.3 
& 0.12258\nl
0.12781 &   41 &  18.790 &   1.740 &  1393 & 18:23:58.19 & 64:26:52.3 
& 0.12258\nl
0.14760 &  229 &  10.627 &   4.111 &  -728 & 18:21: 8.73 & 63:52: 9.0 & 
0.15039\nl
0.15727 &   68 &   9.319 &   1.741 &   679 & 18:22: 4.38 & 64: 8:38.5 & 
0.15465\nl
0.16350 &   54 &   1.263 &   0.882 &   -67 & 18:22:49.76 & 64:19:28.6 & 
0.16376\nl
0.16990 &  523 &   3.340 &   0.373 &  -246 & 18:21:36.61 & 64:21:25.0 
& 0.17086\nl
0.17915 &   71 &   2.278 &   0.447 &   165 & 18:22:20.16 & 64:21:46.7 & 
0.17850\nl
0.18049 &   75 &   3.455 &   3.213 &    94 & 18:19:55.22 & 64:35: 8.4 & 
0.18012\nl
0.19794 &   58 &   9.446 &   3.134 &  -657 & 18:22:14.04 & 64: 3: 5.4 & 
0.20057\nl
0.19905 &   29 &   6.027 &   3.134 &  -380 & 18:22:14.04 & 64: 3: 5.4 & 
0.20057\nl
0.20961 &   32 &  30.669 &   3.134 &  2249 & 18:22:14.04 & 64: 3: 5.4 & 
0.20057\nl
0.21176 &  104 &  37.867 &   3.134 &  2781 & 18:22:14.04 & 64: 3: 5.4 
& 0.20057\nl
0.21326 &  483 &  41.249 &   0.105 & -3034 & 18:21:54.40 & 64:20: 9.3 
& 0.22560\nl
0.21656 &  145 &  30.185 &   0.105 & -2219 & 18:21:54.40 & 64:20: 9.3 
& 0.22560\nl
0.22489 &  739 &   2.367 &   0.105 &  -174 & 18:21:54.40 & 64:20: 9.3 & 
0.22560\nl
0.22621 &  280 &   1.040 &   0.388 &   -71 & 18:21:38.82 & 64:20:31.7 & 
0.22650\nl
0.22782 &   71 &   4.407 &   0.388 &   322 & 18:21:38.82 & 64:20:31.7 & 
0.22650\nl
0.23864 &   46 &  18.822 &   0.462 & -1379 & 18:21:56.28 & 64:22:50.4 
& 0.24435\nl
0.24132 &   79 &   9.987 &   0.462 &  -731 & 18:21:56.28 & 64:22:50.4 & 
0.24435\nl
0.24514 &   79 &   1.882 &   0.625 &  -130 & 18:22:19.11 & 64:18:43.5 & 
0.24568\nl
0.25814 &  134 &  21.884 &   1.950 &  1594 & 18:20:31.50 & 64:20:24.0 
& 0.25147\nl
0.26156 &  163 &  16.547 &   4.016 & -1172 & 18:21: 0.73 & 64:37:54.3 
& 0.26650\nl
0.26660 &  163 &   0.847 &   0.795 &   -21 & 18:22:10.26 & 64:17:15.6 & 
0.26669\nl
0.29674 &  524 &   0.665 &   0.427 &   -37 & 18:22:13.61 & 64:20:14.5 & 
0.29690\nl
\enddata
\tablenotetext{a}{Rest equivalent width.}
\tablenotetext{b}{Absorber-galaxy three-dimensional and projected distances 
assuming $H_{0}$ = 75 km s$^{-1}$ Mpc$^{-1}$ and neglecting any 
departures from a pure Hubble flow.}
\tablenotetext{c}{$\Delta v = c(z_{\rm abs} - z_{\rm gal})/(1 + z_{\rm 
mean})$ where $z_{\rm mean}$ is the mean of $z_{\rm abs}$ and $z_{\rm 
gal}$.}
\end{deluxetable}

\clearpage
\begin{deluxetable}{cccccccc}
\tablewidth{0pc}
\tablecaption{Nearest Galaxies to PG 1116+215 Lyman $\alpha$ 
Absorbers\label{p1116near}}
\tablehead{
$z_{\rm abs}$ & $W_{\rm r}$\tablenotemark{a} & 
Distance\tablenotemark{b} & Projected & $\Delta v$\tablenotemark{c} & 
RA(gal) & Dec(gal) & $z_{\rm gal}$ \nl
    \         & (m\AA )       & (Mpc)    & Distance\tablenotemark{b} \ (Mpc) & 
(km s$^{-1}$) & (2000) & (2000) & \nl
}
\startdata
0.01639 &  110 &  16.261 &   0.557 & -1219 & 11:20:46.06 & 21:11:15.2 
& 0.02053\nl
0.01950 &  167 &   4.075 &   0.557 &  -303 & 11:20:46.06 & 21:11:15.2 
& 0.02053\nl
0.02845 &  168 &   9.085 &   0.531 &   680 & 11:18:46.12 & 21: 1:55.3 & 
0.02612\nl
0.03223 &   92 &   0.588 &   0.359 &   -35 & 11:18:43.36 & 21:27:23.6 & 
0.03235\nl
0.04125 &  164 &   0.991 &   0.746 &    49 & 11:19: 9.67 & 21: 2:43.2 & 
0.04108\nl
0.05905 &  157 &   0.731 &   0.601 &   -31 & 11:19:24.29 & 21:10:30.3 & 
0.05916\nl
0.06079 &   62 &   1.155 &   0.717 &    68 & 11:19:43.67 & 21:26:52.0 & 
0.06055\nl
0.08118 &  110 &   2.572 &   2.046 &   116 & 11:18: 5.82 & 20:59:42.8 & 
0.08076\nl
0.09279 &   64 &   1.402 &   1.398 &    -8 & 11:19:10.35 & 21: 4:31.2 & 
0.09282\nl
0.11910 &  118 &   5.496 &   3.212 &   332 & 11:20: 3.85 & 21:43:48.5 & 
0.11786\nl
0.13159 &   54 &   6.227 &   5.923 &  -143 & 11:17:31.41 & 20:38:53.6 & 
0.13213\nl
0.13861 &  447 &   0.909 &   0.711\tablenotemark{d} &    
42\tablenotemark{d} & 11:18:50.00 & 21:16:11.0 & 0.13845\nl
0.13862 &   84 &   0.932 &   0.711\tablenotemark{d} &    
45\tablenotemark{d} & 11:18:50.00 & 21:16:11.0 & 0.13845\nl
0.16547 &  152 &   1.347 &   1.281 &    31 & 11:19: 3.12 & 21:11: 2.6 & 
0.16535\nl
0.16616 &  104 &   1.149 &   1.084 &   -28 & 11:19:19.33 & 21:12:42.4 & 
0.16627\nl
0.16697 &  269 &   2.658 &   1.084 &   180 & 11:19:19.33 & 21:12:42.4 
& 0.16627\nl
0.17366 &  262 &   2.495 &   2.494 &     3 & 11:18:55.59 & 21: 3:56.6 & 
0.17365\nl
\enddata
\tablenotetext{a}{Rest equivalent width.}
\tablenotetext{b}{Absorber-galaxy three-dimensional and projected distances 
assuming $H_{0}$ = 75 km s$^{-1}$ Mpc$^{-1}$ and neglecting any 
departures from a pure Hubble flow.}
\tablenotetext{c}{$\Delta v = c(z_{\rm abs} - z_{\rm gal})/(1 + z_{\rm 
mean})$ where $z_{\rm mean}$ is the mean of $z_{\rm abs}$ and $z_{\rm 
gal}$.}
\tablenotetext{d}{In the perturbed Hubble flow model (see \S 5.1), the 
nearest neighbor is the galaxy at $z_{\rm abs}$ = 0.13814 with $\rho$ = 
0.127 Mpc and $\Delta v$ = 125 km s$^{-1}$.}
\end{deluxetable}

\clearpage
\begin{deluxetable}{ccccclccl}
\tablewidth{0pc}
\tablecaption{Galaxies Within Projected Distances of 1 Mpc with Associated 
($\Delta v \leq$ 350 km s$^{-1}$) Lyman $\alpha$ 
Absorbers\label{galx600}}
\tablehead{\multicolumn{4}{c}{\underline{\ \ \ \ \ \ \ \ \ \ \ \ \ \ \ \ \ \ \ \ \ \ \ 
Galaxy \ \ \ \ \ \ \ \ \ \ \ \ \ \ \ \ \ \ \ \ \ \ \ }} & \ \ & 
\multicolumn{4}{c}{\underline{\ \ \ \ \ \ \ \ \ \ \ \ \ \ \ \ \ \ \ \ \ \ Nearest Lyman 
$\alpha$ Absorber\ \ \ \ \ \ \ \ \ \ \ \ \ \ \ \ \ \ \ \ \ \ }} \nl
RA & Dec & $z_{\rm gal}$ & $\rho$\tablenotemark{a} & \ \ & $z_{\rm 
abs}$ & $W_{\rm r}$\tablenotemark{b} & $\Delta v$\tablenotemark{c} & 
Metals? \nl
\multicolumn{2}{c}{(J2000)} & \ & (kpc) & \ \ & \ & (m\AA ) & (km s$^{-
1}$) & \ }
\startdata
\multicolumn{8}{l}{\underline{PG 1116+215}} \nl
11:20:46.06 & 21:11:15.2 & 0.02053 & 557 & \ \ & 0.01950 & 
167\tablenotemark{d} & -303 & \nodata \nl
11:18:43.36 & 21:27:23.6 & 0.03235 & 359 & \ \ & 0.03223 & 
92\tablenotemark{e} & -35 & \nodata \nl
11:19:09.67 & 21:02:43.2 & 0.04108 & 746 & \ \ & 0.04125 & 164 & 49 
& \nodata \nl
11:19:24.29 & 21:10:30.3 & 0.05916 & 601 & \ \ & 0.05905 & 157 & 
-31 & \nodata \nl
11:19:42.06 & 21:26:10.7 & 0.06134 & 677 & \ \ & 0.06079 & 62 & -155 
& \nodata \nl
11:19:06.67 & 21:18:28.3 & 0.13814 & 127 & \ \ & 0.13861,0.13862 & 
531 & 125 & \ion{Si}{2}, \ion{Si}{3} \nl
\multicolumn{8}{l}{\underline{H 1821+643}} \nl
18:21:41.17 & 63:51:38.0 & 0.02404 & 783 & \ \ & 0.02454 & 290 & 146 
& \nodata \nl
18:20:42.96 & 64:19:45.8 & 0.08930 & 739 & \ \ & 0.08910 & 47 & -55 & 
\nodata \nl
18:20:53.45 & 64:19:37.1 & 0.11155 & 773 & \ \ & 0.11152 & 66 & -8 & 
\nodata \nl
18:22:02.65 & 64:21:39.3 & 0.12154 & 144 & \ \ & 0.12123,0.12157 & 
846 & -40 & \nodata \nl
18:22:49.76 & 64:19:28.6 & 0.16376 & 882 & \ \ & 0.16350 & 54 & -67 & 
\nodata \nl
18:21:36.61 & 64:21:25.1 & 0.17086 & 373 & \ \ & 0.16990 & 523 & 
-246 & \nodata \nl
18:22:20.16 & 64:21:46.7 & 0.17850 & 447 & \ \ & 0.17915,0.18049 & 
146 & 340 & \nodata \nl
18:21:54.40 & 64:20:09.3 & 0.22560 & 105 & \ \ & 0.22489,0.22621 & 
1019 & -85 & \ion{O}{6}\tablenotemark{f} \nl
18:21:38.82 & 64:20:31.7 & 0.22650 & 388 & \ \ & 0.22782 & 71 & 322 
& \nodata \nl
18:21:56.28 & 64:22:50.4 & 0.24435 & 462 & \ \ & 0.24514 & 79 & 190 
& \nodata \nl
18:22:10.26 & 64:17:15.6 & 0.26669 & 795 & \ \ & 0.26660 & 163 & -21 
& \nodata
\enddata
\tablenotetext{a}{Impact parameter (i.e., projected distance) calculated 
assuming $H_{0}$ = 75 km s$^{-1}$ Mpc$^{-1}$ and $q_{0}$ = 0.}
\tablenotetext{b}{Rest equivalent width. Ly$\alpha$ lines within 350 \kms\ 
of each other are treated as a single absorber. Equivalent widths of individual 
components in these blends are listed in Tables 4 and 5.}
\tablenotetext{c}{$\Delta v \ = \ c(z_{{\rm abs}} - z_{\rm gal})/(1 + z_{\rm 
mean})$ where $z_{\rm mean}$ is the mean of $z_{\rm abs}$ and $z_{\rm 
gal}$. For blended Ly$\alpha$ lines, the equivalent width weighted mean 
redshift was used for $z_{{\rm Ly}\alpha}$.}
\tablenotetext{d}{This line may be blended with the Galactic \ion{Mg}{2} 
1239.9, 1240.4 \AA\ doublet and possibly Galactic \ion{N}{5} 1238.8 \AA\ 
(see \S 3.1.2).}
\tablenotetext{e}{Blended with Galactic \ion{S}{2} 1253.8 \AA .}
\tablenotetext{f}{See Savage, Tripp, \& Lu (1998).}
\end{deluxetable}

\clearpage
\begin{deluxetable}{llllcc}
\tablewidth{0pc}
\tablecaption{Kolmogorov-Smirnov Tests of Nearest Neighbor 
Distributions\label{kstests}}
\tablehead{Ly$\alpha$ Sample\tablenotemark{a} & Hubble Flow & Data 
Set 1 & Data Set 2 & KS D & KS Probability\tablenotemark{b}}
\startdata
Total & Perturbed & Galaxy-Ly$\alpha$ & Galaxy-Random & 0.305 & 
1.4$\times 10^{-5}$ \nl
Total & Pure      & Galaxy-Ly$\alpha$ & Galaxy-Random & 0.250 & 
6.5$\times 10^{-4}$ \nl
Complete & Perturbed & Galaxy-Ly$\alpha$ & Galaxy-Random & 0.340 & 
1.5$\times 10^{-5}$ \nl
Complete & Pure      & Galaxy-Ly$\alpha$ & Galaxy-Random & 0.285 & 
5.2$\times 10^{-4}$ \nl
``1 Mpc'' Complete & Perturbed & Galaxy-Ly$\alpha$ & Galaxy-Random & 
0.368 & 2.4$\times 10^{-5}$ \nl
``1 Mpc'' Complete & Pure & Galaxy-Ly$\alpha$ & Galaxy-Random & 
0.299 & 1.2$\times 10^{-3}$ \nl
Total & Perturbed & Galaxy-Ly$\alpha$ & Galaxy-Galaxy & 0.209 & 
0.016 \nl
Total & Pure      & Galaxy-Ly$\alpha$ & Galaxy-Galaxy & 0.239 & 
3.8$\times 10^{-3}$ \nl
Complete & Perturbed & Galaxy-Ly$\alpha$ & Galaxy-Galaxy & 0.135 & 
0.38 \nl
Complete & Pure      & Galaxy-Ly$\alpha$ & Galaxy-Galaxy & 0.200 & 
0.05 \nl
``1 Mpc'' Complete & Perturbed & Galaxy-Ly$\alpha$ & Galaxy-Galaxy & 
0.118 & 0.67 \nl
``1 Mpc'' Complete & Pure & Galaxy-Ly$\alpha$ & Galaxy-Galaxy & 
0.188 & 0.14 \nl
\enddata
\tablenotetext{a}{See \S 5.1 for a description of the samples.}
\tablenotetext{b}{Probability that Data Set 1 and Data Set 2 are drawn from 
the same distribution.}
\end{deluxetable}

\clearpage
\begin{deluxetable}{llllcc}
\tablewidth{0pc}
\tablecaption{KS Tests of Nearest Neighbor Distributions: Weaker vs. 
Stronger Lines\label{ksweak}}
\tablehead{Ly$\alpha$ Sample\tablenotemark{a} & Hubble Flow & Data 
Set 1 & Data Set 2 & KS D & KS Probability\tablenotemark{b}}
\startdata
Complete & Perturbed & Galaxy-Ly$\alpha \ (W_{\rm r} > $ 100 m\AA ) & 
Galaxy-Random & 0.380 & 3.2$\times 10^{-3}$ \nl
Complete & Perturbed & Galaxy-Ly$\alpha \ (W_{\rm r} < $ 100 m\AA ) & 
Galaxy-Random & 0.308 & 7.6$\times 10^{-3}$ \nl
Complete & Pure      & Galaxy-Ly$\alpha \ (W_{\rm r} > $ 100 m\AA ) & 
Galaxy-Random & 0.345 & 9.9$\times 10^{-3}$ \nl
Complete & Pure      & Galaxy-Ly$\alpha \ (W_{\rm r} < $ 100 m\AA ) & 
Galaxy-Random & 0.288 & 0.015 \nl
Complete & Perturbed & Galaxy-Ly$\alpha \ (W_{\rm r} > $ 100 m\AA ) & 
Galaxy-Galaxy & 0.133 & 0.85 \nl
Complete & Perturbed & Galaxy-Ly$\alpha \ (W_{\rm r} < $ 100 m\AA ) & 
Galaxy-Galaxy & 0.160 & 0.48 \nl
Complete & Pure      & Galaxy-Ly$\alpha \ (W_{\rm r} > $ 100 m\AA ) & 
Galaxy-Galaxy & 0.255 & 0.13 \nl
Complete & Pure      & Galaxy-Ly$\alpha \ (W_{\rm r} < $ 100 m\AA ) & 
Galaxy-Galaxy & 0.229 & 0.11 \nl
\hline
Complete & Perturbed & Galaxy-Ly$\alpha \ (W_{\rm r} > $ 200 m\AA ) & 
Galaxy-Random & 0.544 & 5.5$\times 10^{-3}$ \nl
Complete & Perturbed & Galaxy-Ly$\alpha \ (W_{\rm r} < $ 200 m\AA ) & 
Galaxy-Random & 0.333 & 1.9$\times 10^{-4}$ \nl
Complete & Pure      & Galaxy-Ly$\alpha \ (W_{\rm r} > $ 200 m\AA ) & 
Galaxy-Random & 0.283 & 6.0$\times 10^{-4}$ \nl
Complete & Pure      & Galaxy-Ly$\alpha \ (W_{\rm r} < $ 200 m\AA ) & 
Galaxy-Random & 0.277 & 3.3$\times 10^{-3}$ \nl
Complete & Perturbed & Galaxy-Ly$\alpha \ (W_{\rm r} > $ 200 m\AA ) & 
Galaxy-Galaxy & 0.366 & 0.15 \nl
Complete & Perturbed & Galaxy-Ly$\alpha \ (W_{\rm r} < $ 200 m\AA ) & 
Galaxy-Galaxy & 0.143 & 0.43 \nl
Complete & Pure      & Galaxy-Ly$\alpha \ (W_{\rm r} > $ 200 m\AA ) & 
Galaxy-Galaxy & 0.200 & 0.054 \nl
Complete & Pure      & Galaxy-Ly$\alpha \ (W_{\rm r} < $ 200 m\AA ) & 
Galaxy-Galaxy & 0.213 & 0.064 
\enddata
\tablenotetext{a}{See \S 5.1 for a description of the samples.}
\tablenotetext{b}{Probability that Data Set 1 and Data Set 2 are drawn from 
the same distribution.}
\end{deluxetable}
\clearpage

\begin{figure}
\caption{ (Following page) High signal-to-noise ultraviolet spectrum 
of H 1821+643 obtained with the Goddard High Resolution Spectrograph on 
the {\it Hubble Space Telescope} with a resolution of $\sim$150 \kms\ 
(FWHM). The observed count rate is plotted versus heliocentric wavelength 
(in \AA ). Absorption lines marked with an L are extragalactic \ion{H}{1} 
Lyman $\alpha$ lines, lines marked G are due to the Galactic ISM, and other 
extragalactic lines are explicitly labeled. The continuum placement is 
indicated with a dotted line. The dashed line shows the zero flux level, and 
the solid line near zero is the 3$\sigma$ error spectrum. The flux uncertainty 
increases at $\lambda \ <$ 1251.6 \AA\ and $\lambda \ >$ 1537.9 \AA\ 
because some of the FP-SPLIT subexposures are shifted off the end of the 
detector array in these wavelength regions. \label{h1821grand}}
\end{figure}

\begin{figure}
\figurenum{1}
\plotone{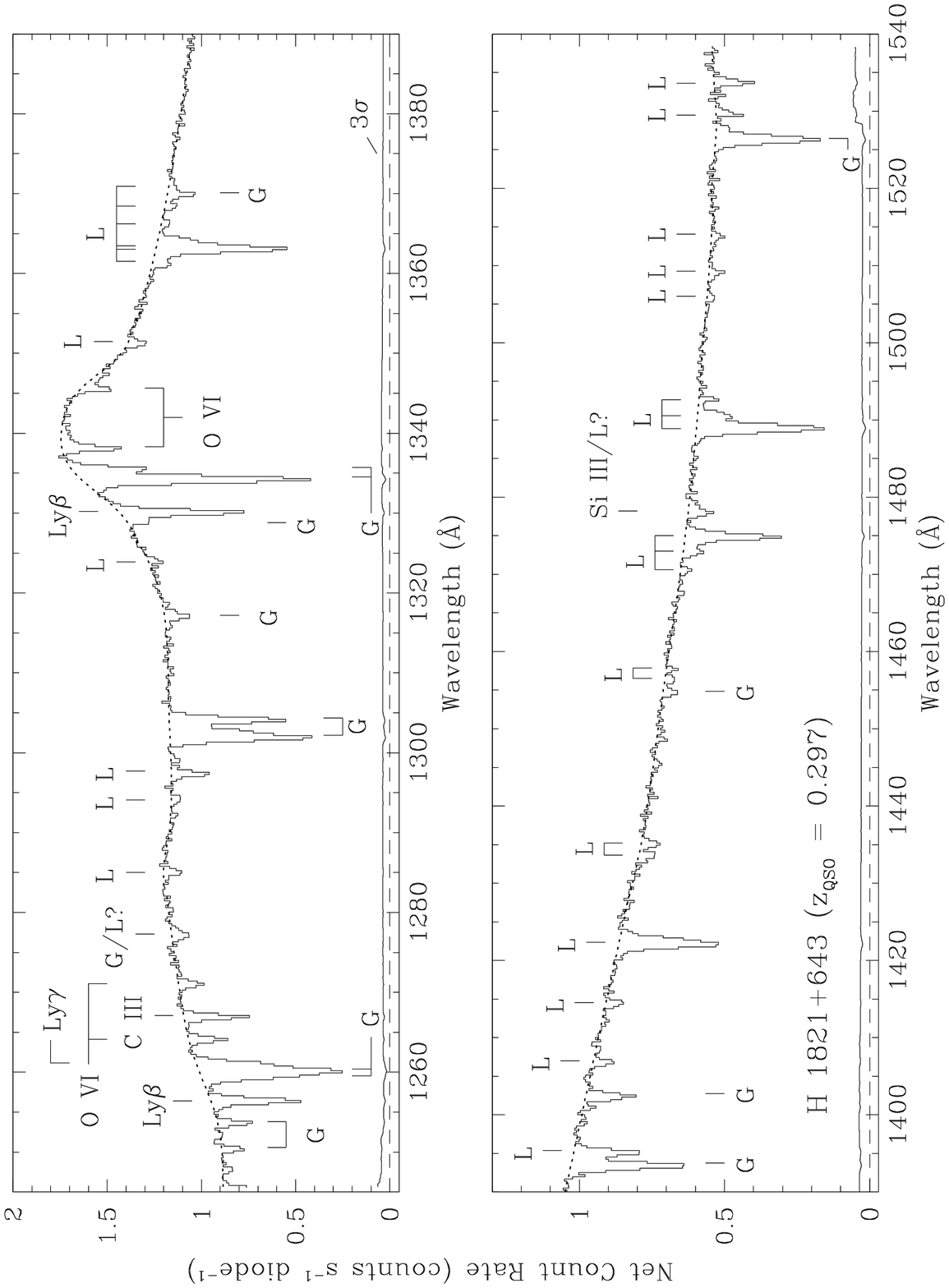}
\caption{ \ }
\end{figure}

\begin{figure}
\plotone{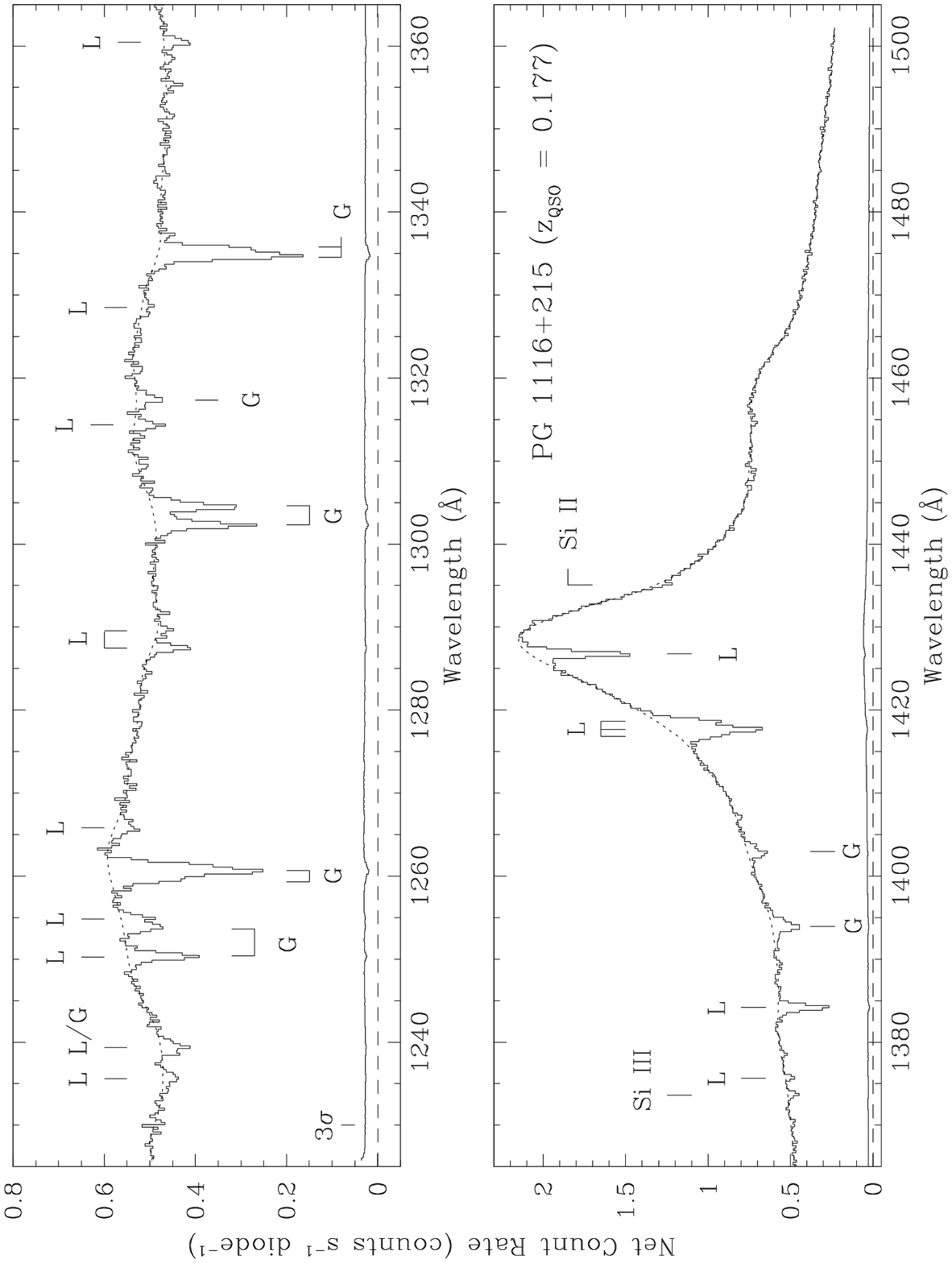}
\caption{High S/N ultraviolet spectrum of PG 1116+215 obtained with the 
G140L grating of the GHRS (see Figure 1 caption).\label{p1116grand}}
\end{figure}

\begin{figure}
\plotone{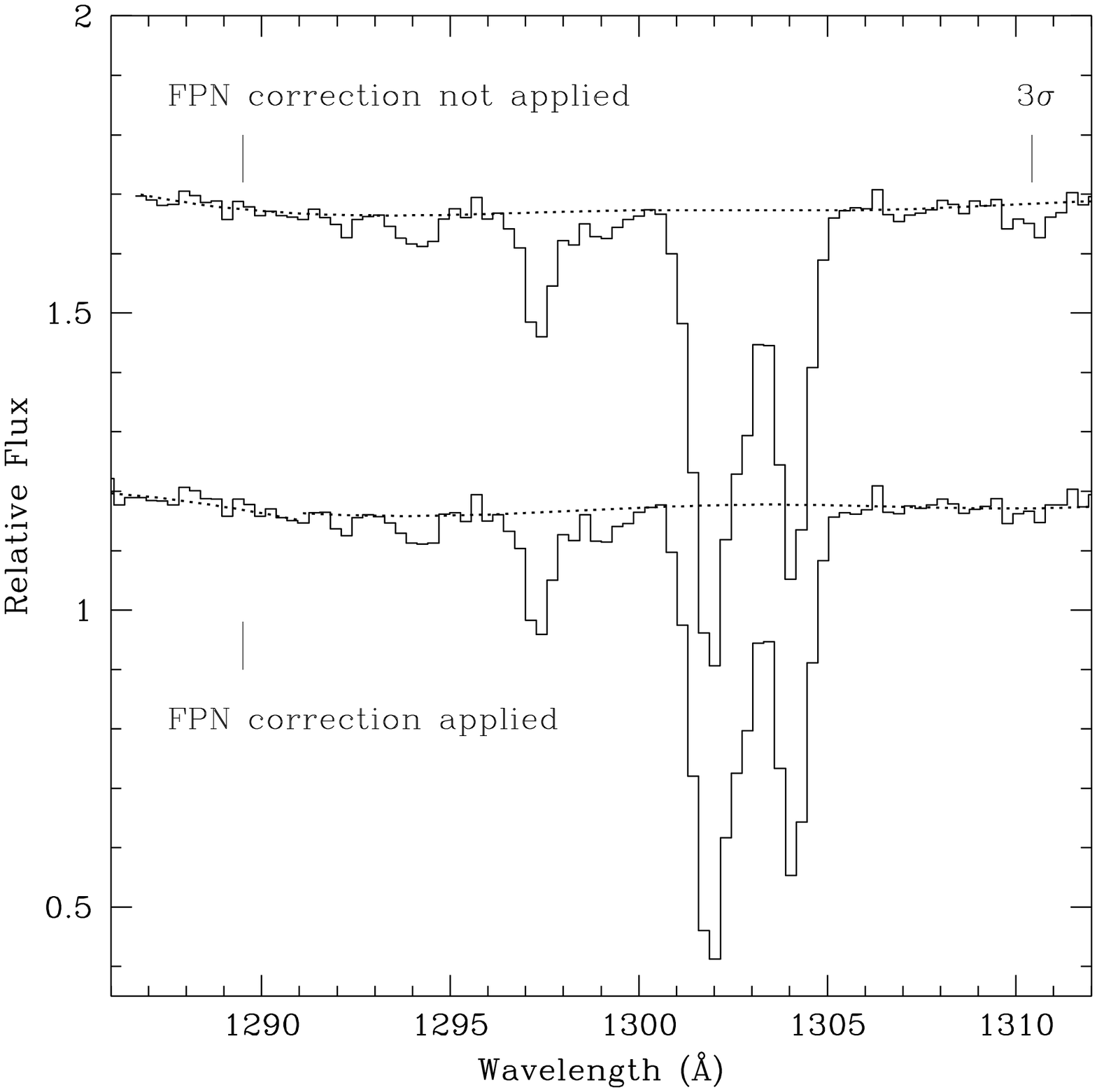}
\caption{Examples of the H 1821+643 spectrum before (upper) and 
after (lower) application of the fixed pattern noise correction; for clarity an 
offset of 0.5 flux units has been added to the upper spectrum. Most 
of the absorption lines are not dramatically affected by the FPN 
correction, but the 3$\sigma$ feature at 1310.4 \AA\ is almost 
entirely removed by the correction.\label{fpndemo}}
\end{figure}

\begin{figure}
\epsscale{1.05}
\plotone{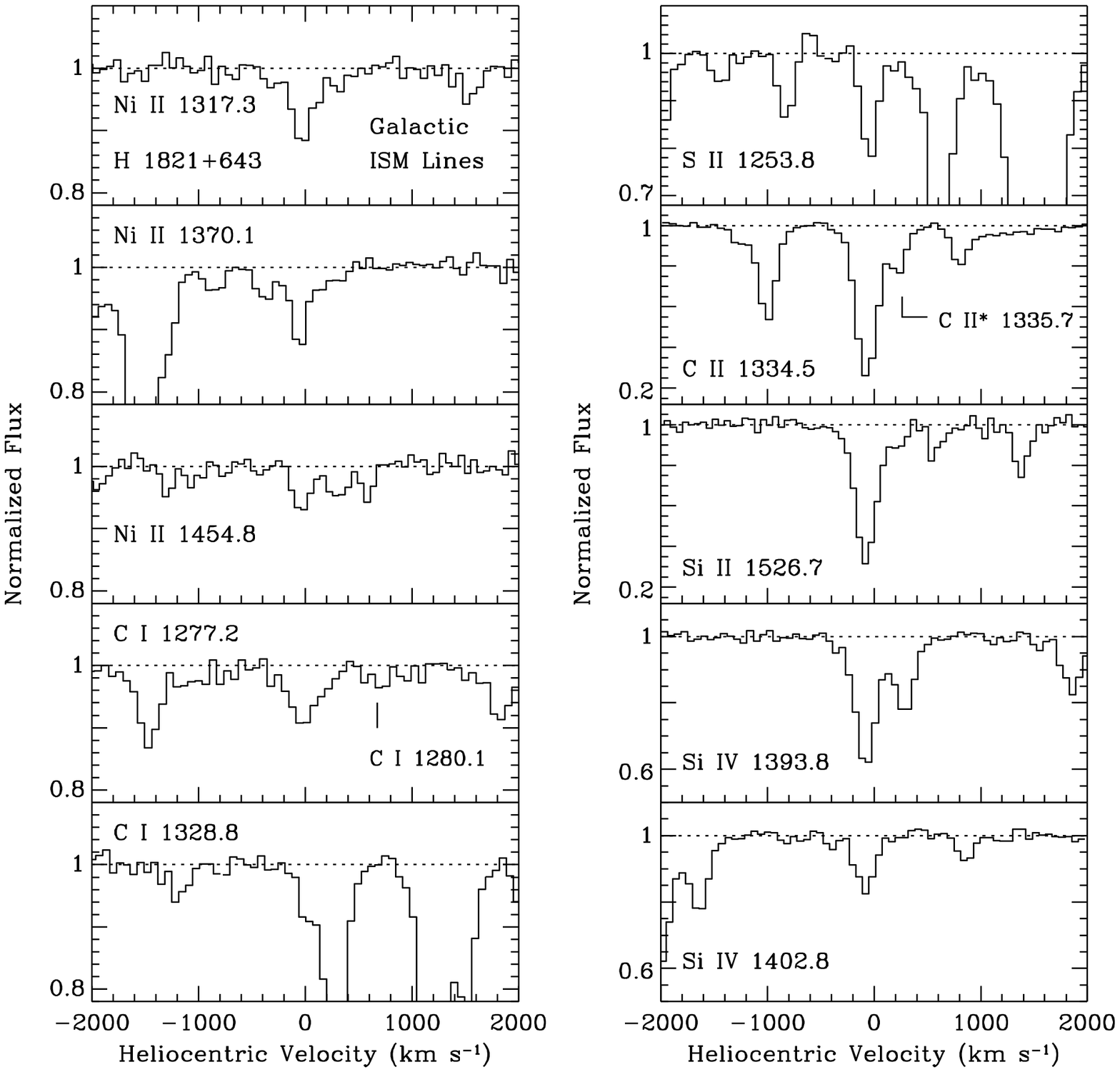}
\caption{Normalized absorption profiles of Galactic interstellar 
medium lines detected in the GHRS G140L spectrum of H 
1821+643. Notice that the y-axis scales are not the same in all of the 
panels. In many cases unrelated lines appear in the same panel due to the 
large velocity range displayed.\label{h1821ism}}
\end{figure}

\begin{figure}
\plotone{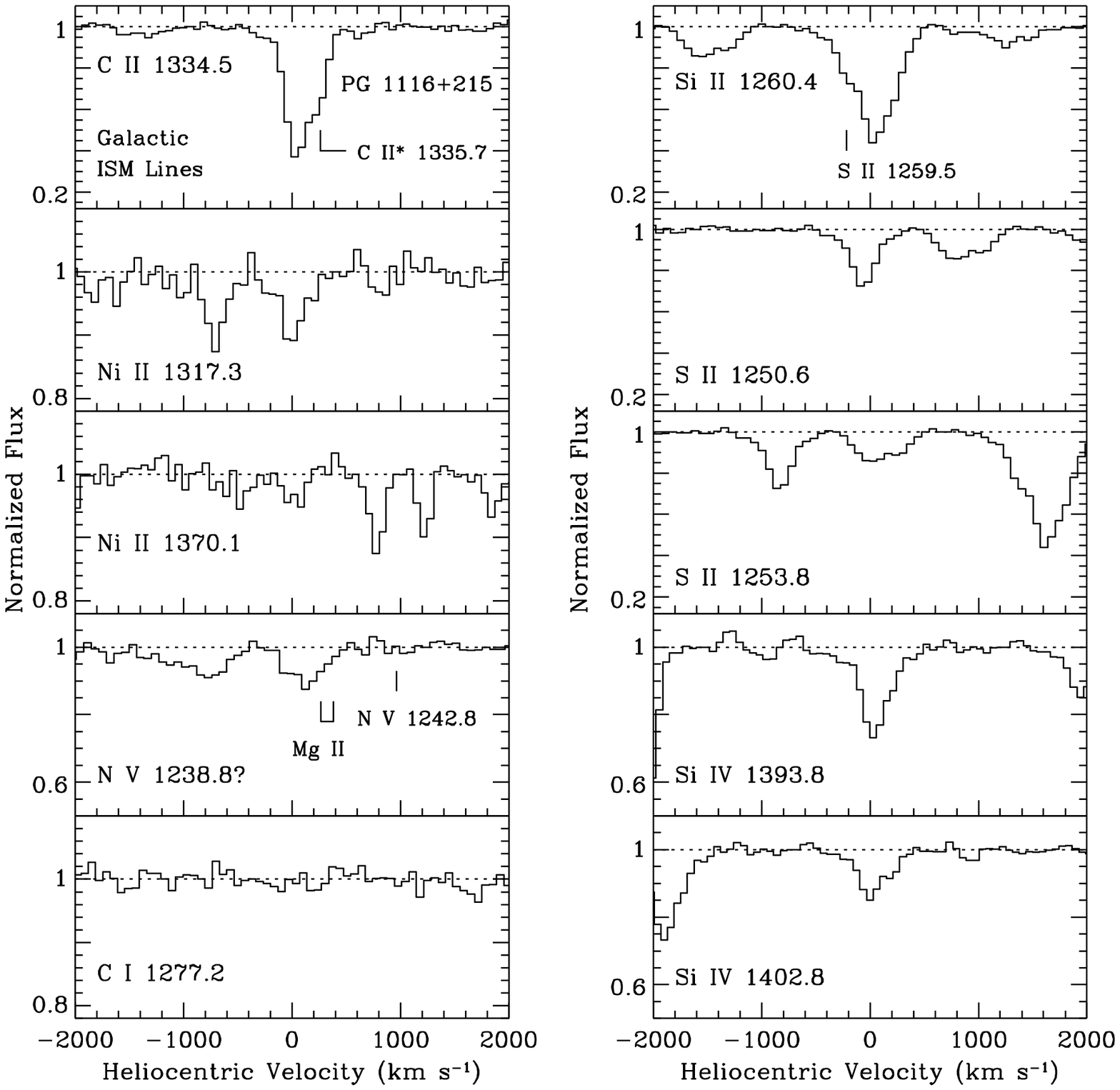}
\caption{Normalized absorption profiles of Galactic interstellar 
medium lines detected in the GHRS G140L spectrum of PG 1116+215. The 
y-axis scales are not the same in all of the 
panels. In many cases unrelated lines appear in the same panel due to the 
large velocity range displayed.\label{pg1116ism}}
\end{figure}

\begin{figure}
\plotone{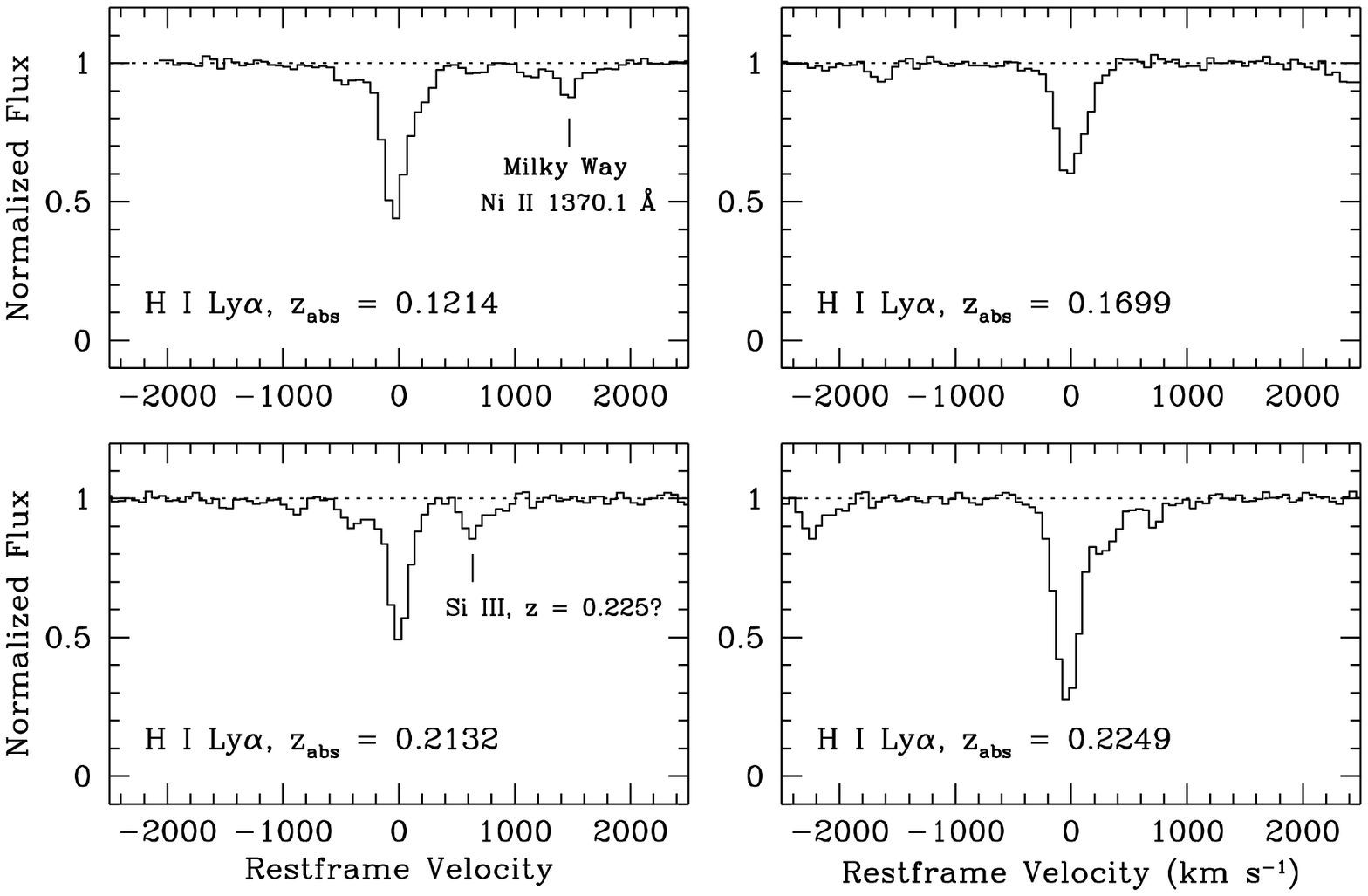}
\caption{Normalized absorption profiles of the four strongest 
\ion{H}{1} \lya lines detected in the GHRS G140L spectrum of H 
1821+643. Note the multiple weak \lya lines clustered around most of the 
strong lines.\label{lyapros}}
\end{figure}

\begin{figure}
\epsscale{1.0}
\plotone{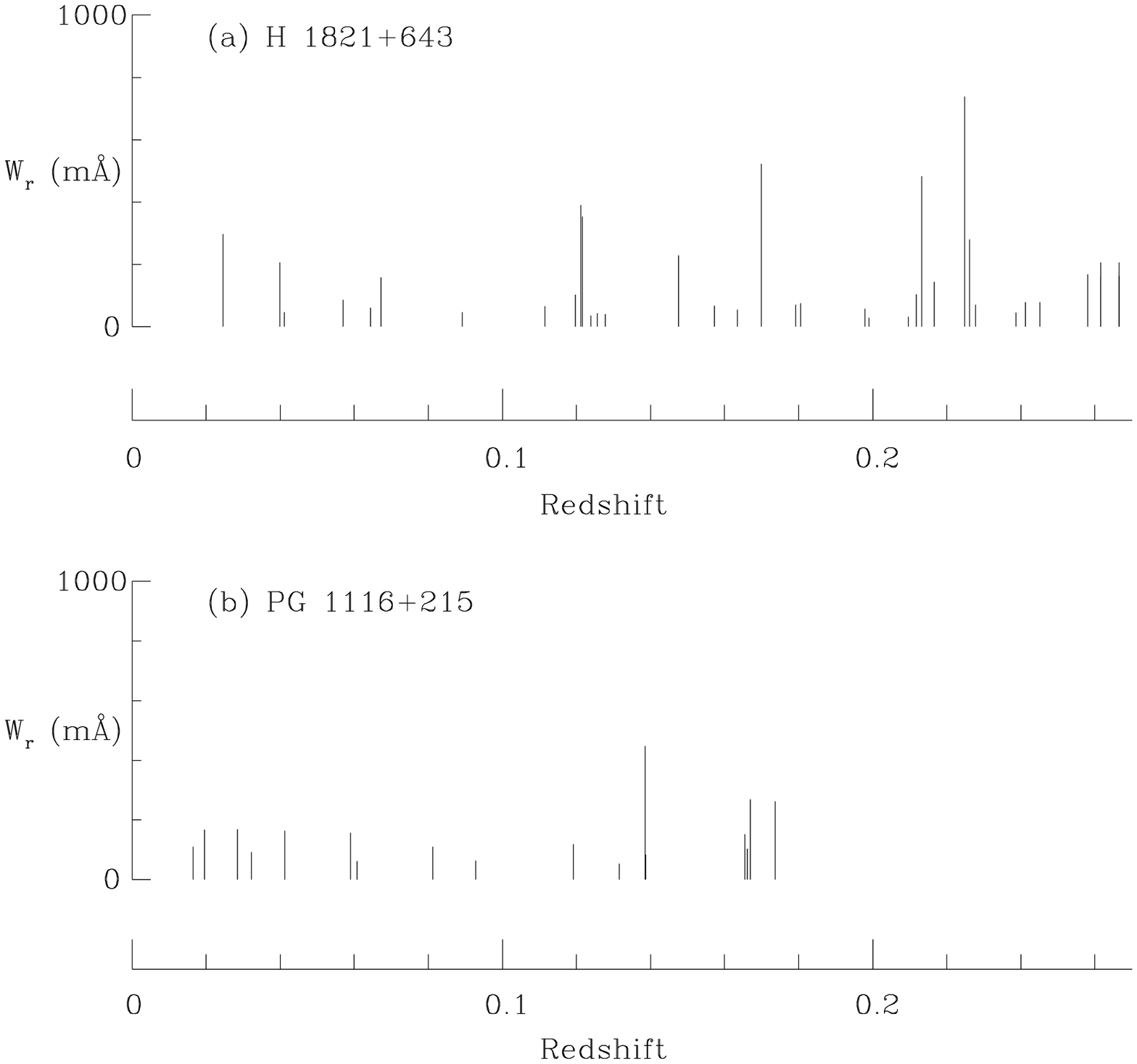}
\caption{Rest equivalent widths versus redshift for the (a) H 1821+643 \lya 
lines, and (b) PG 1116+215 \lya lines. Note the tendency for strong lines to 
have multiple weak lines around them.\label{lyastick}}
\end{figure}

\begin{figure}
\plotone{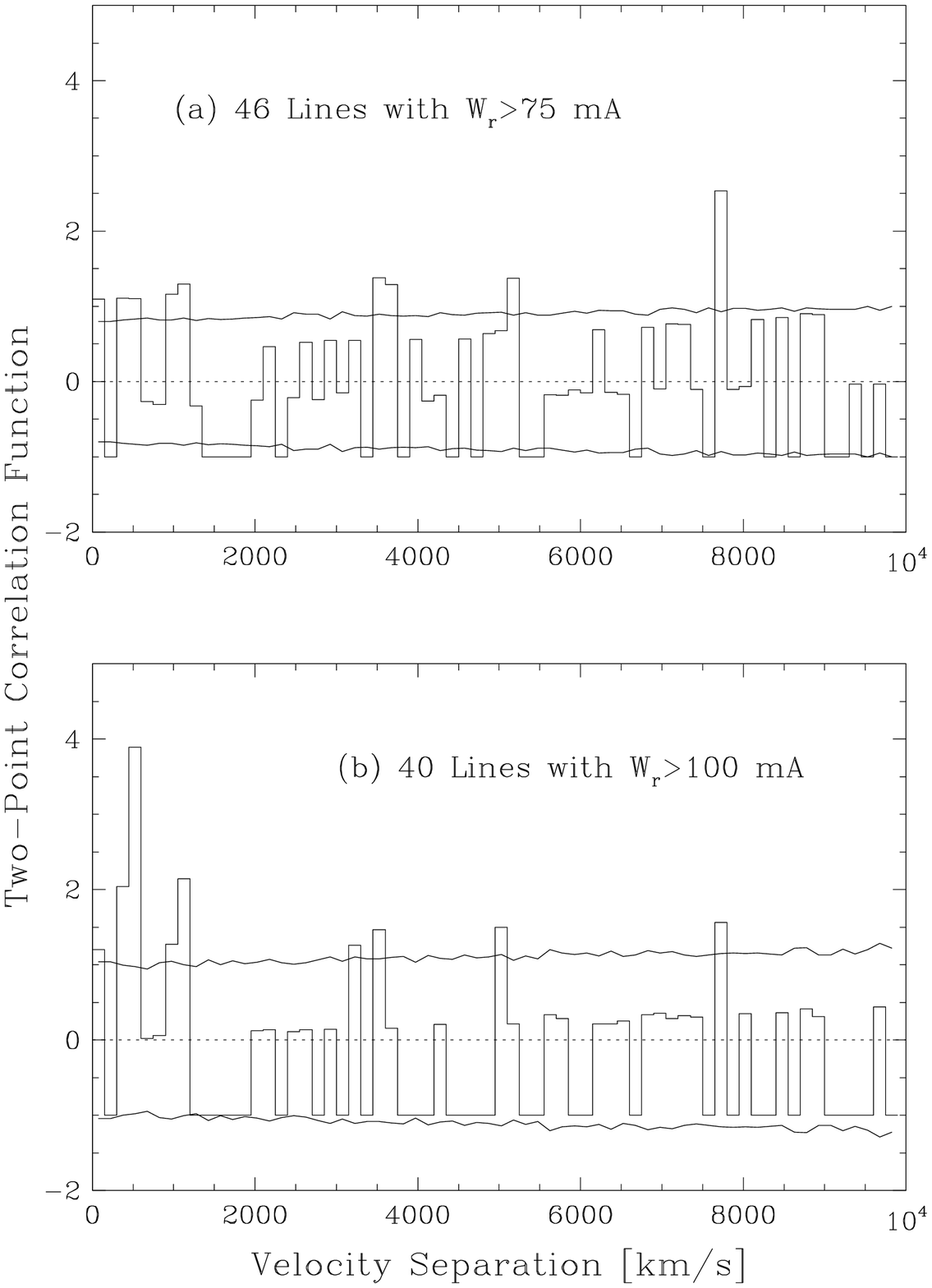}
\vspace{-1.5cm}
\caption{Two point correlation function calculated with (a) all lines with 
$W_{\rm r} >$ 75 m\AA\ (not including some sight lines which have insufficient 
sensitivity to detect lines as weak as 75 m\AA ), and (b) all lines 
with $W_{\rm r} >$ 100 m\AA\ (including all sight lines). The histogram 
shows the correlation function, and the solid lines show the $\pm 1\sigma$ 
fluctuations expected for unclustered lines.\label{twop}}
\end{figure}

\begin{figure}
\epsscale{1.07}
\plotone{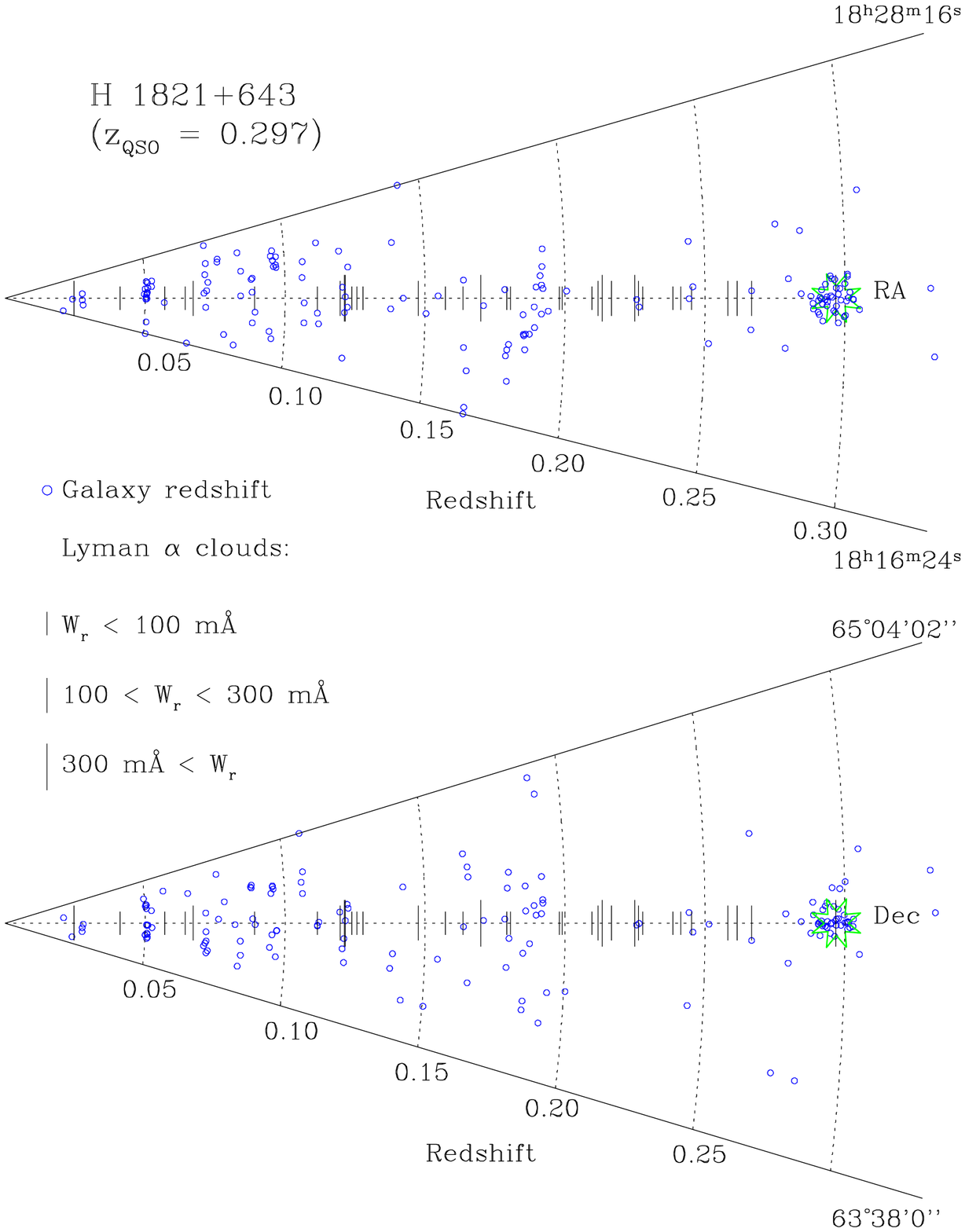}
\vspace{-3cm}
\caption{RA and Dec slices showing the location of the \lya absorption lines 
(vertical lines) with respect to the galaxies with measured redshifts
from the WIYN survey and the literature (open circles) in the direction of H 
1821+643. The QSO is indicated with a large open star, and the angles have 
been exaggerated by a factor of 15.\label{pie1821}}
\end{figure}

\begin{figure}
\epsscale{1.06}
\plotone{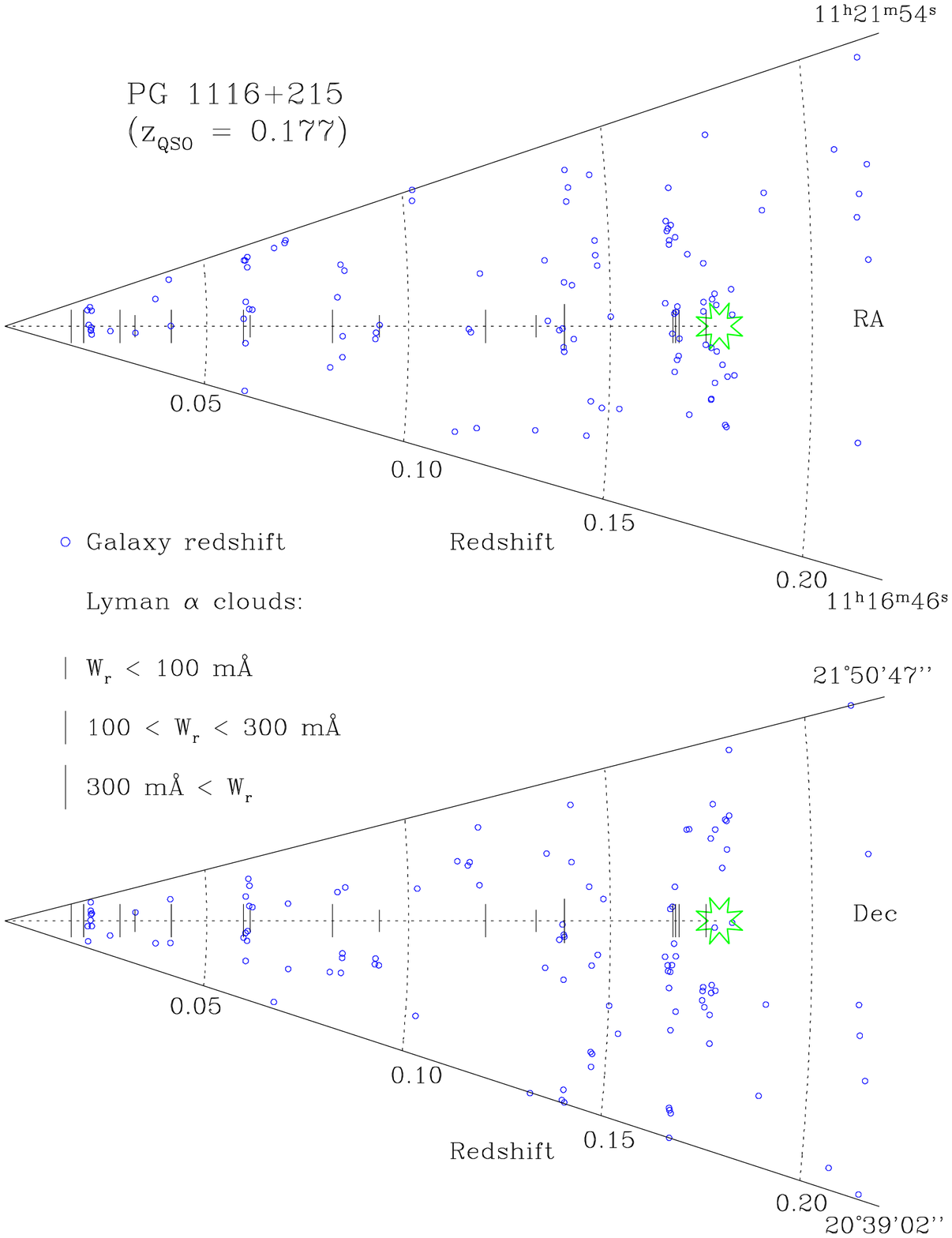}
\vspace{-2.7cm}
\caption{RA and Dec slices showing the location of the \lya absorption lines 
(vertical lines) with respect to the galaxies with measured redshifts
from the WIYN survey (open circles) in the direction of PG 1116+215. The 
QSO is indicated with a large open star, and the angles have been 
exaggerated by a factor of 15.\label{pie1116}}
\end{figure}

\begin{figure}
\epsscale{1.0}
\plotone{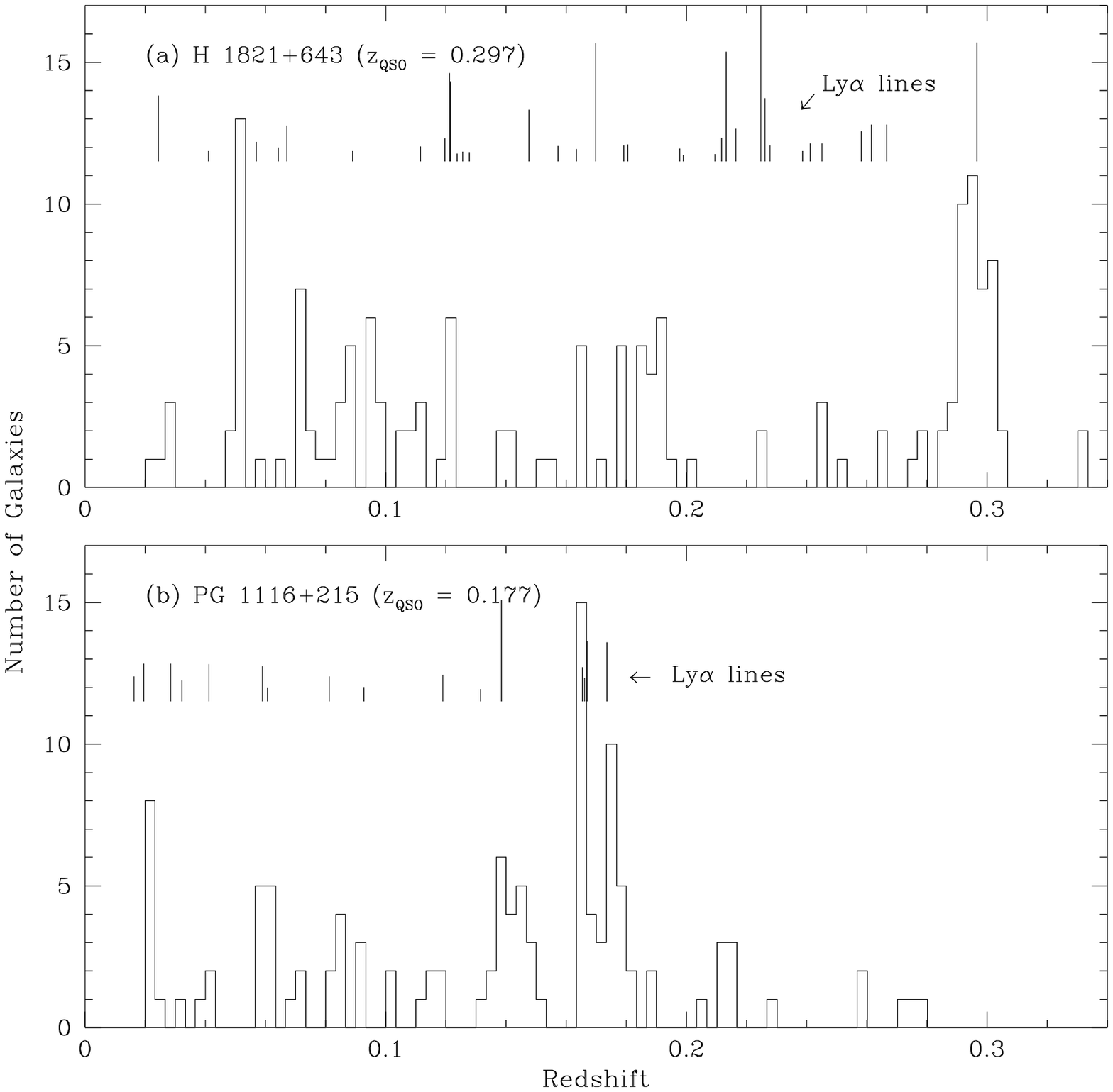}
\caption{Histogram of galaxy redshifts measured in the fields of (a) H 
1821+643, and (b) PG 1116+215. Also shown at the top of each panel are 
the redshifts of the Lyman $\alpha$ clouds, indicated with a vertical line with 
length proportional to the cloud rest equivalent width.\label{histreds}}
\end{figure}

\begin{figure}
\plotone{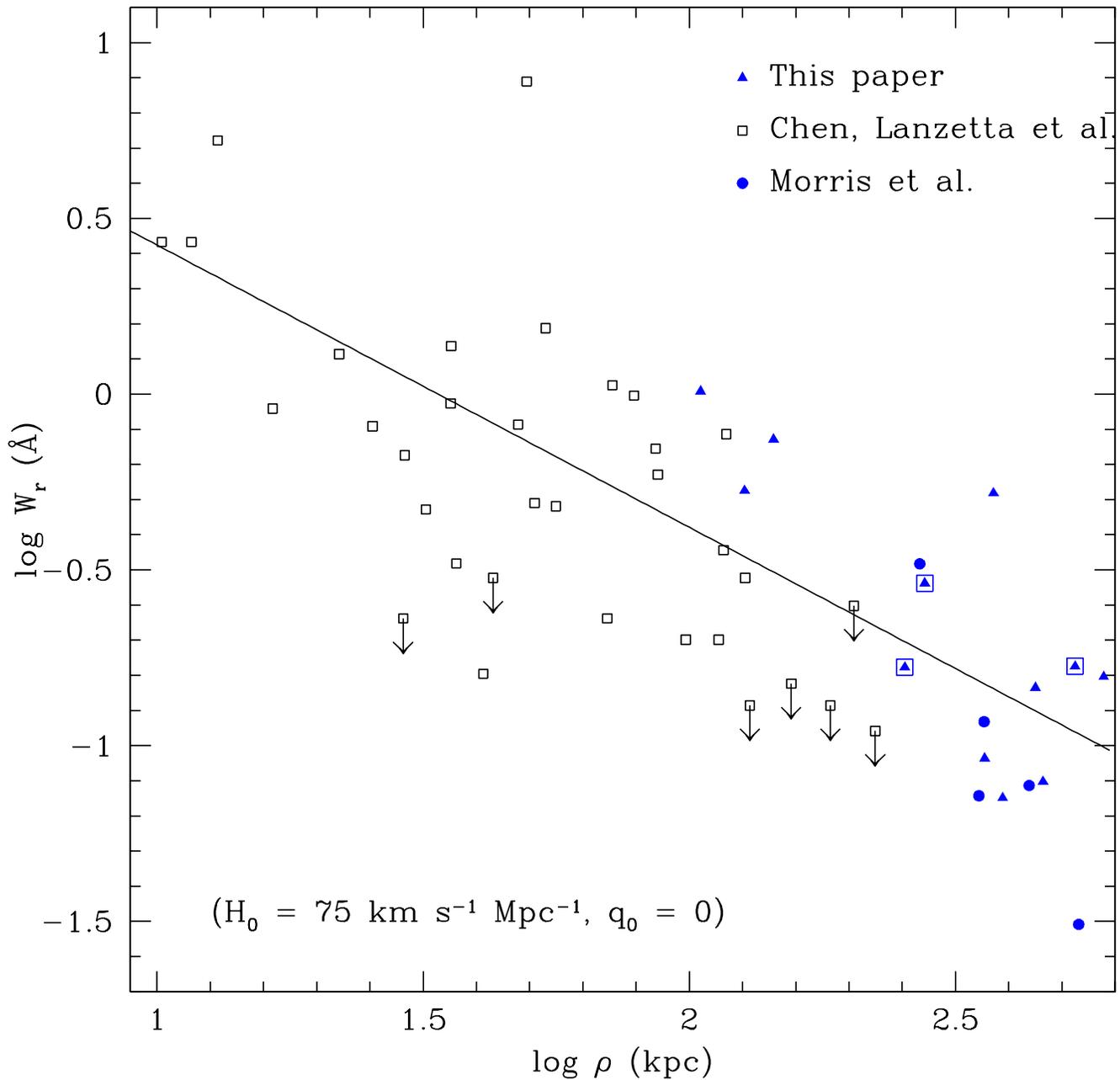}
\caption{Lyman $\alpha$  cloud rest equivalent width vs. galaxy impact 
parameter, for galaxies with impact parameters of 600 kpc or less, from this 
paper, Chen et al. (1998), and Morris et al. (1993). The Spearman 
rank-order correlation coefficient is $r_{s}$ = -0.762, which indicates that 
these 
data are anticorrelated at the 7.6$\sigma$ level. The best-fit power law is 
shown with a solid line. Note that all 17 galaxies with $\rho \ \leq$ 600 kpc 
from this paper and Morris et al. (i.e., the 3C 273, H 1821+643, and PG 
1116+215 sight lines) have an associated \lya line with $\Delta v \ <$ 1000 
\kms . There are only three galaxy-absorber pairs with 500 $< \Delta v <$ 
1000 \kms ; these are encased in squares.\label{rhocor}}
\end{figure}

\begin{figure}
\plotone{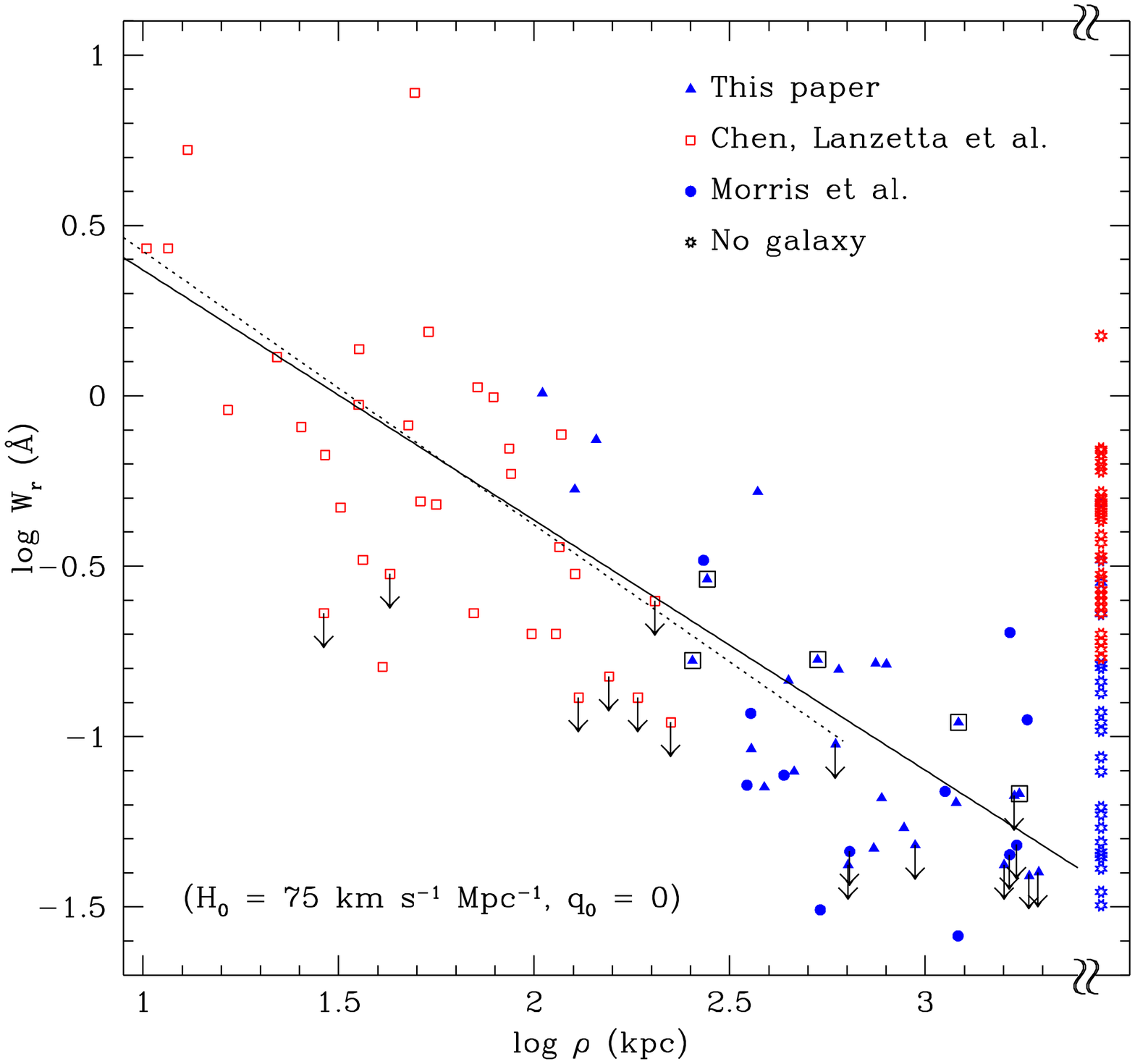}
\caption{Same as Figure 12, but including all galaxies with impact 
parameters of 2100 kpc or less. The best-fit power law is indicated with a 
solid line. The best-fit power law from Figure 12, which is shown with a 
dotted line, is very similar. There are only five galaxy-absorber pairs with 
500 $< \Delta v <$ 1000 \kms ; these are encased in squares. The open stars 
on the right side of the plot are \lya lines from our sight lines and some 
of the 
\lya lines from the Chen et al. (1998) sight lines which do not have associated 
galaxies and consequently are excluded by the selection criteria. These may 
not have associated galaxies because the appropriate redshift survey is 
incomplete, but if these excluded lines are matched with with galaxies at 
large values of $\rho$, then they will significantly weaken the 
anticorrelation.\label{bigrhocor}}
\end{figure}

\begin{figure}
\plotone{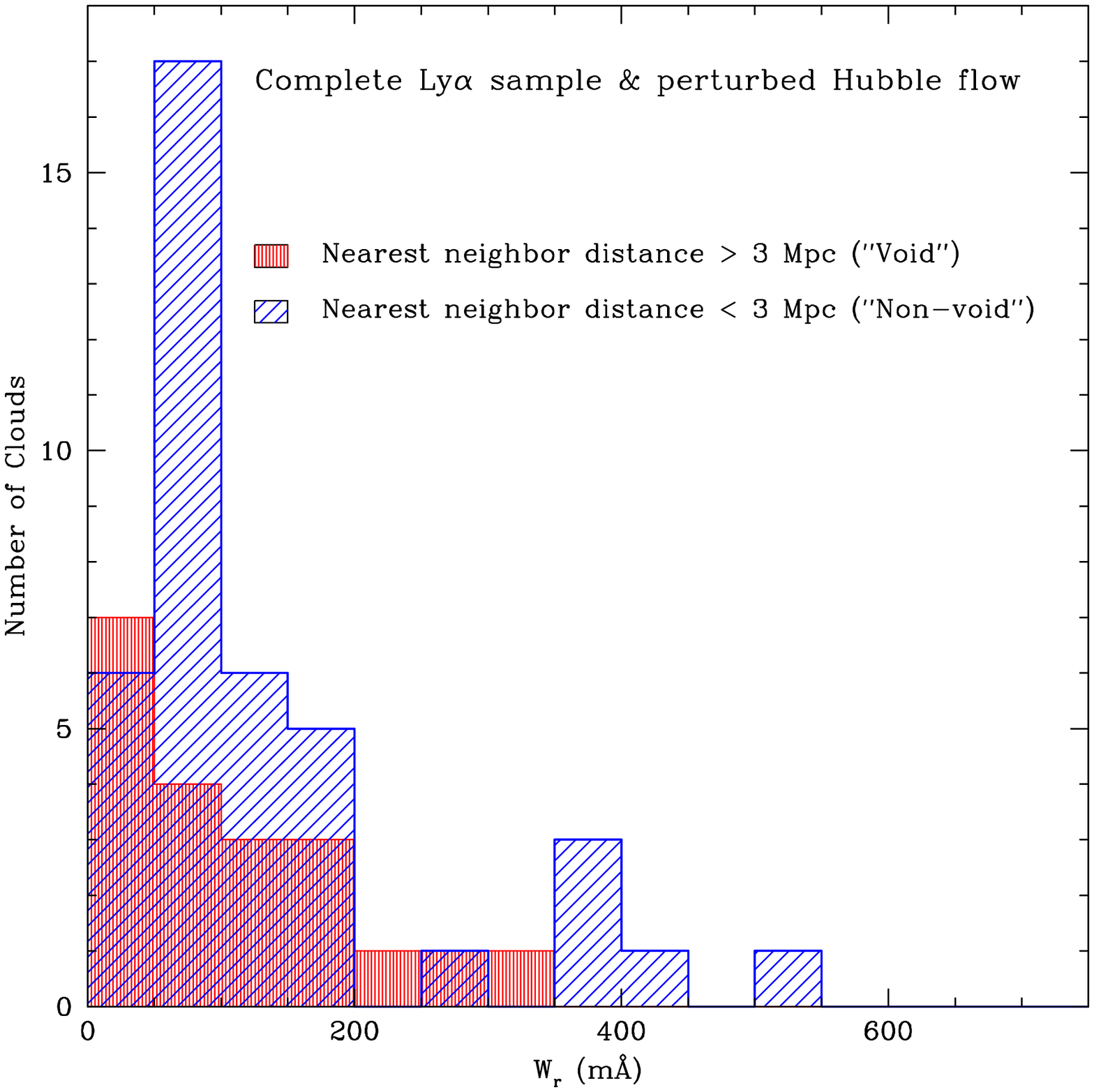}
\caption{Equivalent width distribution of ``void'' \lya absorbers vs. ``non-
void'' \lya absorbers (see text \S 5.2).\label{voidnot}}
\end{figure}

\begin{figure}
\plottwo{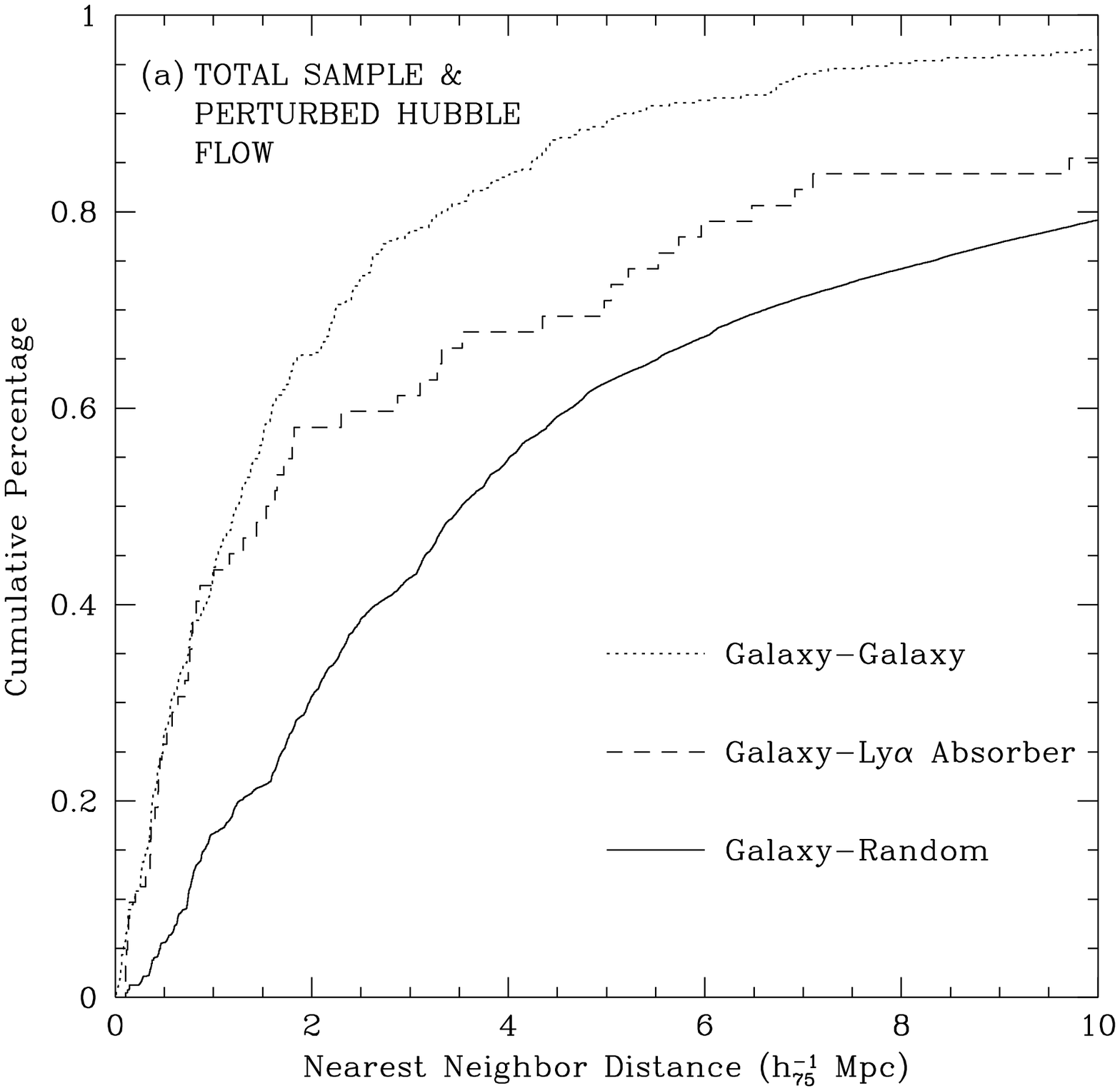}{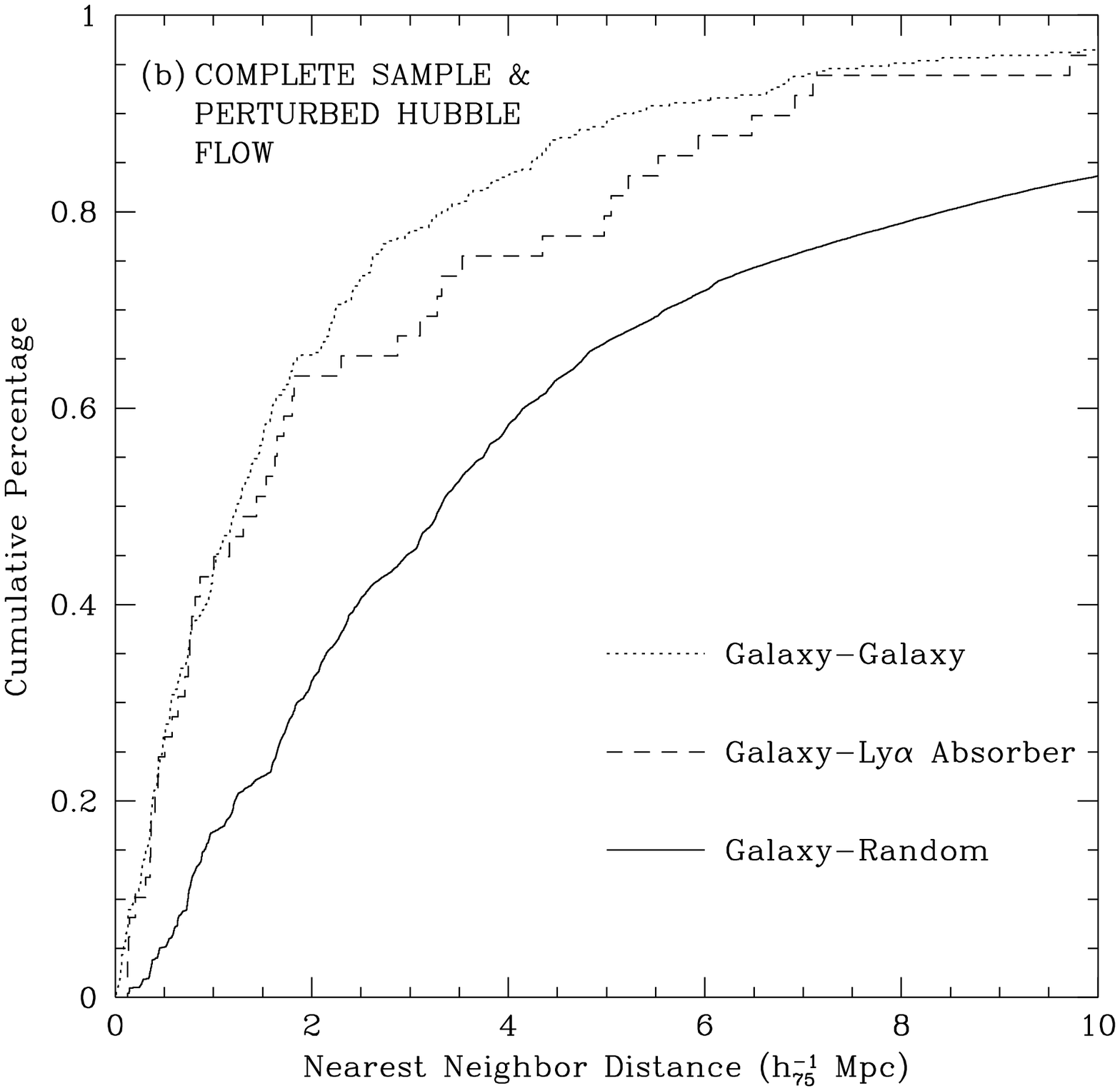}
\caption{The galaxy-Lyman $\alpha$ absorber nearest neighbor distance 
cumulative distribution (dashed line) compared to the galaxy-galaxy 
distribution (dotted line) and the Monte Carlo galaxy-random redshift 
distribution (solid line) for (a) the total \lya sample, and (b) the 
complete 
\lya sample. In both cases the radial distances were calculated with the 
perturbed Hubble flow.\label{cumdist}}
\end{figure}

\begin{figure}
\plotone{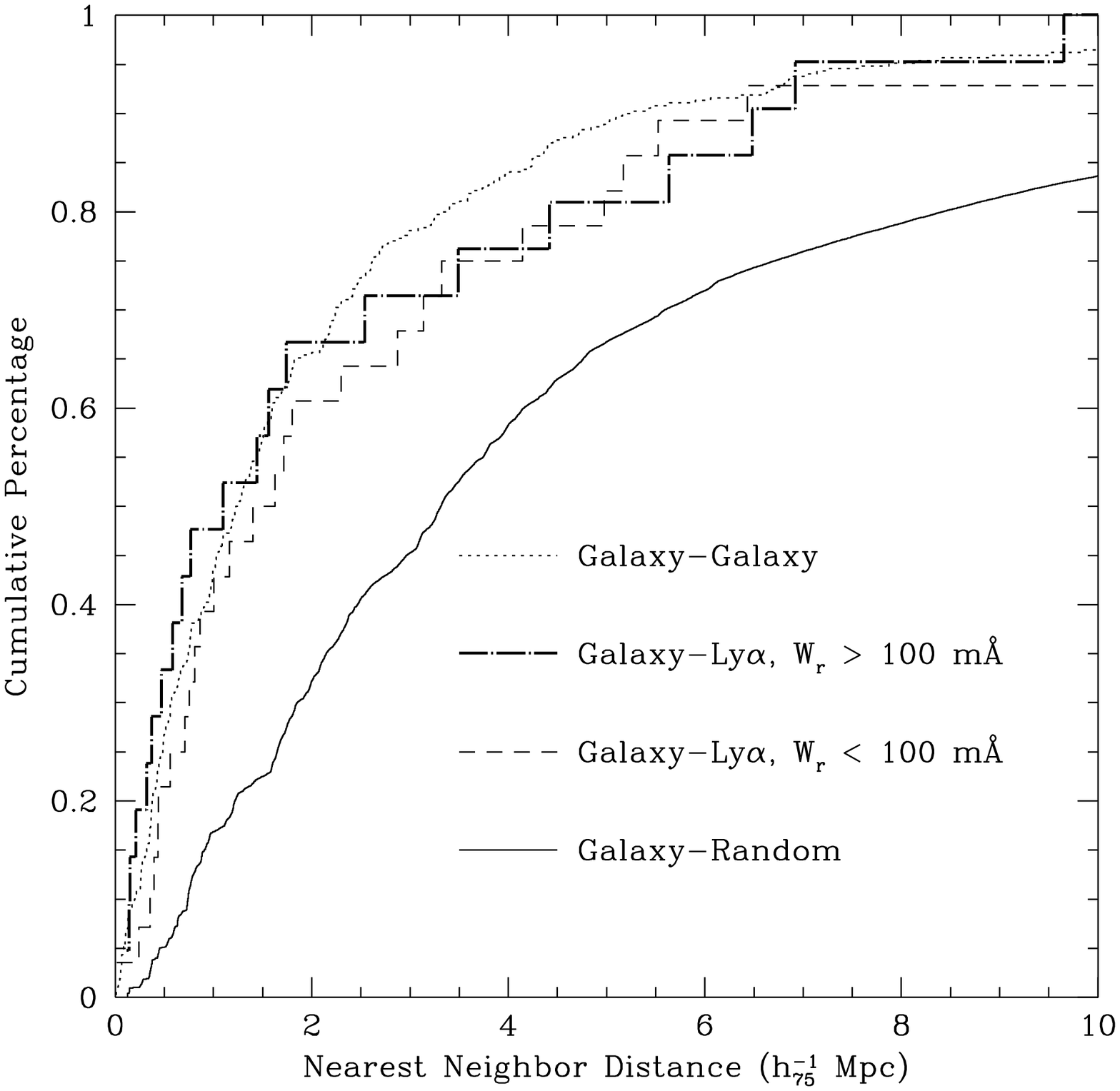}
\caption{Same as Figure 15b but showing the complete sample divided into 
strong clouds with $W_{\rm r} \ >$ 100 m\AA\ (dash-dot line) and weak 
clouds with $W_{\rm r} \ <$ 100 m\AA\ (short dash line).\label{cumweak}}
\end{figure}

\begin{figure}
\plotone{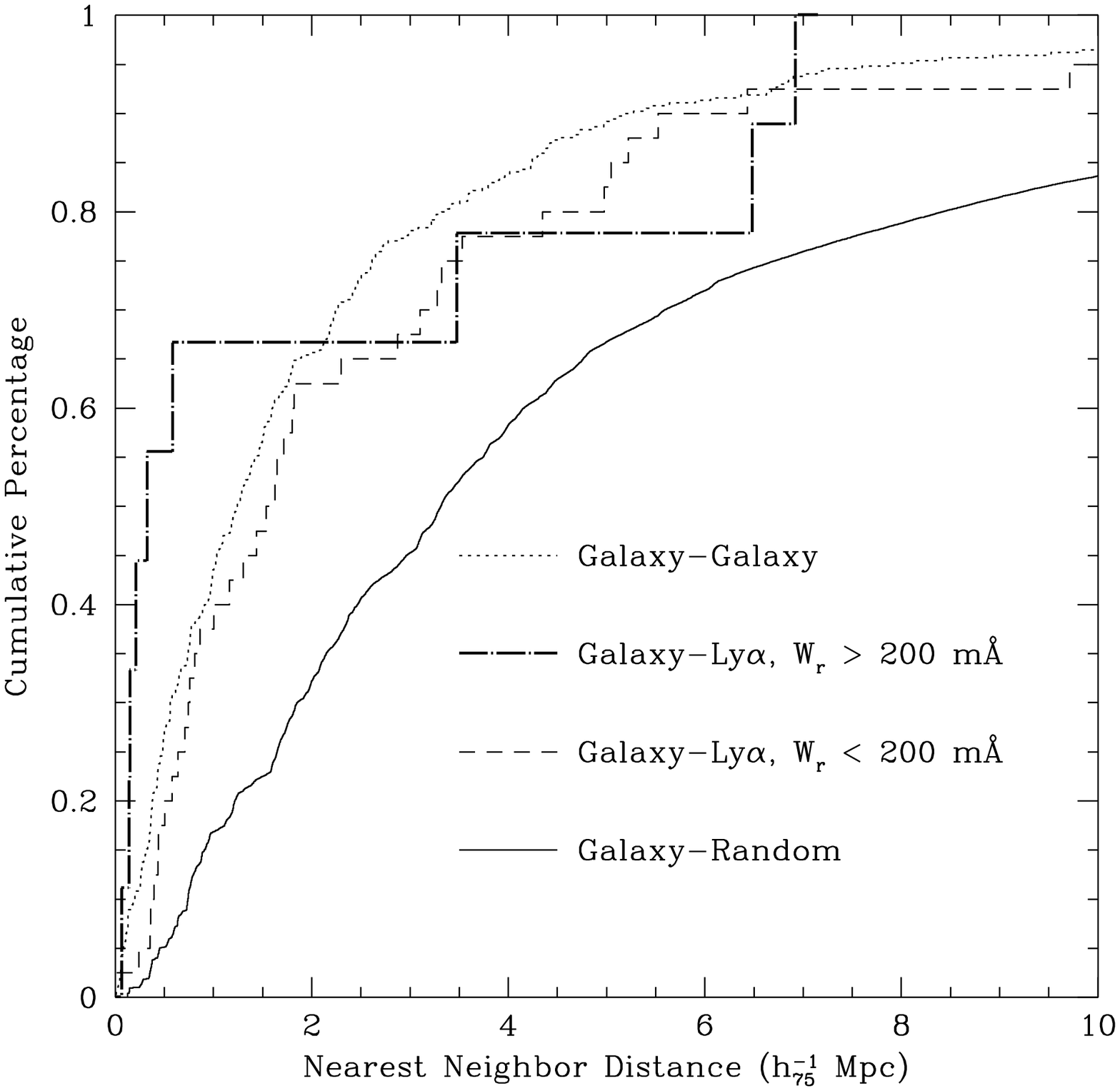}
\caption{Same as Figure 15b but showing the complete sample divided into 
strong clouds with $W_{\rm r} \ >$ 200 m\AA\ (dash-dot line) and weak 
clouds with $W_{\rm r} \ <$ 200 m\AA\ (short dash line).\label{weak200}}
\end{figure}

\end{document}